\documentclass[twocolumn,pra,aps,floatfix,showpacs]{revtex4}
\usepackage{amsmath} \usepackage{amssymb}
\usepackage{graphicx} \usepackage{esint}
\usepackage{doi} \usepackage{hyperref} \usepackage{braket}
\usepackage{lipsum} \usepackage{appendix}

\makeatletter

\usepackage{times} \newcommand{\bk}{\mathbf{k}}
\newcommand{\br}{\mathbf{r}} \newcommand{\ee}{\mathrm{e}}
\newcommand{\ii}{\mathrm{i}}

\makeatother
\begin{document}

\title{
Interplay between Rashba spin-orbit coupling and adiabatic rotation in a two-dimensional Fermi gas
}
\author{E. Doko$^{1}$, A. L. Suba{\c s}{\i}$^{2}$, and M. Iskin$^{1}$}
\affiliation{$^{1}$Department of Physics, Ko{\c c} University, Rumelifeneri Yolu,
34450 Sar{\i}yer, Istanbul, Turkey. \\
$^{2}$Department of Physics, Faculty of Science and Letters, Istanbul
Technical University, 34469 Maslak, Istanbul, Turkey.}

\begin{abstract} 

We explore the trap profiles of a two-dimensional atomic Fermi gas in the presence 
of a Rashba spin-orbit coupling and under an adiabatic rotation. We first consider 
a non-interacting gas and show that the competition between the effects of Rashba 
coupling on the local density of single-particle states and the Coriolis effects caused 
by rotation gives rise to a characteristic ring-shaped density profile that survives 
at experimentally-accessible temperatures. Furthermore, Rashba splitting of the 
Landau levels takes the density profiles on a ziggurat shape in the rapid-rotation limit. 
We then consider an interacting gas under the BCS mean-field approximation for 
local pairing, and study the pair-breaking mechanism that is induced by the Coriolis 
effects on superfluidity, where we calculate the critical rotation frequencies 
both for the onset of pair breaking and for the complete destruction of superfluidity 
in the system. In particular, by comparing the results of fully-quantum-mechanical 
Bogoliubov-de Gennes approach with those of semi-classical local-density approximation, 
we construct extensive phase diagrams for a wide-range of parameter regimes in the trap 
where the aforementioned competition may, e.g., favor an outer normal edge that is 
completely phase separated from the central superfluid core by vacuum. 

\end{abstract}

\pacs{
03.75.Ss, 
03.75.Hh, 
67.85.Lm	
}

\date{\today}
\maketitle

\section{Introduction}
\label{sec:introduction}

Most of the exotic many-body phenomena observed in an atomic Fermi gas are 
triggered by a variety of couplings between single-particle states and 
externally-applied fields, and therefore, they relate directly to the single-particle 
properties of a normal (N) Fermi gas. For instance, appropriate couplings between 
the internal atomic degrees of freedom and laser fields have made it possible 
to create and engineer effective electromagnetic fields, i.e., artificial Abelian 
gauge fields, for neutral atoms~\cite{Dalibard2011}. Alternatively, since the 
effects of rotation are analogous to those of an effective magnetic field on a 
particle, where the Coriolis force on a neutral atom mimics the Lorentz force 
on a charged particle, such a coupling may be used to form Landau levels 
in a trapped Fermi gas exhibiting an integer quantum-Hall effect~\cite{Ho2000}. 
In addition, recent progress in creating effective spin-orbit couplings (SOC), 
i.e., artificial non-Abelian gauge fields, also opens the door for analogous cold-atom 
studies on quantum spin-Hall effect~\cite{Sinova2004, Qi2011} and topological 
insulators~\cite{Hasan2010, Qi2011}. Bringing such couplings together naturally 
generates further novel effects. For example, while the simultaneous presence 
of Rashba SOC and Zeeman field may give rise to an anomalous-Hall 
effect~\cite{Zhang2010}, the interplay of Rashba coupling and adiabatic rotation may lead 
to the formation of a ring-shaped annulus in a trapped Fermi gas~\cite{Doko2016}.

 SOC and related contemporary phenomena have arisen as some of the key 
components in the interdisciplinary contexts of modern many-body quantum 
systems, including the cold atoms, and their understanding premise in the possibility 
of engineering cutting-edge technologies that are based on topological solid-state 
materials. In this respect, the experimental realizations of an effective SOC 
via artificial gauge fields~\cite{Lin2011,Zhang2012,Wang2012,Cheuk2012,Qu2013,Goldman2014,Olson2014}, 
including the possibility of real-time control~\cite{Jimenez-Garcia2015}, extend 
the stage to investigate SOC physics in experimentally-controllable settings. 
Even though early experimental works were limited to a one-dimensional SOC, 
which may be considered as an equal-weight combination of Rashba and 
Dresselhaus couplings, a two-dimensional SOC has recently been 
realized~\cite{Huang2015} by using a three-laser Raman scheme, paving the way 
for the realization of a purely Rashba coupling. There are various other theoretical 
proposals that are based on magnetic or generalized Raman schemes for creating 
a Rashba coupling as well~\cite{Ruseckas2005,Stanescu2007,Dalibard2010,Campbell2011,Xu2012,Anderson2013,Campbell2016}.
Furthermore, motivated by the experimental realizations of a 2D Fermi gas~\cite{Martiyanov2010,Dyke2011,Frohlich2011,Feld2011,Sommer2012,Ries2015},
the effects of SOC on a 2D Fermi gas have recently been the subject of many 
theoretical works~\cite{Zhou2011a,He2012,Gong2012,Yang2012,Takei2012,Ambrosetti2014,Zhang2013,Iskin2013,Cao2014}. 

An essential signature of a trapped atomic superfluid (SF) is the appearance 
of quantized vortices when the system is rotated~\cite{Abo2001,Zwerlein2005},
i.e., the vortex cores consist of rotating N atoms with quantized angular 
momenta, as the rotation gradually destroys the SF phase by breaking
the time-reversal symmetry. In particular, when the rotation is introduced 
adiabatically without exciting vortices in a SF Fermi gas, some of the 
Cooper pairs may be broken due to the Coriolis effects and form a rotating 
N edge carrying the resultant angular momentum. This possibility 
was first proposed for a 3D resonant Fermi gas using the energy densities 
obtained from the Monte-Carlo simulations together with a local-density 
approximation (LDA) for the trap~\cite{Bausmerth2008a,Bausmerth2008}. 
It was shown that an adiabatic rotation gives rise to a phase separation 
between the non-rotating SF at the center and a rigidly-rotating N gas at 
the edge, which was further supported by the results of both LDA~\cite{Urban2008} 
and Bogoliubov-de Gennes (BdG)~\cite{Iskin2009} approaches that are
based on the microscopic BCS mean-field theory. The latter works also showed that 
the central SF core and the outer N edge are connected by a coexistence 
region, i.e., a partially-rotating gapless SF (gSF) phase in between. Furthermore, 
such a pair-breaking scenario has shown to be energetically preferred against 
the vortex formation in a sizeable parameter regime even in the absence of 
the adiabaticity assumption~\cite{Warringa2011,Warringa2012}.

By assuming a BCS mean-field approximation for local pairing and an LDA 
for trap, we earlier this year have reported our initial results for the effects of 
an adiabatic rotation on a Rashba-coupled 2D Fermi gas~\cite{Doko2016}. 
In contrast to the non-rotating case, we showed that the pairing can either
be enhanced or suppressed via Rashba coupling in a rotating system, 
and that the gSF region may disappear entirely from the trap forming an outer 
ring-shaped N edge that is completely phase separated from the central SF 
core by vacuum. Here, we not only extend this LDA analysis to a wider 
parameter regime but also compare its results with those of BdG approach 
showing a perfect agreement for the most parts.

The rest of the paper is organized as follows. The details of the theoretical 
framework are given in Sec.~\ref{sec:for}, where we introduce the BCS 
mean-field formalism for pairing, and BdG and LDA approaches for the 
harmonic trap. Through a thorough analysis of the resultant self-consistency 
equations, we characterize the trap profiles of first a non-interacting Fermi 
gas in Sec.~\ref{sec:NI} and then an interacting one in Sec.~\ref{sec:MFT},
with a special emphasis on the formation of a characteristic ring-shaped 
N edge. In Sec.~\ref{sec:NAD}, we calculate the critical rotation frequencies 
both for the onset of pair breaking and for the complete destruction of superfluidity, 
and construct extensive phase diagrams for a wide-range of parameter regimes in the trap
demonstrating all possible phase profiles. We end the paper with a brief summary 
of our conclusions and outlook in Sec.~\ref{sec:conclusions}, followed by a short 
Appendix~\ref{sec:angmombasis} on the details of the BdG approach.

\section{Theoretical framework}
\label{sec:for}

To study the interplay between Rashba coupling and adiabatic rotation in a 2D Fermi 
gas, we may consider a harmonic-confinement potential that is isotropic in space 
for its simplicity, and a short-ranged (i.e., contact) attractive interaction that is most 
relevant in the cold-atom context. For this purpose, we start with the introduction 
of the parameters of the model Hamiltonian, and then derive the self-consistency 
equations for the fully quantum-mechanical BdG as well as the semi-classical LDA 
approaches, by restricting ourselves to the BCS mean-field approximation for pairing.

\subsection{Hamiltonian}
\label{sec:ham}

In the rotating frame of reference, the non-interacting part $H_\mathrm{ni}$ 
of the total grand-canonical Hamiltonian
$
H=H_\mathrm{ni}+H_\mathrm{int}
$
can be written as a sum of three terms
$
H_\mathrm{ni}=H_\mathrm{sho}+H_{\mathrm{rot}}+H_{\mathrm{soc}},
$
corresponding, respectively, to the contributions of the 
simple-harmonic-oscillator potential, adiabatic rotation and Rashba coupling. 
In particular, by denoting $\psi_{\br \sigma}^\dagger$ and $\psi_{\br \sigma}$
as the creation and annihilation operators for a pseudo-spin 
$\sigma \equiv \{\uparrow,\downarrow\} \equiv \{1/2,-1/2\}$ fermion at position 
$\mathbf{r} \equiv (x,y)$, the harmonic-oscillator term can be expressed as
\begin{equation}
 H_\mathrm{sho}=\sum_{\sigma}\int d^2\br\,
\psi_{\br \sigma}^{\dagger} \left( \frac{\mathbf{p}^{2}}{2M}
+V_{r}-\mu \right) \psi_{\br \sigma}, 
\end{equation} 
where $\mathbf{p}=-i\mathbf{\nabla}$ is the linear-momentum operator 
in units of $\hbar=1$, $M$ is the mass of the particles, $V_{r}=M\omega^{2}r^{2}/2$ 
is the harmonic potential with $\omega$ the trapping frequency, and 
$\mu$ is the chemical potential. Likewise, choosing the perpendicular ($z$) 
direction as the rotation axis, the adiabatic-rotation term can be expressed as
\begin{equation} 
H_\mathrm{rot}=-\Omega\sum_{\sigma}\int d^2\br\,
\psi_{\br \sigma}^{\dagger} L^{z}_{\br}\psi_{\br \sigma}, 
\end{equation}
where $\Omega \geq 0$ is the rotation frequency and $L_\br^{z}$ is the 
$z$-projection of the angular-momentum operator 
$\mathbf{L}_\br=\mathbf{r}\times \mathbf{p}$. Note that there is a 
well-known upper bound on $\Omega$ as the harmonic potential can only
trap the particles for $\Omega < \omega$. Lastly, the Rashba-coupling 
term can be expressed as 
\begin{equation}
H_\mathrm{soc}=\alpha\sum_{\sigma\sigma'}\int d^2\br\, \psi_{\br\sigma}^{\dagger} 
\left( \mathbf{p}\cdot\vec{\mathbf{\sigma}}\right)_{\sigma \sigma'} \psi_{\br \sigma'}, 
\end{equation} 
where $\alpha\ge 0$ is the strength of the spin-momentum coupling, and 
$\vec{\mathbf{\sigma}} \equiv (\sigma^x, \sigma^y)$ is a vector of Pauli 
spin matrices.

Lastly, the interacting part $H_\mathrm{int}$ of the total Hamiltonian can be 
expressed as
\begin{equation} 
H_\mathrm{int}=-g\int d^2\br\, \psi_{\br \uparrow}^{\dagger}\psi_{\br
\downarrow}^{\dagger}\psi_{\br \downarrow}\psi_{\br \uparrow},
\end{equation} 
where $g\geq0$ is the strength of the bare attraction between $\uparrow$ 
and $\downarrow$ particles. To make connection with the literature, 
we follow the usual convention, and relate $g$ to the two-body binding 
energy $E_b\ge0$ of $\uparrow$ and $\downarrow$ particles in vacuum 
via the relation
$
1/g=(1/A) \sum_{\textbf{k}}1/(2\epsilon_{\textbf{k}}+E_b),
$
where $A$ is the area of the system, and $\epsilon_{\bk}=k^{2}/(2M)$ is the 
free-particle dispersion with $k=|\bk|$ the magnitude of momentum $\mathbf{k}$. 
This leads to 
$
g = 4\pi/[M \ln(1+2E_c/E_b)]
$
in two dimensions, where $E_c$ is 
the energy cut-off used in the $\mathbf{k}$-space sum. We note that the
ultraviolet dependence on $E_c$ is a direct reflection of the zero-ranged 
nature of the contact interaction, and that none of our numerical results 
depend strongly on its specific value as long as it is chosen sufficiently 
high. See Sec.~\ref{sec:MFT} for more details on its numerical 
implementation.

\subsection{Mean-field theory}
\label{sec:mft}

To make further progress with the interacting term, we adopt the BCS 
mean-field approximation for pairing, and introduce the pair potential 
$
\Delta_{\br}=g\langle\psi_{\br \uparrow}\psi_{\br\downarrow}\rangle,
$ 
which serves as the order parameter for pairing characterizing the SF phase. 
Here, $\langle \cdots \rangle$ is a thermal average. This approximation 
reduces the interaction part of the Hamiltonian to
\begin{equation} 
H_\mathrm{int}^\mathrm{mf}=\int d^2\br\, 
\left( \Delta_{\br}\psi_{\br \uparrow}^\dag \psi_{\br \downarrow}^\dag
+\Delta^{*}_{\br}\psi_{\br \downarrow} \psi_{\br \uparrow}
+\frac{|\Delta_{\br}|^2}{g} \right),
\end{equation}
and therefore, the total mean-field Hamiltonian 
$
H^\mathrm{mf}= H_\mathrm{ni} + H_\mathrm{int}^\mathrm{mf}
$
has effectively the form of a single-particle one. In order to obtain self-consistent 
solutions, one needs to solve $\Delta_{\br}$ and $H^\mathrm{mf}$ together 
with the number density
$
n_{\br}=\sum_\sigma\langle\psi_{\br\sigma}^{\dag}\psi_{\br \sigma}\rangle,
$  
in such a way that the total number of particles $N=\int d^2\br n_{\br}$
is fixed to a specified value through the parameter $\mu$. 
Furthermore, while a vanishing/non-zero $|\Delta_\br|$ is a characteristic
property of N/SF phase in general, the SF phase may further be classified 
as being gapped or gapless depending on its excitation spectrum in 
momentum space. Having this purpose in mind, we are also interested in the 
mass-current density $\mathbf{J}_\br$ in this paper, which can be extracted
from the continuity equation
$
M\partial_t n_\br+\nabla \cdot \mathbf{J}_\br=0.
$
Next we derive explicit expressions for the resultant self-consistency equations 
using both BdG and LDA approaches.

\subsubsection{Bogoliubov-de Gennes approach}
\label{sec:bdg}

Using a generalized Bogoliubov-Valatin transformation, we first diagonalize 
$H^\mathrm{mf}$, leading to the matrix-eigenvalue equation
$
H^\mathrm{BdG}_{\br}\Psi_{\br\eta}=E_{\eta}\Psi_{\br \eta},
$
where the BdG Hamiltonian can be expressed as
\begin{equation} 
H^\mathrm{BdG}_{\br}\!\!=\!\!\begin{pmatrix} K_\br
&S_\br & 0 & \Delta_{\br}\\ S^{\dagger}_{\br} & K_\br &  -\Delta_{\br} &
0\\ 0 & -\Delta^{*}_{\br}  & -K_\br  & -S^{\dagger}_{\br}\\
\Delta^{*}_{\br} & 0 & -S_{\br} & -K_\br \end{pmatrix}-\Omega L^{z}_{\br}.
\end{equation}
Here, $K_\br=-\nabla^2/(2M)+V_{r}-\mu$ and $L^{z}_{\br}=x\partial_y-y\partial_x$ 
are the spin-conserving single-particle terms, and $S_{\br}=\alpha(\partial_x-i\partial_y)$ 
is the spin-flipping Rashba one. The eigenfunctions are formed by 
a four-component Nambu spinor
$
\Psi_{\br\eta}= [u_{\br\uparrow\eta},u_{\br\downarrow\eta},v_{\br
\uparrow\eta},v_{\br\downarrow\eta}]^\mathrm{T},
$  
and the associated quasi-particles have energy $E_{\eta}>0$, where 
the creation (annihilation) operator $\gamma_\eta^\dag$ ($\gamma_\eta$) 
is such that
$
H^\mathrm{mf} = E_\mathrm{gs} + \sum_\eta E_\eta \gamma_\eta^\dag \gamma_\eta,
$
with $E_\mathrm{gs}$ the ground-state energy of the system.

We then make use of the inverse transformations 
$
\psi_{\br\sigma} =\sum_{\eta }\left(u_{\br\sigma\eta}\gamma_{\eta}
+v^{*}_{\br\sigma\eta}\gamma_{\eta}^{\dagger}\right),
$
and determine the self-consistency equations for $\Delta_{\br}$ and $\mu$ as 
\begin{align} 
\Delta_{\br}&=g\sum_{\eta}
\left[u_{\br\uparrow\eta}v^{*}_{\br\downarrow\eta}f(-E_{\eta})
+u_{\br\downarrow\eta}v^{*}_{\br\uparrow\eta}f(E_{\eta}) \right],
\label{eq:BdGgap} \\
n_{\br}&=\sum_{\eta\sigma}
\left[\left|u_{\br\sigma\eta}\right|^{2}f(E_{\eta})
+\left|v_{\br\sigma\eta}\right|^{2}f(-E_{\eta})\right],
\label{eq:BdGnum} 
\end{align}
where $f(x)=1/[1+\exp(\beta x)]$ is the Fermi function with the inverse temperature 
$\beta=1/(k_BT)$. In addition, $\mathbf{J}_\br$ has a non-vanishing component
in the azimuthal direction, which can be written as a sum of two terms
$
J^\theta_{r}=\sum_\sigma J^{\theta\sigma}_{r} + 
2M\alpha J^{\theta\uparrow\downarrow}_{r},
$
corresponding, respectively, to the usual contribution and the Rashba one, 
where
\begin{eqnarray} 
J^{\theta\sigma}_{r}&
=&\sum_{\eta} \left[u^{*}_{\br\sigma\eta} \frac{\partial}{r\partial \theta}u_{\br\sigma\eta} f(E_{\eta}) 
+ v_{\br\sigma\eta} \frac{\partial}{r\partial \theta}v^{*}_{\br\sigma\eta} f(-E_{\eta}) \right], \nonumber \\
J^{\theta\uparrow\downarrow}_{r}&
= & \sum_\eta\left[|u^{*}_{\br\uparrow\eta} u_{\br\downarrow\eta}| f(E_{\eta})
+|v_{\br\uparrow\eta}   v^{*}_{\br\downarrow\eta}| f(-E_{\eta}) \right].
\label{eq:currdens}
 \end{eqnarray}
Here, while all of the $\eta$ sums are restricted to $E_\eta<E_c$, none of 
our results depend strongly on the specific value of the cut-off energy $E_c$
 as noted above in Sec.~\ref{sec:ham}. 

Lastly, by expanding the components of $\Psi_{\br\eta}$ in the angular-momentum 
basis of a simple-harmonic oscillator, it is possible to obtain closed-form expressions
for all of these equations as briefly summarized in Appendix~\ref{sec:angmombasis}.

\subsubsection{Local-density approximation}
\label{sec:lda}

Unlike the BdG approach where the harmonic-oscillator potential is taken 
exactly into account in a fully-quantum-mechanical manner, the LDA approach 
is a semi-classical one, as it amounts to treating the local system at 
$\boldsymbol{r}$ as a uniform gas with the local chemical potential 
$\mu_r=\mu-V_{r}$ and a rotation 
$
\Omega L_{\br\bk}^{z}=\mathbf{v}_{\br}\cdot \bk
$ 
term, where 
$
\mathbf{v}_{\br}=\Omega\mathbf{\hat{z}}\times\br.
$ 
Note that the trap center is immune to the direct-effects of rotation. Within this 
approach, we expand the field operators in the plane-wave basis, i.e.,
$
\psi_{\br\sigma}=(1/\sqrt{A})\sum_{\bk}e^{i\bk\cdot\br}a_{\bk\sigma}
$
and its Hermitian conjugate, where $a_{\bk\sigma}$ is the annihilation
operator for a pseudo-spin $\sigma$ fermion at momentum $\bk=(k_x, k_y)$, 
and obtain the local Hamiltonian density
$
H^{\mathrm{LDA}}_\br=(1/2) \sum_{\bk}\Psi_{\bk}^{\dagger}H_{\br\bk}^{\mathrm{LDA}}\Psi_{\bk}+C_\br.
$
Here, the LDA Hamiltonian can be expressed as
\begin{equation} 
H^{\mathrm{LDA}}_{\br\bk}=\begin{pmatrix}
\xi_{\br\bk}& S_{\bk} & 0 & \Delta_\br\\ S_{\bk}^{*} & \xi_{\br\bk}&
-\Delta_\br & 0\\ 0 & -\Delta^{*}_\br & -\xi_{\br\bk} & S_{\bk}^{*}\\
\Delta^{*}_\br & 0 & S_{\bk} & -\xi_{\br\bk} \end{pmatrix}
-\Omega L_{\br\bk}^{z}\,, 
\label{eqn:ham} 
\end{equation}
where $\xi_{\br\bk}=\epsilon_{\bk}-\mu_r$ with the free-particle dispersion 
$
\epsilon_{\bk}=k^2/(2M),
$ 
$
S_{\bk}=\alpha(k_{x}-ik_{y})
$ 
is the Rashba coupling,
$
\Delta_\br=(g/A)\sum_{\bk}\langle a_{\bk\uparrow}a_{-\bk\downarrow}\rangle
$ 
is the local order parameter,
$
\Psi_{\bk}=[a_{\bk\uparrow},a_{\bk\downarrow},
a_{\mathbf{k}\uparrow}^{\dagger},a_{\mathbf{-k}\downarrow}^{\dagger}]^\textrm{T}
$
is the spinor operator, and 
$
C_\br=\sum_{\bk}(\xi_{\br\bk}+\Omega L_{\br\bk}^{z})+A|\Delta_\br|^{2}/g
$ 
is a local constant. This Hamiltonian can be written in its diagonal form as
$
H^{\mathrm{LDA}}_\br=\sum_{\bk s} 
(E_{\br\bk s}\gamma_{\bk s}^{\dag}\gamma_{\bk s}-E_{\br\bk s}/2)+C_\br,
$
where the operator $\gamma_{\bk s}^{\dag}$ ($\gamma_{\bk s}$) creates 
(annihilates) a quasi-particle with momentum $\bk$, helicity $s=\pm$,
and local excitation energy
\begin{equation} 
E_{\br \bk s}=\sqrt{(\xi_{\br\bk}+s\alpha k)^{2}
+|\Delta_\br|^{2}}-\Omega L_{\br\bk}^{z}
\label{eq:EXC}.
\end{equation}
We note that while a locally-gapped SF has a non-zero $E_{\br\bk s}>0$ 
everywhere in $\bk$ space, the locally-gapless SF (i.e., gSF) has 
$E_{\br\bk s}=0$ for some $\bk$-space points even though $|\Delta_\br|>0$.

In order to determine the self-consistency equations, we first calculate the 
local thermodynamic potential 
$
G_\br = - (1/\beta) \mathrm{Tr}\{\ln [\exp(-\beta H^{\mathrm{LDA}}_\br)]\},
$
leading to
$
G_\br=(1/2) \sum_{\bk s}\lbrace (1/\beta) \ln[1-f(E_{\br\bk s})]-E_{\br\bk s} \rbrace + C_\br,
$
and then minimize it, i.e., $\partial G_\br/\partial|\Delta_\br|=0$ together with 
$n_\br= - (1/A) \partial G_\br/\partial\mu_r$. This procedure gives rise to
the following closed-form expressions
\begin{align} 
\label{eq:Gap}
 \frac{1}{g} &=\frac{1}{4 A}\sum_{\bk s}\frac{1-2 f(E_{\br\bk s})}{E_{\br\bk s}
+\Omega L_{\br\bk}^{z}}, \\
\label{eq:Number} 
n_{\br} &= \frac{1}{2 A} \sum_{\bk s}
\left\{1-\frac{\xi_{\br\bk}+s\alpha k}{E_{\br\bk s}
+\Omega L_{\br\bk}^{z}}  \left[1-2 f(E_{\br\bk s})\right]\right\},
\end{align}
for $\Delta_\br$ and $\mu$. Furthermore, the components of the mass-current 
density $\mathbf{J}_{\br}=(J_{\br}^x, J_{\br}^y)$ can again be written as a sum
of two terms
\begin{equation} 
(J_{\br}^x, J_{\br}^y) = \frac{1}{A} \sum_{\bk \sigma}
(k_x,k_y)n_{\br\bk \sigma}  +   \left(  P^{x}_{\br},P^{y}_{\br}
\right)M\alpha n_\br.
\end{equation}
We note that while the usual contribution is related directly to the local 
momentum distribution 
$
n_{\br\bk \sigma} =  \langle a^{\dag}_{\bk \sigma}a_{\bk \sigma} \rangle 
$
of particles which can be extracted from the summand of Eq.~(\ref{eq:Number}) 
as $n_\br=(1/A) \sum_{\bk \sigma} n_{\br\bk \sigma}$, the Rashba one is 
related directly to the local average spin polarization 
$
P^i_\br=[1/(A n_\br)] \sum_{\sigma \sigma'} \langle
a^\dag_{\bk \sigma} \vec{\mathbf{\sigma}}^i_{\sigma \sigma'}a_{\bk
\sigma'}\rangle
$
of particles with its components determined by
$
P^{x}_{\br}+i P^{y}_{\br}=[2/(A n_\br)] \sum_{\bk} \langle a^{\dag}_{\bk\uparrow} 
a_{\bk \downarrow} \rangle.
$

Having presented the details of the BdG and LDA approaches, next we analyze 
the resultant self-consistency equations for a non-interacting Fermi gas, and show 
that the competition between the effects of Rashba coupling on the local density 
of single-particle states and the Coriolis effects caused by rotation gives rise 
to a characteristic ring-shaped number density that survives at experimentally 
accessible temperatures.

\section{Non-interacting Fermi gas} 
\label{sec:NI}

Given that the LDA results are in very good agreement with those of numerically-exact 
quantum-mechanical ones for a wide-range of parameter regimes and with the 
additional advantage that they permit analytical insights into the limiting 
cases~\cite{Doko2016}, we rely mostly on the LDA approach throughout this section 
and analyze the generic trap profiles of a non-interacting Fermi gas in the 
presence of Rashba coupling and under slow or moderate rotations. We note in 
passing that, since the LDA approach does not capture the correct physics in 
the Landau regime of a rapidly-rotating Fermi gas, we rely only on the BdG approach 
in this extreme regime as discussed towards the end in Sec.~\ref{sec:LLL}.

By setting $\Delta_r=0$ in Eq.~(\ref{eq:EXC}), we get the local dispersion relation
$
\varepsilon_{\br \bk s}=k^{2}/(2M)+s\alpha k
-r\Omega k\sin(\theta_{\bk}-\theta_{\br})-\mu_r
$
for the non-interacting particles, where $\theta_{\bk}$ and $\theta_\br$ are, respectively, 
the polar angles in $\mathbf{k}$ and $\mathbf{r}$ spaces. Since the trapping potential 
is assumed to be isotropic in space in this paper, and limiting ourselves only to the 
rotationally-symmetric solutions, we may take $\theta_{\br}=0$ corresponding to the 
positive $x$-direction in space without the loss of generality. Given the local dispersion, we 
may express the total local energy density of states (LDOS) as 
$
D_r(\epsilon)=\sum_s D_{rs}(\epsilon),
$ 
where 
$
D_{rs}(\epsilon)=\sum_{\bk}\delta(\epsilon-\varepsilon_{r\bk s})
$
is the LDOS with helicity $s$. Similarly, the total number density may also 
be expressed as 
$
n_r=\sum_s n_{rs},
$ 
where
$
n_{rs}=(1/A)\sum_{\bk}f\left(\varepsilon_{r \bk s}\right)
=(1/A) \int\mathrm{d}\epsilon D_{rs}(\epsilon)f(\epsilon)
$
is the number density with helicity $s$. Furthermore, the local Fermi surfaces 
are defined by $\varepsilon_{r \bk s}=0$, leading to the following curves
\begin{equation} 
k_{1,2}^{s}= M(\varOmega r\sin\theta_{\bk}-s\alpha)
\pm\sqrt{M^{2}(\varOmega r\sin\theta_{\bk}-s\alpha)^{2}+2M\mu_r}.
\label{eq:fermisurf} 
\end{equation}
To gain as much insight as possible into the basic properties of a non-interacting
Fermi gas, we first discuss these quantities in a few analytically-tractable limits, 
prior to the presentation of our numerical results for the generic case with 
arbitrary $\Omega$ and $\alpha$.

\begin{figure} 
\includegraphics[scale=0.125]{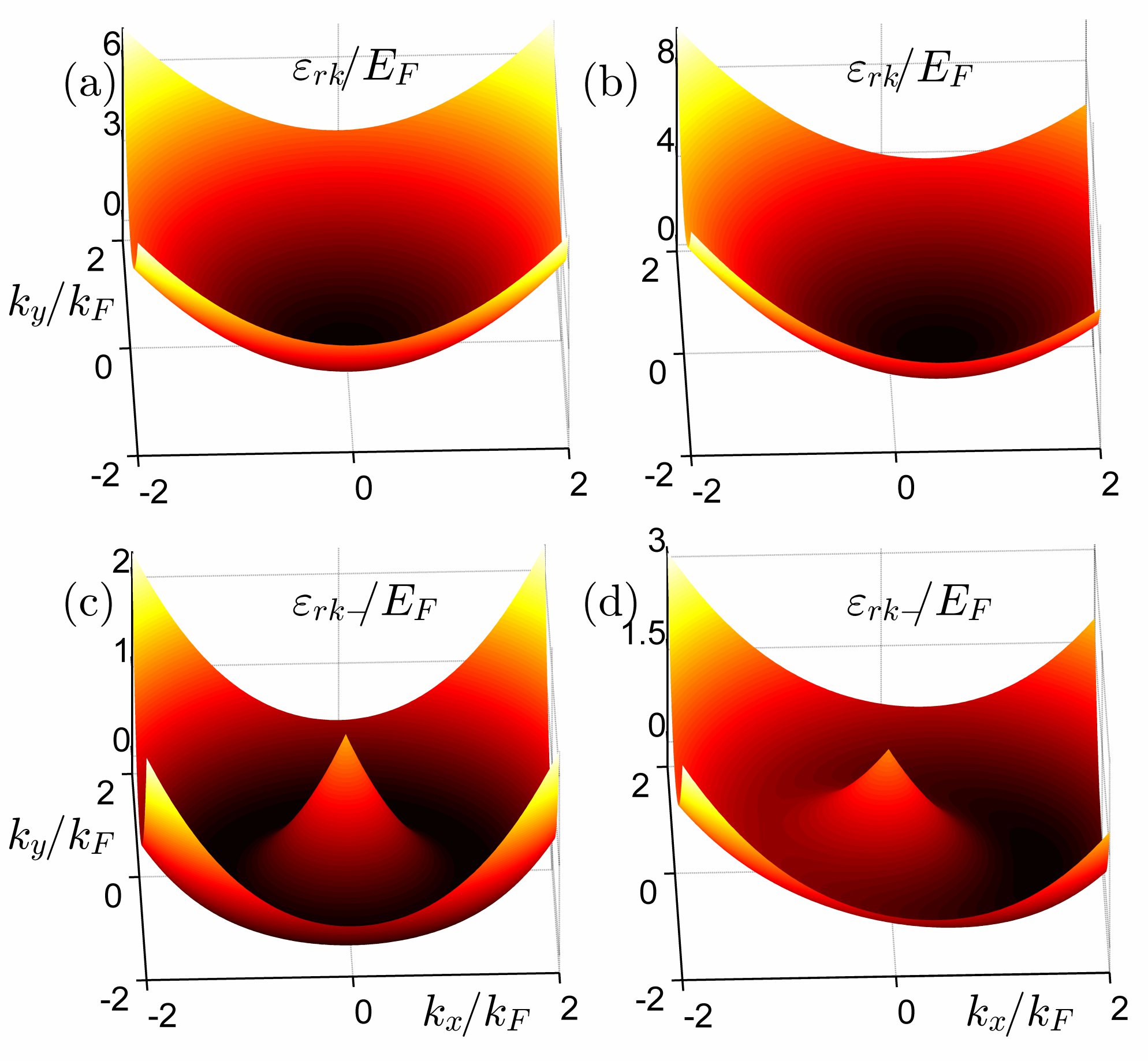} 
\caption{(Color online) 
Local dispersion relations for
(a) $\Omega=0$ and $\alpha=0$: the usual paraboloid,
(b) $\Omega\neq0$ and $\alpha=0$: finite momentum states that are energetically 
favored by rotation result in a shifted paraboloid,
(c) $\Omega = 0$ and $\alpha\protect\neq0$: the negative-helicity 
band has a degenerate circular minima, and
(d) $\Omega\protect\neq0$ and $\alpha\protect\neq0$:  
rotation causes an asymmetric minimum.
\label{fig:ES}
} 
\end{figure}
\subsection{Trapped Fermi gas 
\\ ($\omega\protect\neq0$, $\alpha=0$ and $\Omega=0$)}
\label{sec:NItrap}

\begin{figure}[ht!]
\includegraphics[scale=0.085]{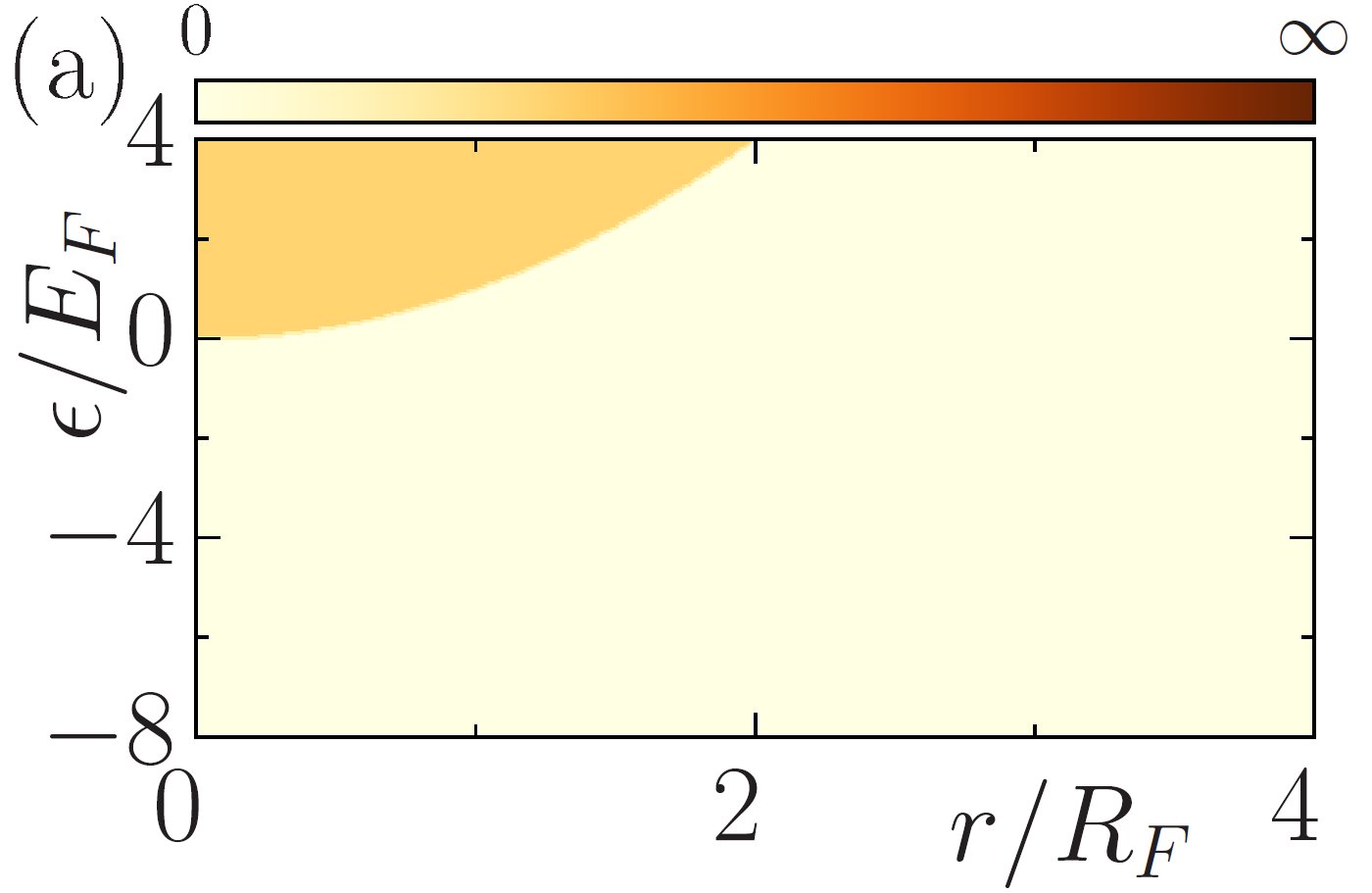} 
\includegraphics[scale=0.085]{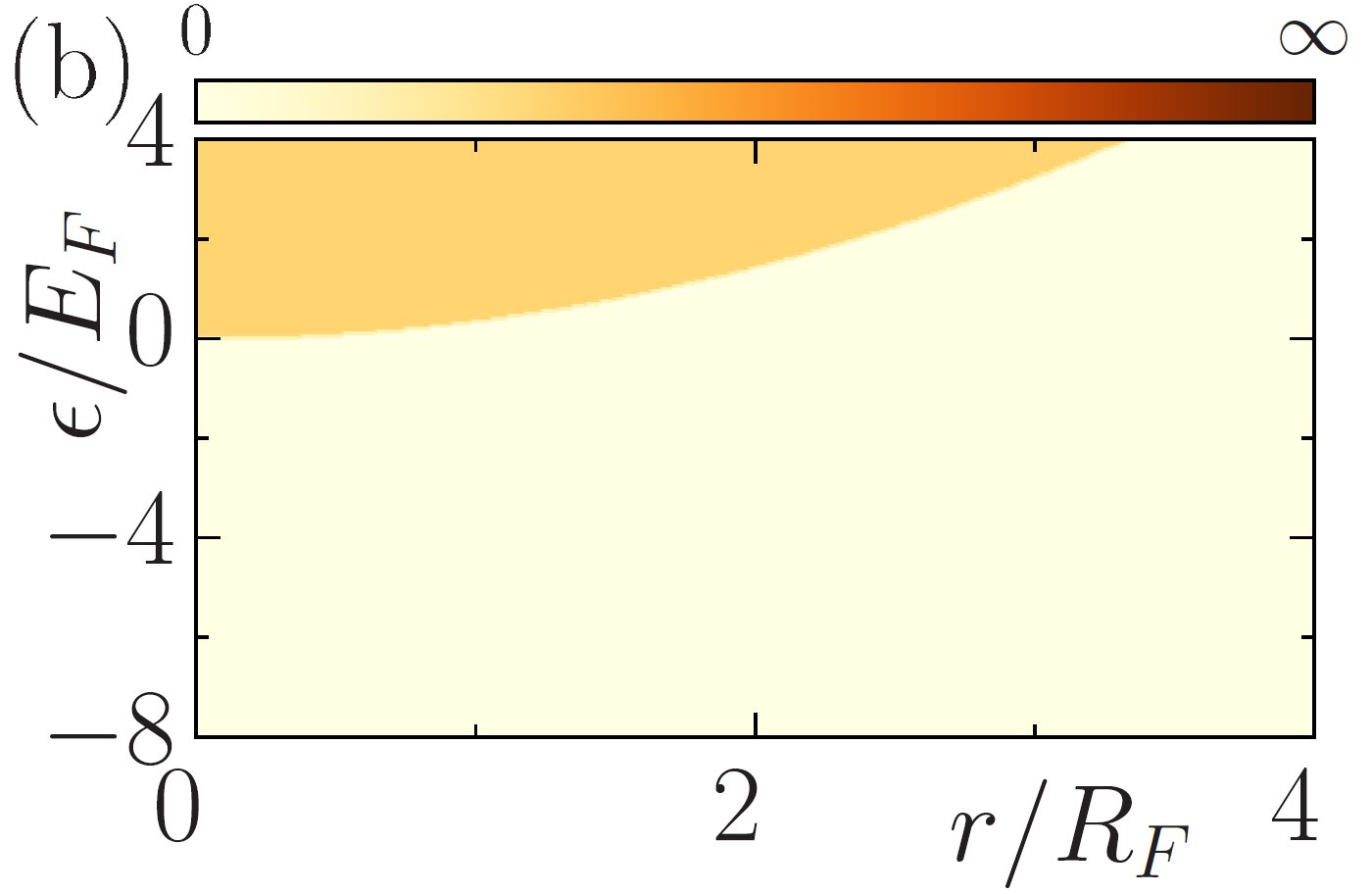} \\
\includegraphics[scale=0.085]{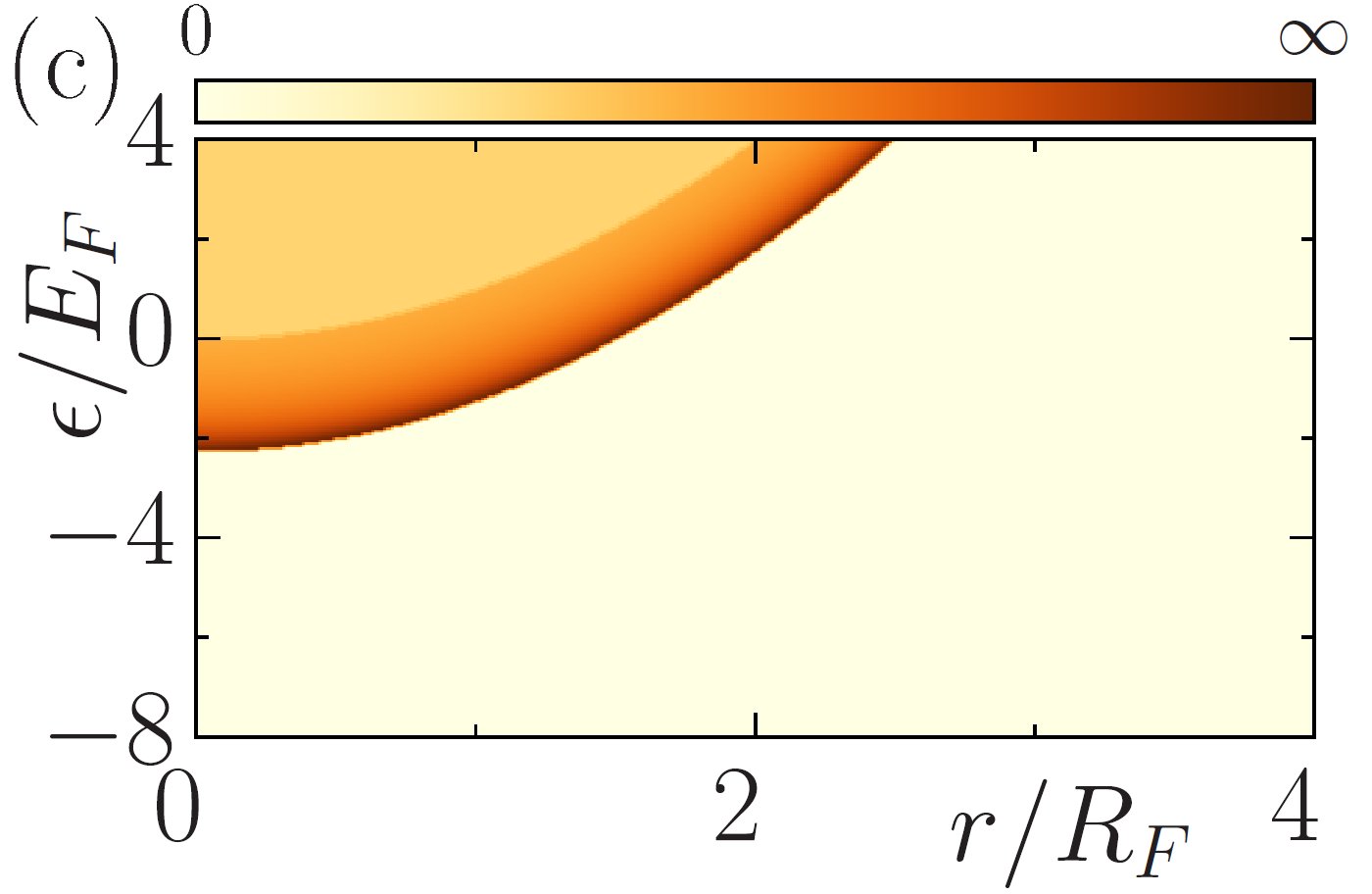} 
\includegraphics[scale=0.085]{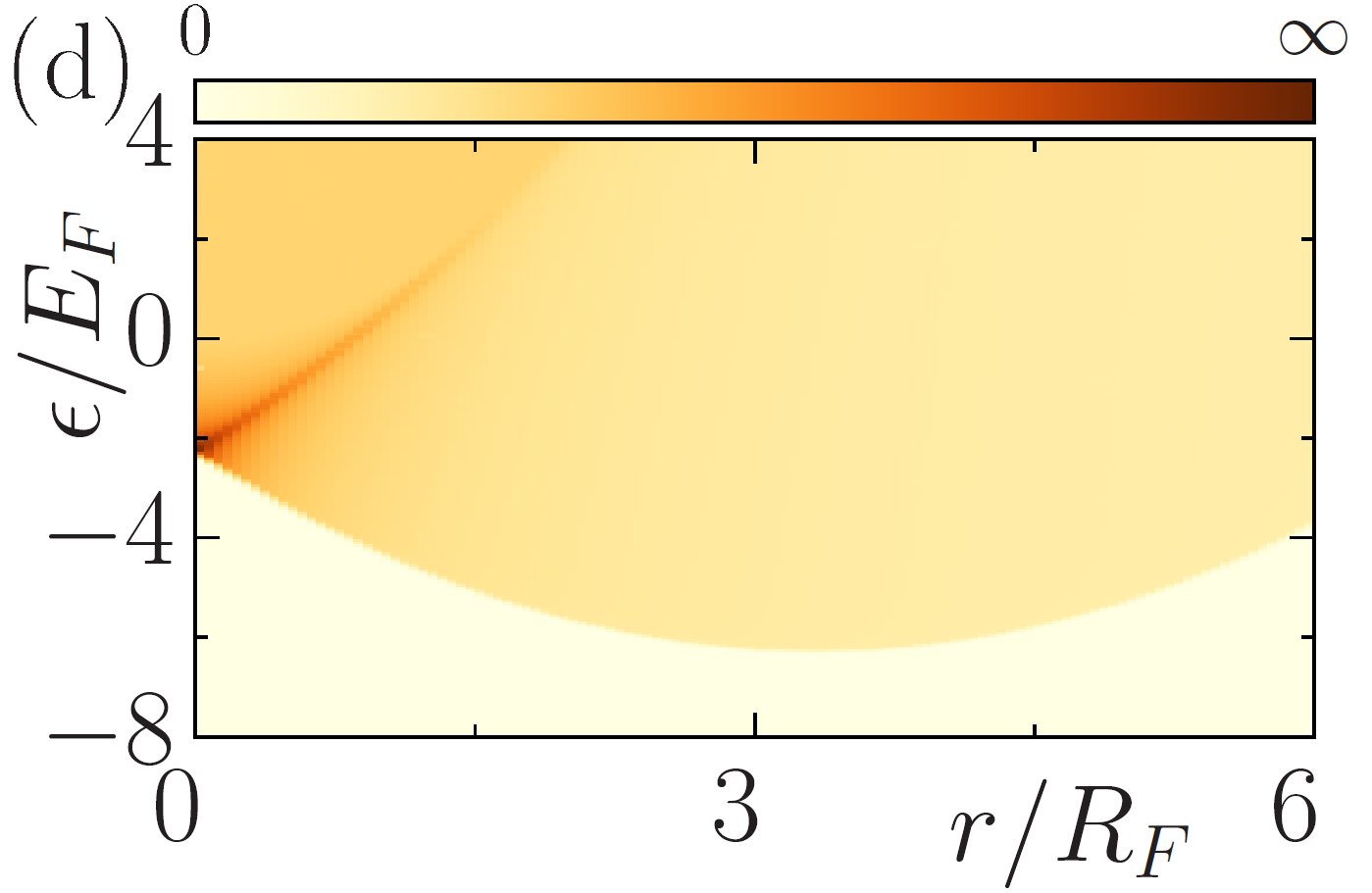}
\caption{(Color online) 
Local energy density of states (LDOS) in the trap within the LDA approach. 
In the absence of a Rashba coupling, the LDOS is a constant except for the 
outward extension due to rotation as shown in    
(a) for $\Omega=0$ and $\alpha=0$ and  
(b) for $\Omega=0.4\omega$ and $\alpha=0$.
In the absence of rotation, Rashba coupling enhances the LDOS for 
the lowest energies as shown in 
(c) for $\Omega=0$ and $\alpha=3 E_F/k_F$.  
In the generic case shown in 
(d) for $\Omega=0.8\omega$ and $\alpha=3 E_F/k_F$, 
rotation not only pushes the minimum of the LDOS away from the trap center 
but it also removes the divergence for $r\neq 0$ as discussed in the main text.  
\label{fig:LDOS} 
}
\end{figure}

The first analytically-tractable limit is a usual 2D Fermi gas with neither Rashba 
coupling nor rotation, for which case the local dispersion relation is simply a 
paraboloid with its minimum at the origin $\mathbf{k}=\mathbf{0}$. This is shown in 
Fig.~\ref{fig:ES}(a) for completeness, and the Fermi surface is trivially
a circle around the origin. Since the DOS of a uniform Fermi gas is a constant 
in 2D, the LDOS of a trapped gas can be written within the LDA approach as 
$
D_r(\epsilon)=MA\Theta(\epsilon_r)/\pi,
$ 
where $\Theta(x)$ is the Heaviside-step function and $\epsilon_r=\epsilon-V_r$, 
and it is shown in Fig.~\ref{fig:LDOS}(a). The resulting number density is an 
inverted parabola, 
$
n_r=M(E_{F}-M\omega^{2}r^{2}/2)/\pi,
$ 
where $E_{F}$ is the Fermi energy at the trap center, and the Thomas-Fermi 
radius $R_{F}$ is given by definition $n_{R_F} = 0$ as
$
R_{F}=\sqrt{2E_{F}/(M\omega^{2})}.
$ 
In addition, the central density can be written as $n_0=k_F^2/(2\pi)$, where 
$k_F$ is the Fermi momentum in such a way that 
$
E_F = k_F^2/(2M) = \omega \sqrt{N}.
$ 
We use $E_F$, $k_F$ and $R_F$ as, respectively, the energy, momentum and 
length scales in our numerical calculations.

\subsection{Trapped Fermi gas with Rashba coupling 
\\ ($\omega\protect\neq0$, $\alpha\protect\neq0$ and $\Omega=0$)}
\label{sec:NIsoc}

The second analytically-tractable limit is a 2D Fermi gas with Rashba 
coupling, for which case the main effect of this coupling on the trap profiles 
is to increase the number density at the trap center through the increased 
low-energy LDOS. To see this effect, we first note that, by breaking the 
spin-rotation symmetry, the Rashba coupling splits the local dispersion 
relation into two ($s = \pm$) local helicity branches. Here, the spin is oriented 
parallel to the momentum $\bk$ in the higher-energy $+$ branch 
and anti-parallel in the lower-energy $-$ one. While the energy of the 
$+$ branch increases monotonically with $k$, the minimum of the spectrum 
is shifted to finite momentum for the $-$ branch forming a circle with radius 
$k=M\alpha$ as shown in Fig.~\ref{fig:ES}(c). Thus, in the local regions 
with $\mu_r>0$, there are two circular Fermi surfaces around the origin 
in $\bk$ space corresponding to $+$ and $-$ branches. When the $+$ 
branch disappears for $\mu_r\leq 0$, an additional circular Fermi surface 
appears in the $-$ branch, in which case, however, all of the $\mathbf{k}$ 
states that are below the Fermi energy have finite momentum. 

Then, within the LDA approach, it is easy to show that the LDOS are given by
$
D_{rs}(\epsilon)=MA(1-sM\alpha/ \sqrt{M^2\alpha^2+2M\epsilon_r})/(2\pi)
$ 
in the local regions with $\epsilon_r>0$, and
$
D_{rs}(\epsilon)=(1-s)M^2A\alpha\Theta(M^2\alpha^2+2M\epsilon_r)/
(2\pi\sqrt{M^2\alpha^2+2M\epsilon_r})
$
in the local regions with $\epsilon_r \leq 0$. This indicates that, in sharp 
contrast to the $\epsilon_r>0$ regions where the total LDOS is clearly 
unaffected by the Rashba coupling, it displays a 1D-like energy dependence 
in the $\epsilon_r \leq 0$ regions arising solely from the $-$ branch,
i.e., the divergence of LDOS in the local regions satisfying the condition 
$\epsilon_r=-M\alpha^2/2$ is very much like that of a uniform Fermi gas in 1D. 
This behavior is a reflection of the degenerate minima discussed above, 
and it gives rise to the enhanced low-energy LDOS shown in Fig.~\ref{fig:LDOS} (c). 
This effect reduces the radius of the gas as the number density increases 
around the trap center due to the increased low-energy LDOS. 

Furthermore, since the Rashba coupling indirectly affects the density profile 
through the depletion of particles from the $+$ branch, we determine the critical radius 
$
r_c=[-2\alpha^{2}/\omega^{2}+\sqrt{4\alpha^{4}/(3\omega^{4})+R_{F}^{4}}]^{1/2}
$
for the complete depletion by setting $\mu_{r_c} = 0$. In addition, we find
$
n_{rs}=M^{2}(-s\alpha+\sqrt{\alpha^{2}+2\mu_r/M})^{2}/(4\pi)
$
in the local regions with $\mu_r>0$ or equivalently $r<r_c$, and $n_{r+}=0$ and
$
n_{r-}=M^{2}\alpha\sqrt{\alpha^{2}+2\mu_r/M}/\pi
$  
in the local regions with $\mu_r<0$ or equivalently $r>r_c$. 
Since $r_c > 0$ for weak couplings, there is an outer ring-shaped region in the trap 
where the local $+$ branch is completely empty. Setting $r_c = 0$, we find the 
critical Rashba coupling $\alpha_c=\sqrt[4]{6}E_{F}/k_{F}$, beyond which the 
$+$ branch is never occupied in the entire trap. We note that even though the 
number density acquires a relatively simple form when $\alpha > \alpha_c$, it is 
quite different from the usual inverted-parabola dependence of a trapped Fermi gas 
discussed in Sec.~\ref{sec:NItrap}. Furthermore, the edge of the gas can be extracted 
from the number density as 
$
R_{O}^{0}=[-\alpha^{2}/\omega^{2}+\sqrt{4\alpha^{4}/(3\omega^{4})+R_{F}^{4}}]^{1/2}
$
for $\alpha<\alpha_c$, and 
$
R_{O}^{0}=R_{F} [3E_{F}/(4\alpha k_{F})]^{1/3}
$ 
for $\alpha \geq \alpha_c$, showing explicitly that the edge of the gas $R_{O}^{0}$ 
moves inward with increasing $\alpha$ and the gas contracts. 
By integrating the number density, we also find
$
\mu=-\alpha^{2}M+\sqrt{\alpha^{4}M^{2}/3+E_{F}^{2}}
$
for $\alpha<\alpha_c$, and 
$
\mu = -\alpha^{2}M/2+ [6E_{F}^{2}/(\alpha\sqrt{2M})]^{2/3}/4
$
for $\alpha>\alpha_c$.

\subsection{Trapped Fermi gas with adiabatic rotation 
\\ ($\omega\protect\neq0$, $\alpha=0$ and $\Omega\protect\neq0$)}
\label{sec:NIrot}

Another analytically-tractable limit is a 2D Fermi gas with adiabatic rotation, 
for which case the main effect of rotation on the trap profiles is to spread out 
the number density supported by the imparted centripetal acceleration on the particles.
To see this effect, we first note that, by breaking the inversion symmetry of the 
dispersion relation, i.e., tilting of the excitation spectrum, rotation causes an 
asymmetry in $\bk$ space. This shifts the minimum of the paraboloid from the 
origin shown in Fig.~\ref{fig:ES}(a) to a finite $k=r\Omega M$ as 
shown in Fig.~\ref{fig:ES}(b) where the Fermi surface is a circle centered around 
this finite momentum.  

We find that the LDOS is simply given by
$
D_r(\epsilon)=MA\Theta(\epsilon_r+M\Omega^2 r^2/2)/\pi,
$ 
showing that it remains to be a constant except for the radial extention,
and that its parabolic shape is retained as shown in Fig.~\ref{fig:LDOS}(b). 
As a consequence, 
$
\mu=E_{F}\sqrt{1-\Omega^{2}/\omega^{2}},
$ 
and the resultant number density is still an inverted parabola
$
n_r=M^{2}[2\mu/M-(\omega^{2}-\Omega^{2})r^{2}]/(2\pi).
$  
We note that since the curvature of the trap profile decreases with increasing 
$\Omega$, the edge of the gas
$
R_{O}^{0}=R_{F}(1-\Omega^{2}/\omega^{2})^{-1/4}
$ 
expands with $\Omega$ until $\Omega = \omega$, beyond which the trap 
cannot supply the necessary centripetal acceleration. In addition, the 
associated mass-current density $J^{\theta}_r=Mn_r\Omega r$ is exactly 
of the form of a rigidly-rotating gas. We note in passing that the asymmetric 
occupation of the finite-angular-momentum states with respect to the rotation 
along with and opposite to the azimuthal direction causes a pair-breaking effect 
on the Cooper pairs with zero center-of-mass momentum, as further discussed in 
Sec.~\ref{sec:Irot}.

\begin{figure}
\includegraphics[scale=0.60]{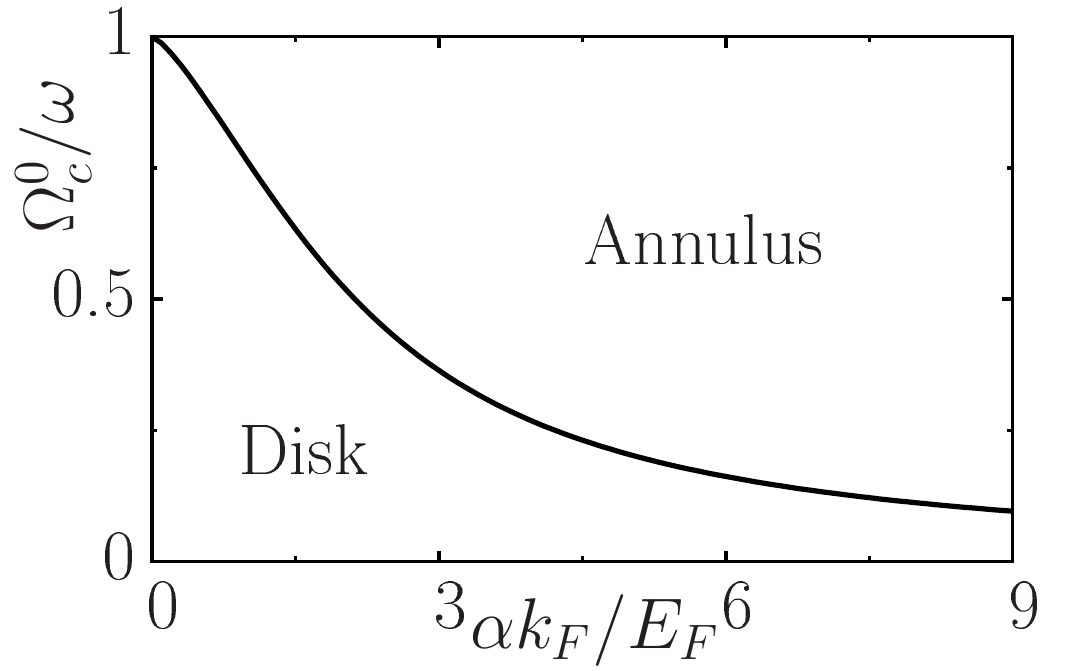} 
\caption{
The critical rotation frequency $\Omega_c^{0}$ for the depletion of the
non-interacting number density at the trap center at $T = 0$.
Increasing the Rashba coupling beyond a critical value transforms 
the disk-shaped density of the Fermi gas into a ring-shaped annulus.  
While an annulus may ultimately form for any $\Omega<\omega$ as long as $\alpha$ 
is sufficiently high, $\Omega_c^{0} \to \omega$ in the $\alpha\to 0$ limit signifying 
the crucial interplay between Rashba coupling and adiabatic rotation for this effect.
\label{fig:Om_c0}
} 
\end{figure}
\subsection{Trapped Fermi gas with Rashba coupling and rotation 
\\ ($\omega\protect\neq0$, $\alpha\protect\neq0$ and $\Omega\protect\neq0$)}
\label{sec:NIgen}

Having shown analytically that the Rashba coupling and adiabatic rotation 
have competing effects on the trap profiles, we are ready to discuss the generic 
case with arbitrary $\Omega$ and $\alpha$, for which case the main effect of 
their interplay is to change the number density from the shape of a disk to a 
ring-shaped annulus. To see this effect, we first note that, by breaking the degeneracy 
in the lowest-energy states, rotation tilts the minima of the $-$ branch as shown 
in Fig.~\ref{fig:ES}(d). Since the tilting-effect is proportional to $k$, it is further 
enhanced by the Rashba coupling, giving rise to three topologically-distinct Fermi 
surfaces for the $-$ branch and one for the $+$ one. For instance, in the local regions 
with $\mu_r>0$, the local Fermi surface of the $-$ branch is a circle around some 
finite $k$. In the local regions with $\mu_r<0$, however, while the local Fermi 
surface is a deformed ring centered around the origin when
$
\sqrt{-2\mu_r/M}-\alpha \leq -2\Omega r,
$ 
it is of the crescent shape when
$
|\sqrt{-2\mu_r/M}-\alpha| \leq \Omega r.
$
In contrast, the local Fermi surface of the $+$ branch is a deformed circle 
centered at some finite $k$. Note that, unlike the non-rotating case discussed in 
Sec.~\ref{sec:NIsoc}, the $+$ branch is locally occupied even in the regions with 
$\mu_r\leq0$ as long as $\sqrt{-4\mu_r}+\alpha \leq 2\Omega r$.

\begin{figure}[ht!]
\includegraphics[scale=0.087]{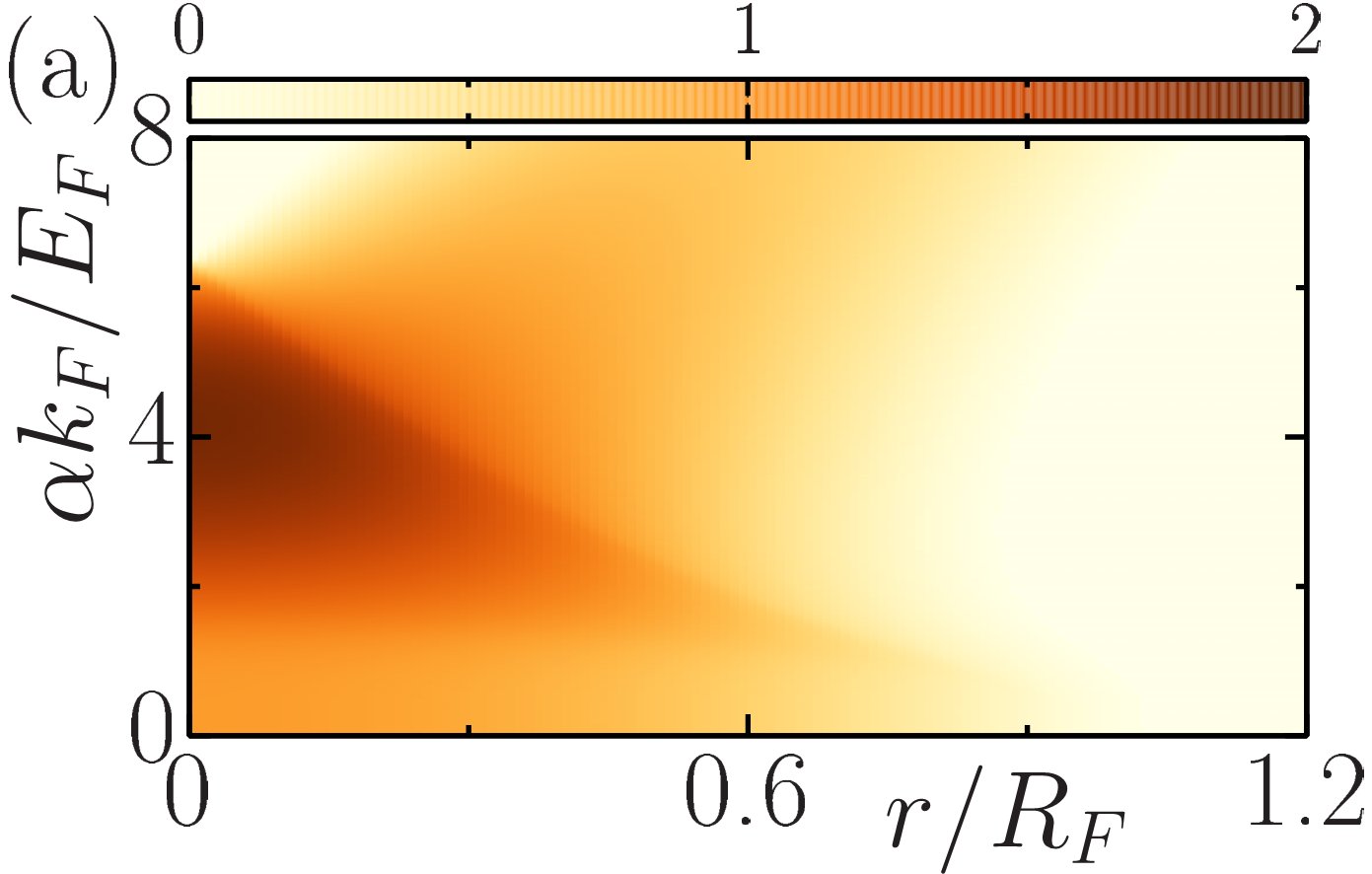}
\includegraphics[scale=0.087]{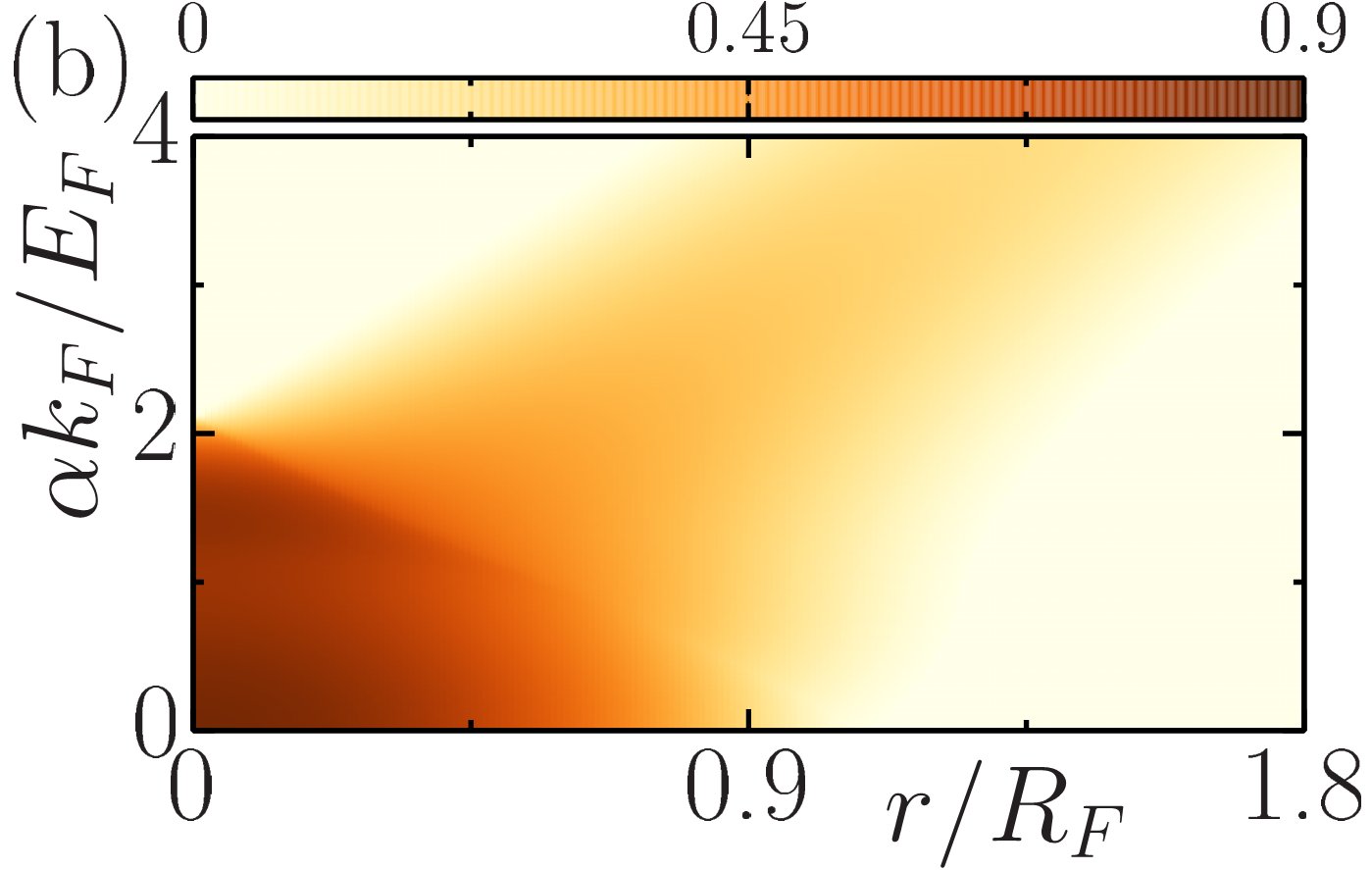} \\
\includegraphics[scale=0.46]{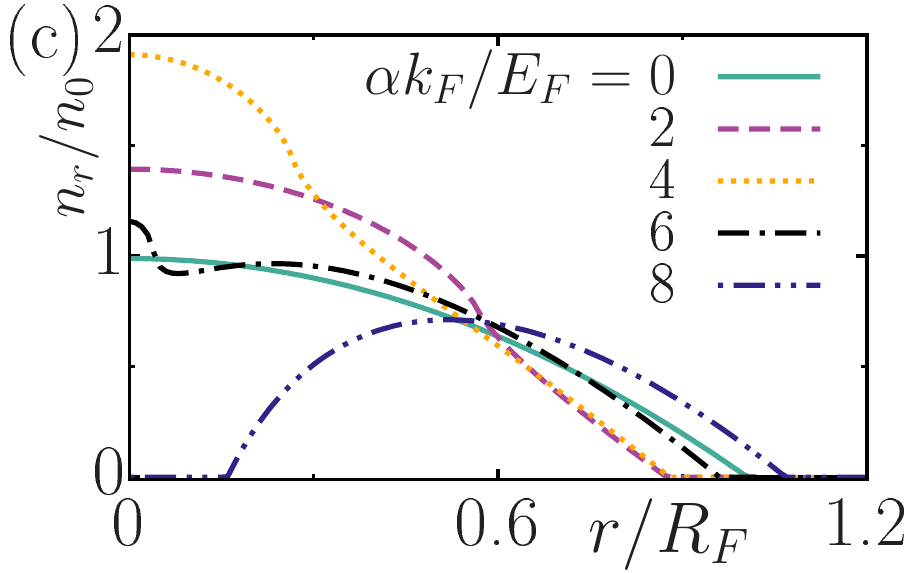}
\includegraphics[scale=0.46]{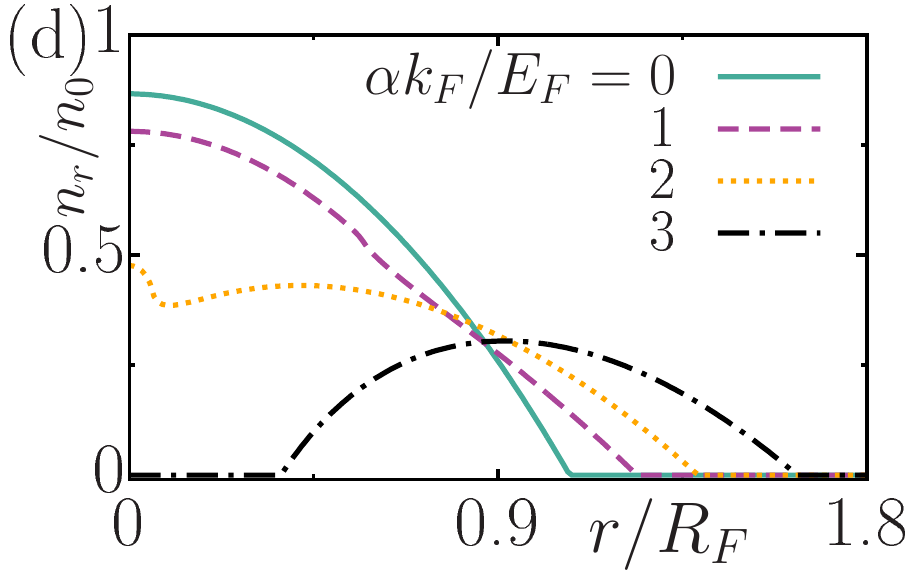}
\caption{(Color online) 
Non-interacting number-density maps at $T = 0$ with changing Rashba coupling for 
(a) $\Omega=0.15\omega$ and 
(b) $\Omega=0.5\omega$.
A few exemplary radial density profiles are plotted in (c) and (d), showing explicitly that the 
rotating Fermi gas eventually takes on a ring-shaped annulus with increasing $\alpha$.  
\label{fig:freeN}
} 
\end{figure}

The radial position of the lowest-energy state in the trap can be determined by minimizing 
$\varepsilon_{r \bk -}$ with respect to both $\bk$ and $r$, leading to 
$
k=\alpha M \omega^2/(\omega^2-\Omega^2)
$ 
with $\theta_\bk=\pi/2$, i.e., opposite to the direction of the mass-current density, and
$
r=\alpha \Omega/ (\omega^2-\Omega^2).
$ 
Note that $k \to 0$ and $r \to 0$ is recovered for the usual case when $\alpha \to 0$ 
and $\Omega \to 0$. To gain further insight, we calculate the LDOS via the following
representation of the Dirac-delta function 
$
\delta(x) \approx (1/\pi) \lim_{\varsigma \to 0} \varsigma/ (x^2+\varsigma^2)
$ 
with $\varsigma=10^{-3}$, and the results are shown in Fig.~\ref{fig:LDOS}(d).
We see that, by breaking the degeneracy of the lowest-energy states 
in the $-$ branch, rotation removes the 1D-like divergence from the LDOS 
profile for all $r \neq 0$. Recall that $r = 0$ is immune to the direct-effects 
of rotation. In addition, since the higher the angular momentum of
the single-particle state the further away its localization distance from the 
trap center, we conclude that the lowest-energy states have finite angular 
momentum, and that a comparison between Figs.~\ref{fig:LDOS}(c) 
and~\ref{fig:LDOS}(d) reveals that the lowest-energy of the 
finite-angular-momentum states is lower than the lowest-energy of the 
non-rotating gas. Thus, depending on the strengths of Rashba coupling 
and adiabatic rotation, it may be energetically more costly for any of the 
particles to occupy the trap center, in which case the number density forms 
a ring-shaped annulus.

\begin{figure}[ht!]
\includegraphics[scale=0.086]{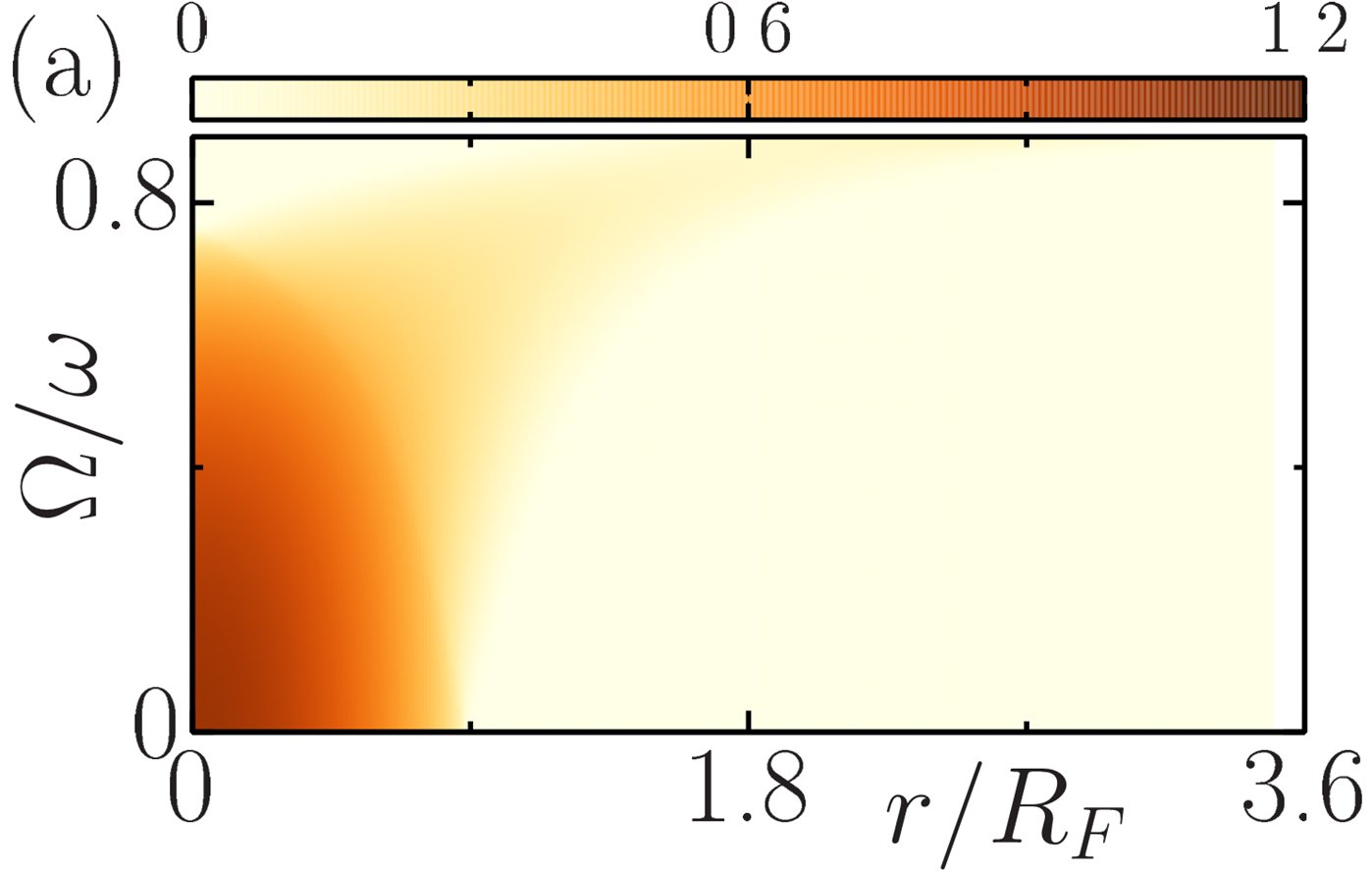}
\includegraphics[scale=0.086]{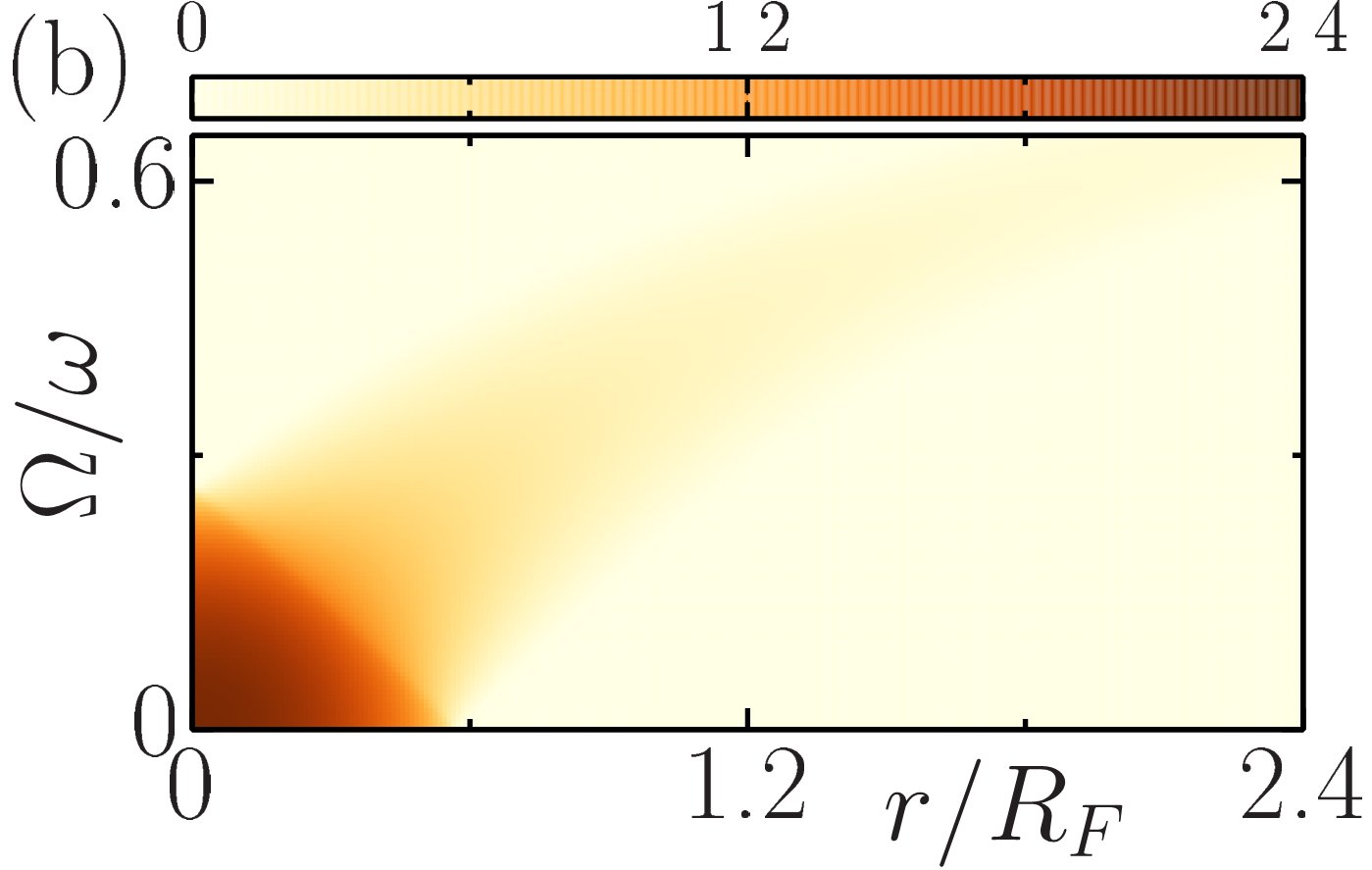} \\
\includegraphics[scale=0.46]{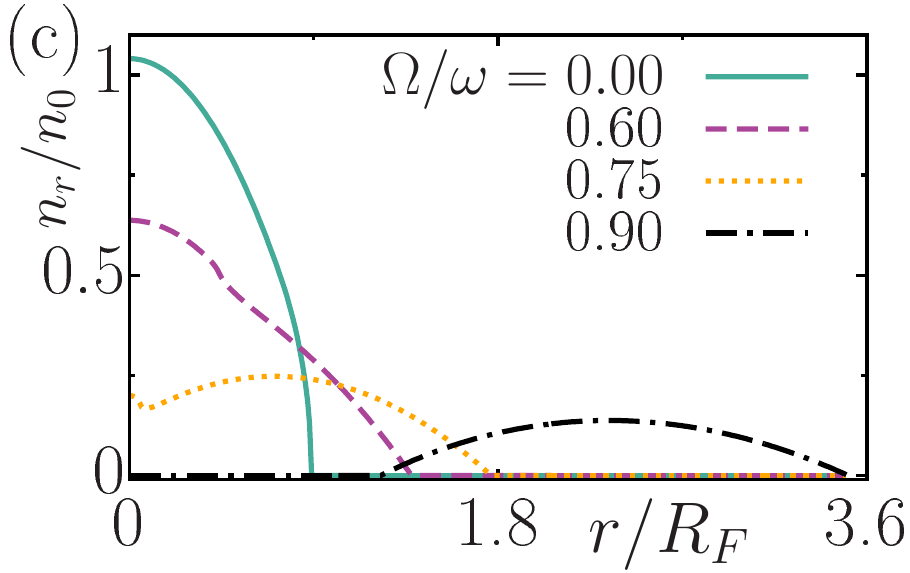}
\includegraphics[scale=0.46]{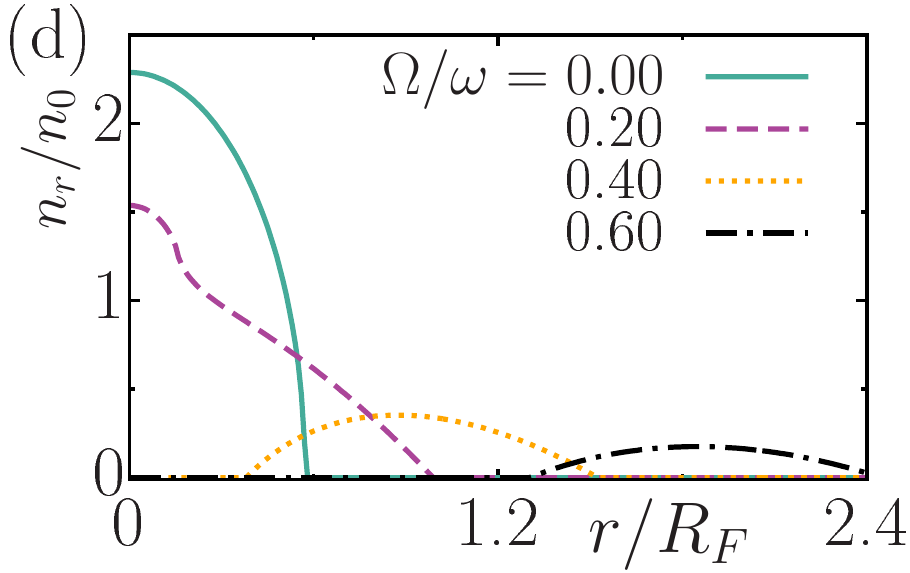}
\caption{(Color online) 
Non-interacting number-density maps at $T = 0$ with changing rotation frequency for 
(a) $\alpha=1k_{F}/E_{F}$ and 
(b) $\alpha=4k_{F}/E_{F}$.  
A few exemplary radial density profiles are plotted in (c) and (d), showing explicitly that the 
Rashba-coupled Fermi gas eventually takes on a ring-shaped annulus with increasing $\Omega$.  
\label{fig:freeN2}
} 
\end{figure}

Alternatively, the depletion of the number density at the trap center and the 
accompanying formation of a ring-shaped annulus can also be deduced 
analytically as follows. First of all, the central density turns out to be
$
n_{r=0}=M(\alpha^{2}M+\mu)/\pi
$ 
for $\mu\geq0$, and 
$
n_{r=0}=\alpha M^{2}\sqrt{\alpha^{2}+2\mu/M}/\pi
$ 
for $\mu<0$, showing explicitly that $n_{r=0} = 0$ when the parameters
satisfy the critical condition $\alpha^{2}+2\mu/M\leq 0$. Here, $\Omega$ 
enters into this condition implicitly through its dependence on $\mu$. 
Then, we note that the locations at which the number density vanishes, 
i.e., the local Fermi surface disappears, are exactly the inner and outer radii 
forming the edges of the gas. Since the presence of a local Fermi surface implies 
real solutions for $k^{s}_{1,2}$, we find the edges by setting the square root to 
zero in Eq.~(\ref{eq:fermisurf}), i.e., $M(\Omega r-s\alpha)^{2}+2\mu_r=0$,
leading to
\begin{equation}
R_{I,O}^{0}=R_{F}\frac{\omega\Omega\alpha\pm\omega\sqrt{\alpha^{2}\omega^{2}
+2\mu(\omega^{2}-\Omega^{2})/M}}{k_{F}(\omega^{2}-\Omega^{2})/M},
\label{eq:RE} 
\end{equation} 
where $I$ ($O$) denotes the inner (outer) edge. The gas may form an annulus 
only if the inner radius satisfies $R_{I}^{0} \geq 0$, leading again to the critical 
condition $\alpha^{2}+2\mu/M\leq0$. Thus, the critical rotation frequency 
$\Omega_c^{0}$ for the emergence of such an annulus can be calculated 
by a self-consistent solution of the number equation together with the critical
condition $\alpha^{2}+2\mu/M=0$. The resultant phase diagram is shown in 
Fig.~\ref{fig:Om_c0}, where $\Omega_c^{0}$ decreases monotonically with 
increasing $\alpha$, and it is in perfect agreement with our fully-numerical 
solutions for the trap profiles as illustrated below. We also note in passing that 
while the gas never forms an annulus in the $\alpha\to0$ limit given the
upper bound on $\Omega$ as the harmonic potential can only trap the particles 
for $\Omega < \omega$, it may form an annulus at any finite $\alpha$ with 
$\Omega_c^{0}<\omega$.

In Figs.~\ref{fig:freeN} and~\ref{fig:freeN2}, the trap profiles are shown for a wide range 
of parameters. For instance, we set $\Omega=0.15\omega$ in Figs.~\ref{fig:freeN}(a) 
and~\ref{fig:freeN}(c), and plot the number-density maps in the entire trap as a 
function of $\alpha$ together with a few exemplary radial density profiles. As discussed 
above, the Rashba coupling not only increases the LDOS around the trap center but it also 
favors energetically some finite-angular-momentum states, causing simultaneously 
an increase in the central density due to the former effect and an expansion of the edge 
due to the latter one as a function of $\alpha$. This competition sharply decreases 
the number density away from the trap center. Once $\alpha$ approaches to the 
critical value $\alpha=\sqrt{-2\mu/M}$, the latter effect gradually dominates leading 
to a reduction in the central density as the gas continues to expand. There is an 
intriguing appearance of an additional local maximum in the number density in the 
vicinity of critical $\alpha$, beyond which it is completely depleted at the trap center, 
and the radius $R_{I}^{0}$ of the depleted region grows linearly with $\alpha$. 
We consider a higher $\Omega=0.5 \omega$ in Figs.~\ref{fig:freeN}(b) 
and~\ref{fig:freeN}(d), showing that the central density decreases and the edge expands 
immediately with increasing $\alpha$, leading to the depletion of the trap center at a 
much lower critical $\alpha$. Similarly, these effects are also seen in Fig.~\ref{fig:freeN2}, 
where $\Omega$ is increased at fixed $\alpha$ values. In contrast to Fig.~\ref{fig:freeN}, 
here we see that a faster rotation leads to a monotonic reduction of the central 
density and a monotonic expansion of the edge.

\begin{figure*}[ht!]
\includegraphics[scale=.54]{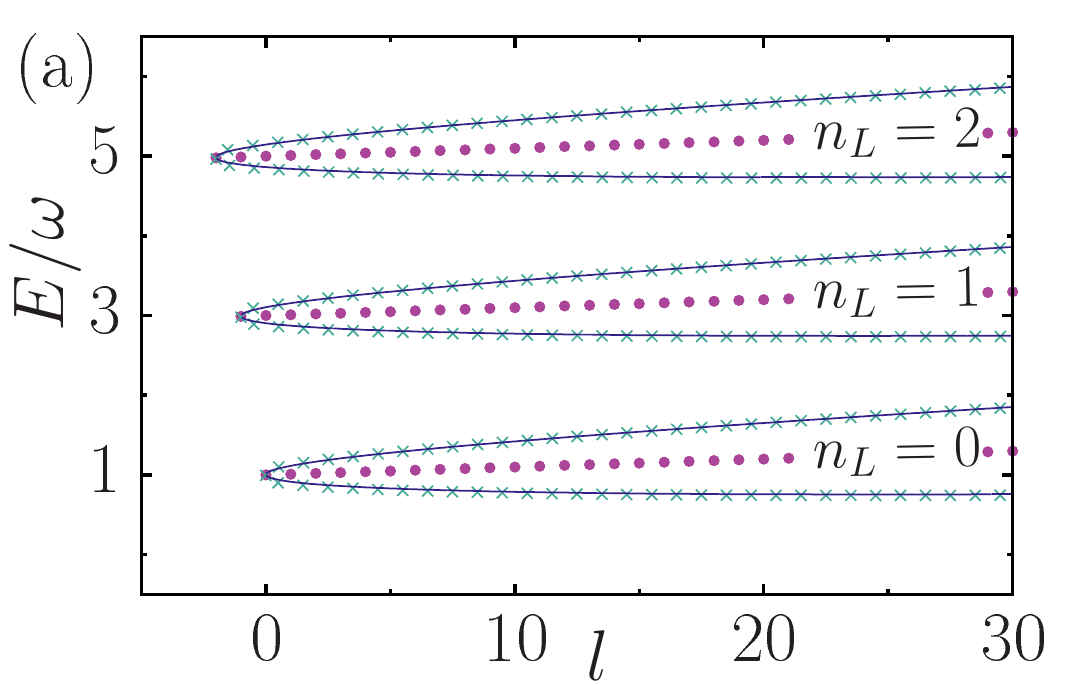}
\includegraphics[scale=.54]{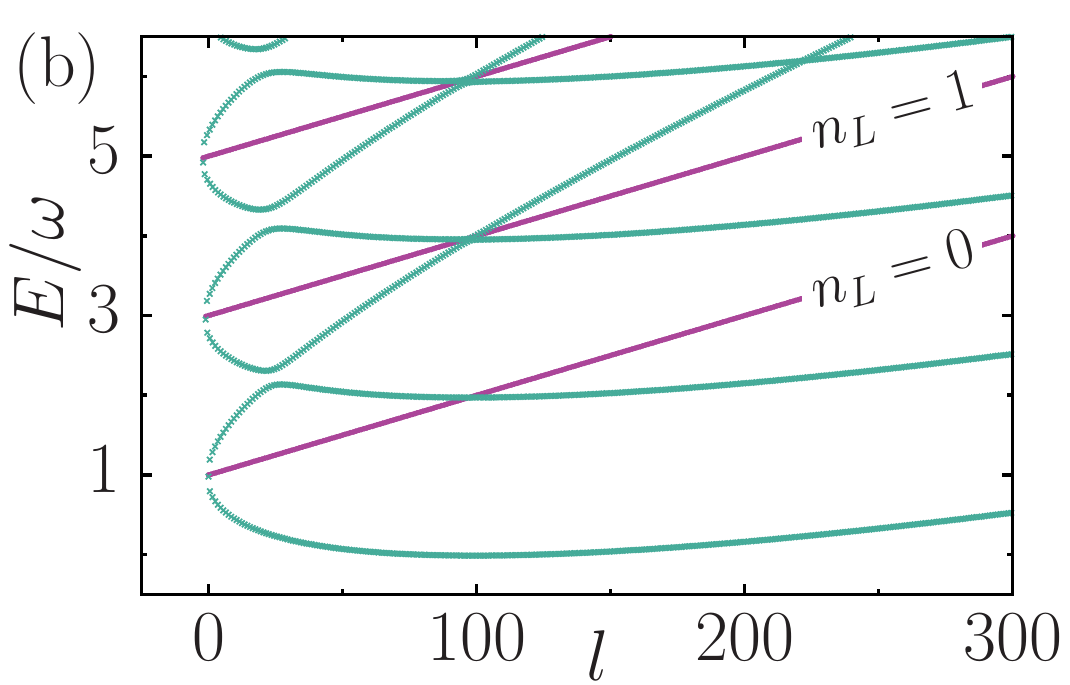}
\includegraphics[scale=.54]{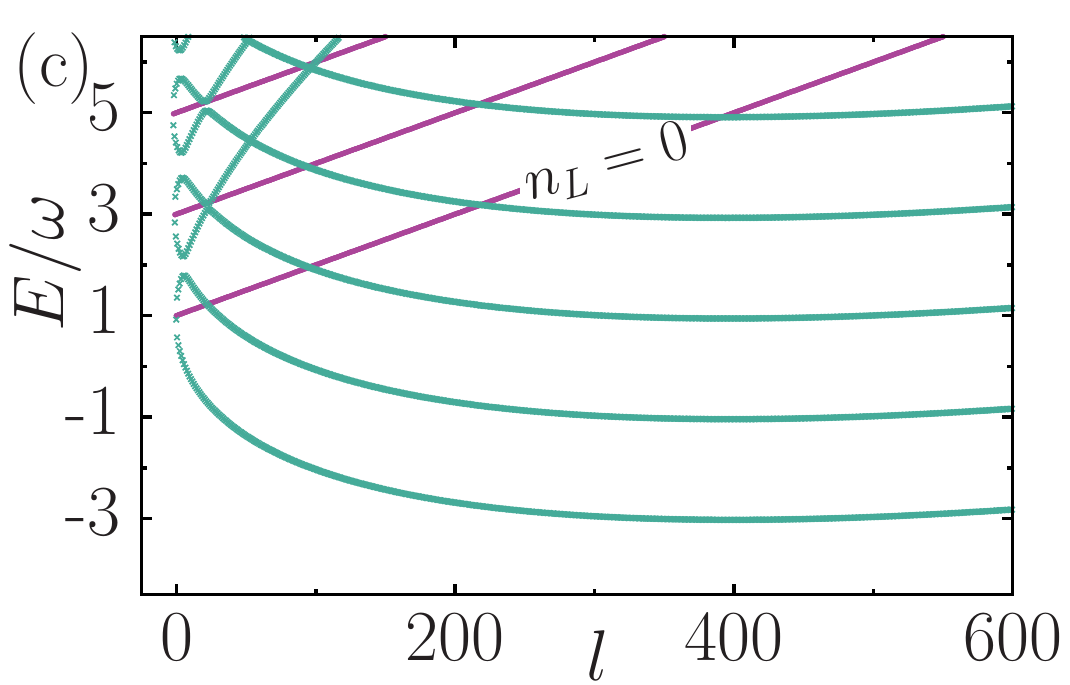}
\caption{(Color online) 
Energy levels as a function of angular momentum for a rapidly-rotating 
Fermi gas with Rashba coupling in the Landau regime when 
$\Omega=0.99\omega$. Here, $\alpha = 0$ limit is shown in purple dots
for the first three Landau levels as a reference.
(a) $\alpha=0.2 \sqrt{\omega/M}$:
Rashba coupling splits each of the Landau levels into two helicity branches 
with a widening energy gap in between as a function of angular momentum.
(Blue solid line corresponds to the perturbation expression given in the text.)
(b) $\alpha=0.4 \sqrt{\omega/M}$: the helicity branches display avoided
level crossings. 
(c) $\alpha=0.8 \sqrt{\omega/M}$: all of the negative-helicity branches 
not only occupy lowest energies but they also develop minimum at 
finite angular momentum. 
\label{fig:LE_fig}
}
\end{figure*}
\begin{figure*}[ht!]
\includegraphics[scale=.54]{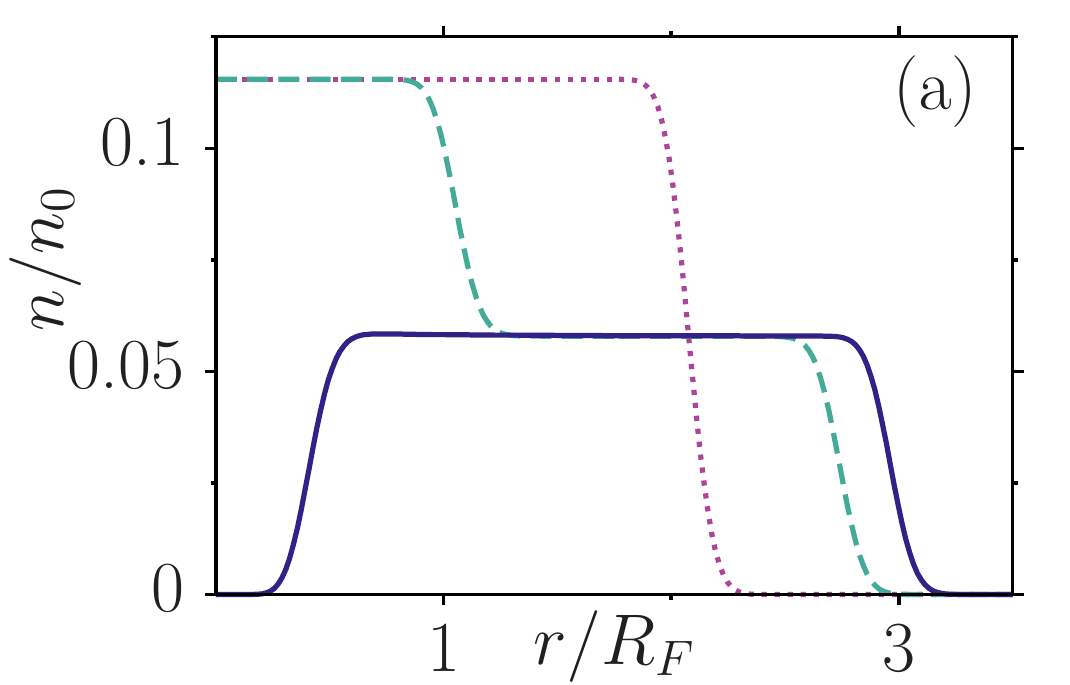}
\includegraphics[scale=.54]{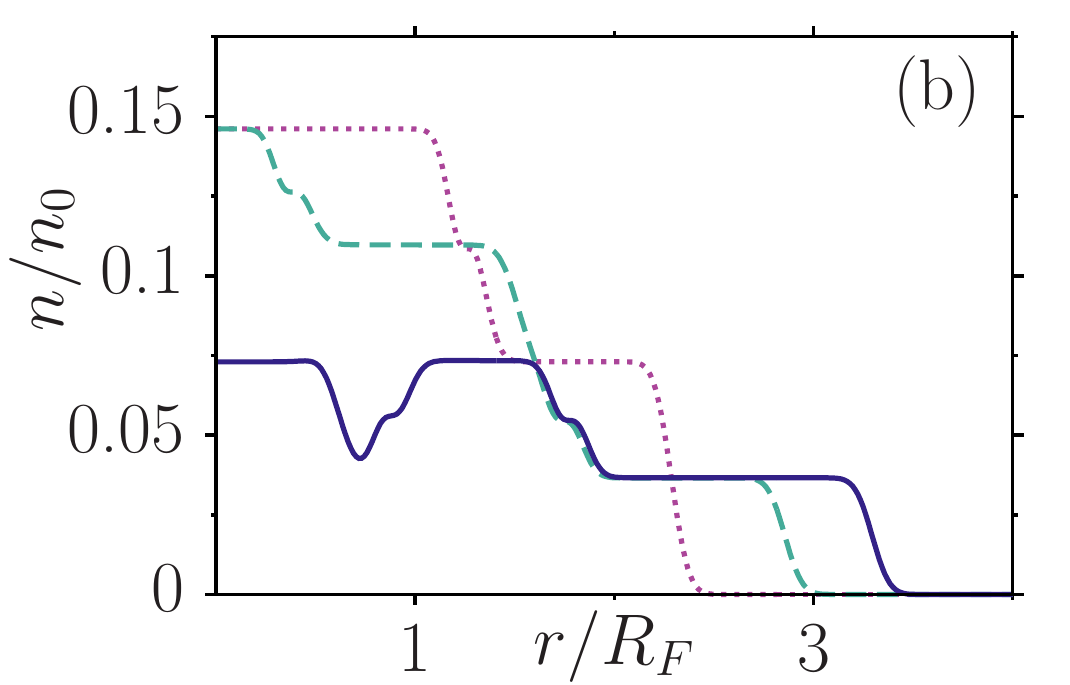}
\includegraphics[scale=.54]{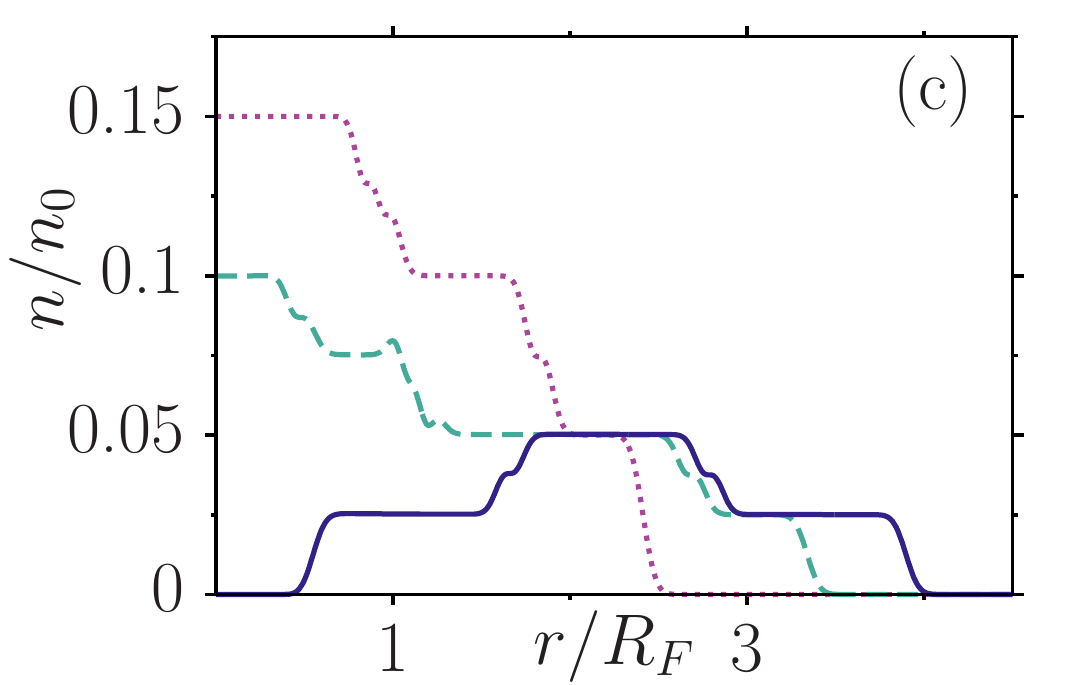}
\caption{(Color online) 
Radial density profiles at $T = 0$ in the Landau regime when $\Omega=0.99\omega$. 
The purple-dotted, green-dashed and blue-solid lines correspond, respectively, 
to higher Rashba couplings (in units of $\sqrt{\omega/M}$) where 
$\alpha=\{0,0.2,0.4\}$ in (a) with $N=750$ particles in the lowest-Landau level, 
$\alpha=\{0,0.2,0.4\}$ in (b) with $N=900$ particles in the first two Landau levels, and 
$\alpha=\{0,0.4,0.8\}$ in (c) with $N=1600$ particles in the first three Landau levels.
These intriguing profiles directly reflect the corresponding energy-level 
structures shown in Fig.~\ref{fig:LE_fig}, where the higher the angular momentum 
of the single-particle state the further away its localization distance from the 
trap center.
\label{fig:landau_dens}
}
\end{figure*}
\subsection{Rapidly-rotating Fermi gas with Rashba coupling 
\\ ($\omega\protect\neq0$, $\alpha\protect\neq0$ and $\Omega\protect \to \omega$)}\label{sec:LLL}

Since the effects of rotation are analogous to those of an effective magnetic 
field on a particle, where the Coriolis force on a neutral atom mimics the Lorentz 
force on a charged particle, a rapidly-rotating Fermi gas may form 
highly-degenerate Landau levels in the $\Omega\to\omega$ limit and 
exhibit an integer quantum-Hall effect~\cite{Ho2000}. 
Since the LDA approach fails to capture the correct physics in the Landau 
regime of a rapidly-rotating Fermi gas, we resort to exact quantum-mechanical 
calculations in the following discussion.

In the absence of a Rashba coupling, and assuming $\omega_-=\omega-\Omega$ 
is small, it is convenient to label the single-particle states with $n_L=(n-l)/2$ 
the Landau-level index and $l$ the angular momentum, leading to the
dispersion relation
$
\varepsilon_{n_L l}=\omega (2n_L+1)+\omega_- l.
$ 
Here, $n_L = \{0,1,2,\cdots\}$ and $l = \{0,1,2,\cdots\}$, so that $\varepsilon_{n_L l}$ 
increases linearly with $l$. Note that all of the consecutive Landau levels are 
separated by an equal gap $2\omega$ for any given $l$, and that each of these 
energy states are two-fold degenerate due to the pseudo-spin $\sigma$ of the 
particles. We set $\Omega=0.99\omega$ in Fig.~\ref{fig:LE_fig}, and show 
this spectrum in purple for the first three Landau levels as a reference. 
Since each fully-filled Landau level below a given $\mu$ results in a uniform
density, this spectrum gives rise to a ziggurat-shaped density profile in the 
trap, where the number of plateaus directly reflects the number of underlying 
Landau levels involved. For instance, such staircase-looking number densities 
are clearly visible in Figs.~\ref{fig:landau_dens}(a),~\ref{fig:landau_dens}(b) 
and~\ref{fig:landau_dens}(c), where $\alpha = 0$ limits are shown as the 
purple-dotted lines corresponding, respectively, to one, two and three fullly-filled 
Landau levels. 

When $\alpha \ne 0$, by breaking the spin-rotation symmetry, the Rashba 
coupling lifts the spin degeneracy, leading to two ($s=\pm$) helicity branches 
for each Landau level. For instance, in the perturbative regime when 
$\alpha\ll 2 a_0 \omega$ with $a_0=1/\sqrt{M\omega}$ the characteristic 
harmonic-oscillator length scale, we approximately find
$
\varepsilon_{n_L,l,s}=\omega (2n_L+1)+\omega_- l
+\omega_-/ [ 2 + 2s \sqrt{1+\alpha^2(n_L+l+1)/(a_0 \omega_-)^2} ].
$
We checked that this expression is in excellent agreement with all of the exact 
results presented in Fig.~\ref{fig:LE_fig}, e.g., our perturbative results (solid blue lines) 
are shown to lie right on top of the exact ones (green crosses) in Fig.~\ref{fig:LE_fig}(a).
As $l$ gets higher, Fig.~\ref{fig:LE_fig}(b) shows that the helicity branches ultimately 
display avoided level crossings. In addition, as $\alpha$ is increased to 
$\alpha\gtrsim a_0 \omega_-$, all of the negative-helicity branches develop a 
minimum near some finite $l$ given by
$
l_\textrm{min}=\alpha^2/(4 a_0 \omega_-)^2.
$ 
This is shown in Fig.~\ref{fig:LE_fig}(c) for $\alpha=0.8 \sqrt{\omega/M}$, 
where all of the energy gaps between the two consecutive same-helicity branches
are approximately $2\omega$ in this perturbative regime.
Thus, the main effect of weak Rashba coupling on the trap profiles is expected to 
be doubling of the number of plateaus in the number density as a reflection 
of the lifted spin degeneracy of the Landau levels. In addition, the interplay of 
Rashba coupling and rapid rotation may ultimately lead to the formation of a 
ring-shaped annulus with a ziggurat texture as shown in Fig.~\ref{fig:landau_dens}(c).

\begin{figure}
\includegraphics[scale=.7]{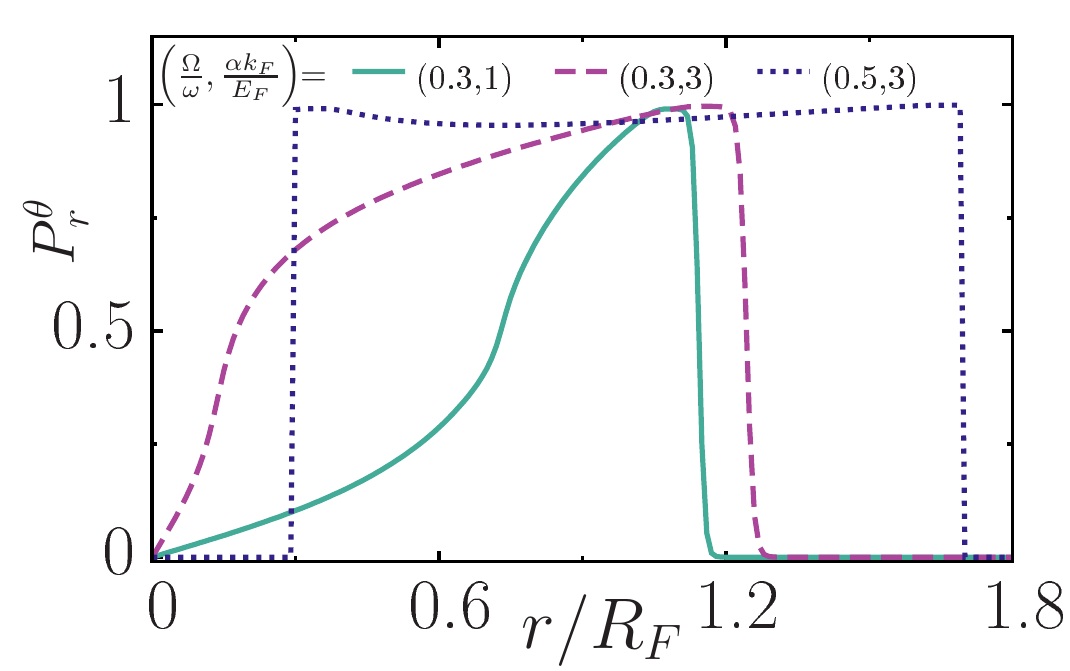}
\caption{(Color online) 
The azimuthal component of the average spin-polarization increases 
monotonically from zero at the trap center to unity at the edge, and the 
gas is almost fully polarized within the ring-shaped annulus.  
\label{fig:pol_fig}
}
\end{figure}
\subsection{Spin-polarization textures}
\label{sec:NIpol}

Since the Rashba coupling splits the spin degeneracy of the free-particle energy 
bands into $\pm$-helicity branches with spins oriented parallel/anti-parallel 
to the momentum $\bk$, and the rotation favors momentum states that are 
parallel to the direction of mass-current density, their interplay polarizes 
the average spin of the particles in the azimuthal direction. The azimuthal 
component of the local average spin-polarization texture can be written as
$
P_{\br}^{\theta}=-P_{\br}^{x} \sin \theta_{\br} +P_{\br}^{y} \cos \theta_{\br},
$
with its components defined in Sec.~\ref{sec:lda}. In the absence of a Zeeman 
field as considered in this paper, we note that the average spin of the system is 
unpolarized not only in the $\alpha\to0$ as the spin and $\bk$ are uncoupled, 
but also in the $\Omega \to 0$ limit as the contributions of $\pm \bk$ states are 
equal in magnitude but opposite in direction. Therefore, the interplay between 
Rashba coupling and adiabatic rotation is proved to be crucial for the appearance 
of spin textures.

For instance, the radial spin-polarization profiles are shown in Fig.~\ref{fig:pol_fig} 
for three sets of $\alpha$ and $\Omega$. The azimuthal polarization increases 
from zero at the trap center as $r = 0$ is immune to the direct-effects of 
rotation, and reaches unity at the edge as only the non-degenerate lowest-energy 
states of the negative-helicity branch having anti-parallel spin orientations with 
respect to the angular direction are occupied. As a result of the disappearance 
of the positive-helicity band and increasing asymmetry in the energy dispersion,
we find that increasing $\alpha$ and $\Omega$ gradually increases the 
polarization in the intermediate region between the trap center and the edge as well.
Furthermore, we also find that the polarization approaches to unity everywhere in 
the trap once the number density forms a ring-shaped annulus, and this is shown 
in Fig.~\ref{fig:pol_fig} when $\Omega=0.5 \omega$ and $\alpha=3 E_F/k_F$.

\begin{figure}[ht!]
\includegraphics[scale=0.085]{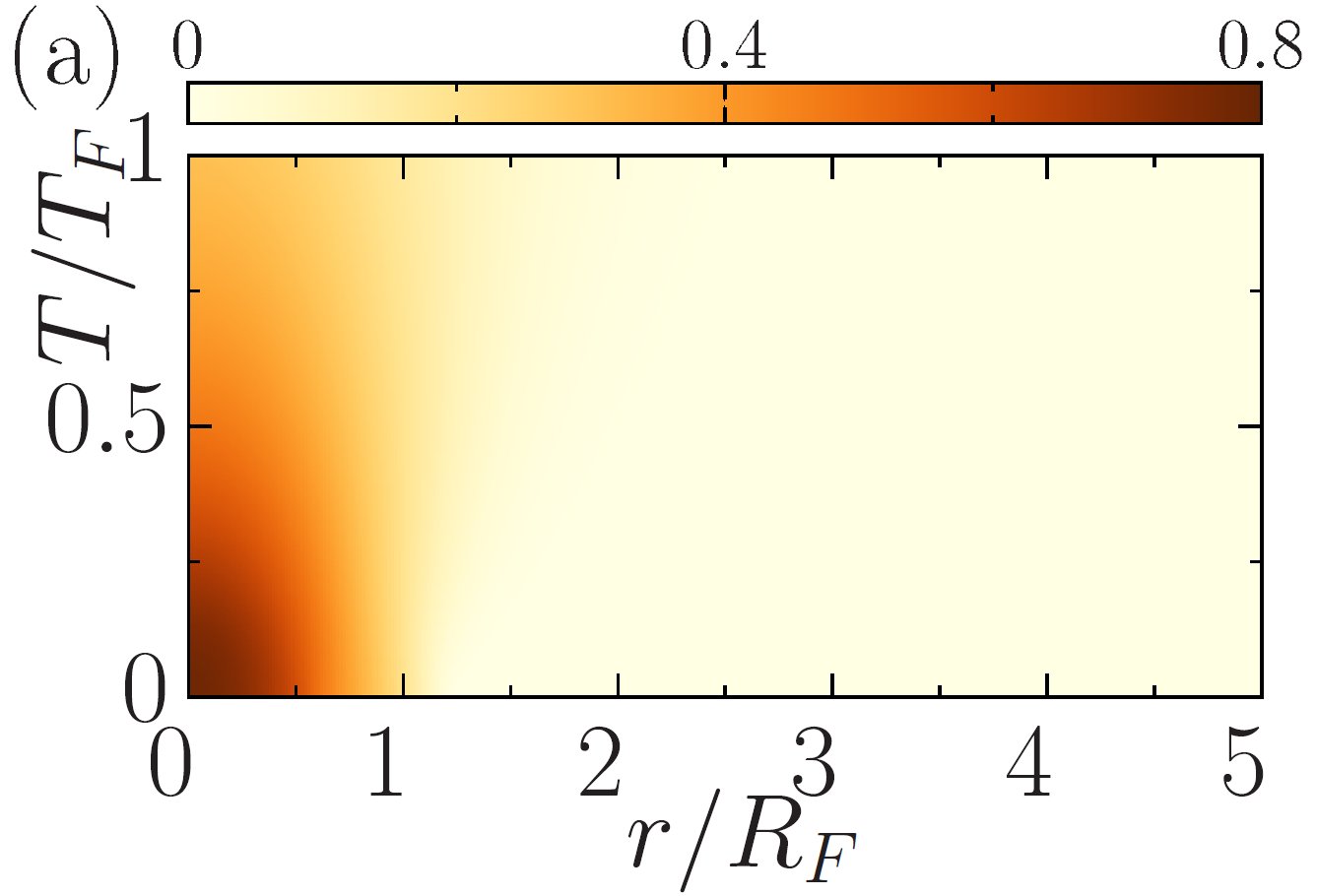}
\includegraphics[scale=0.085]{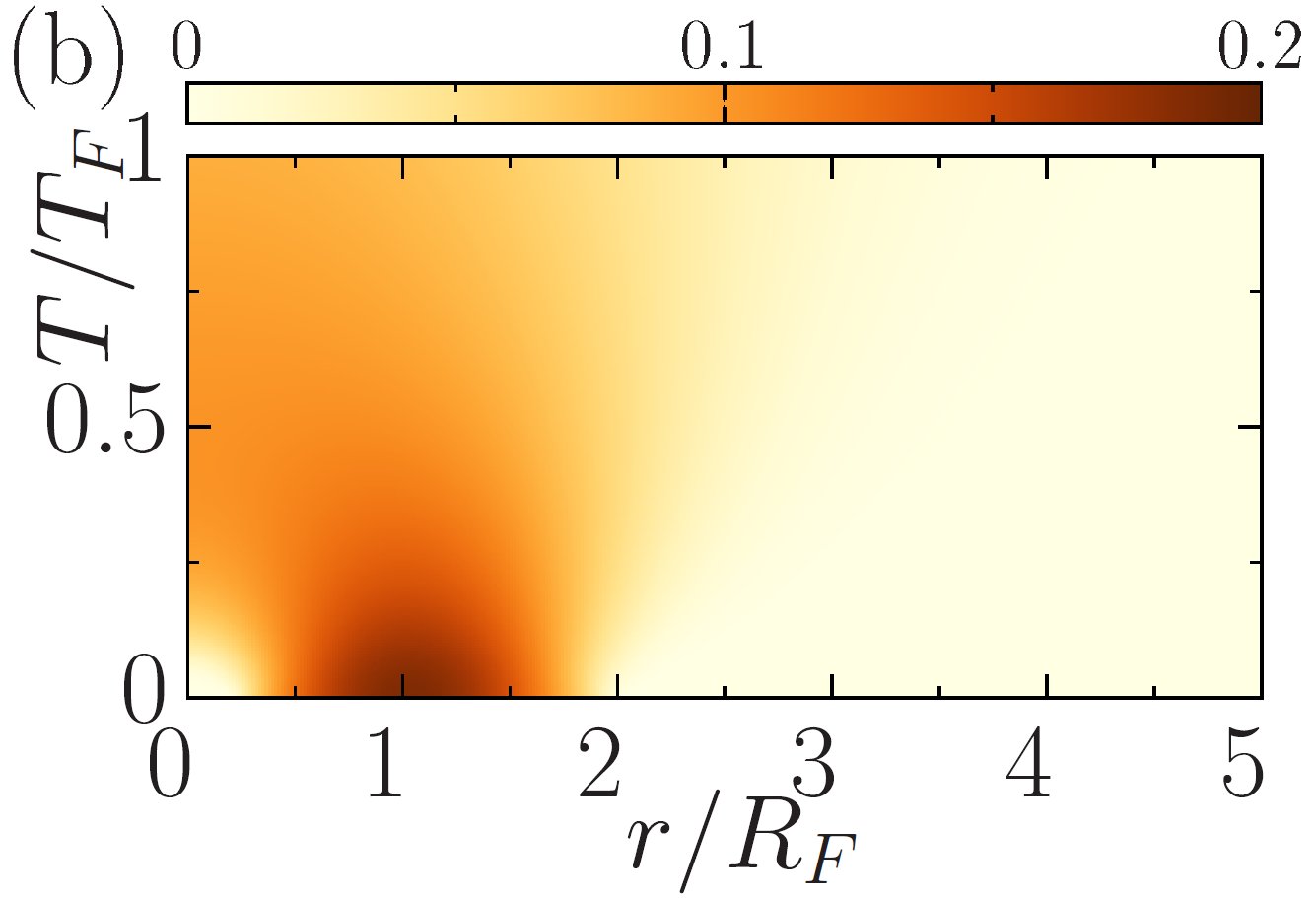} \\
\includegraphics[scale=0.085]{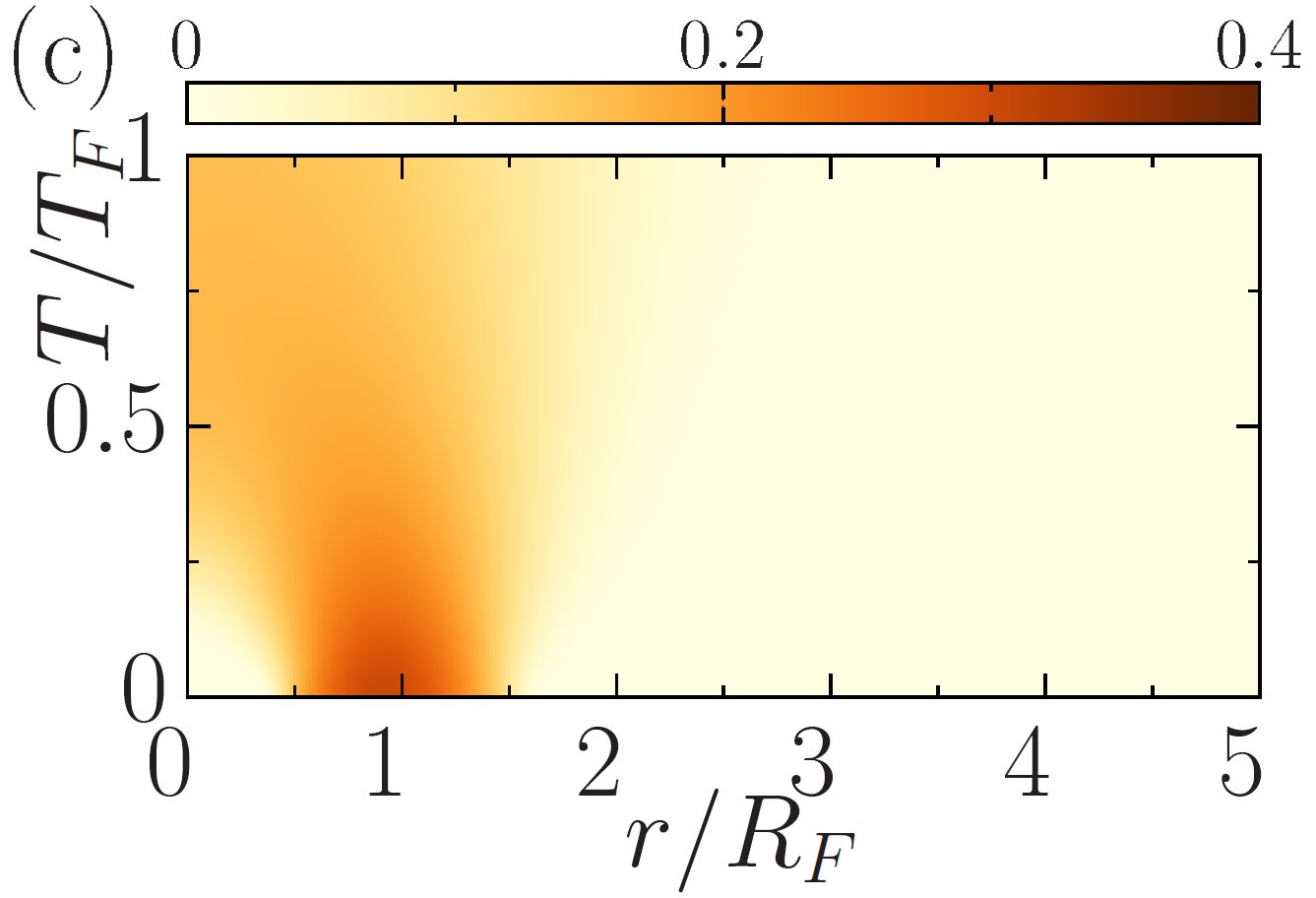}
\includegraphics[scale=0.085]{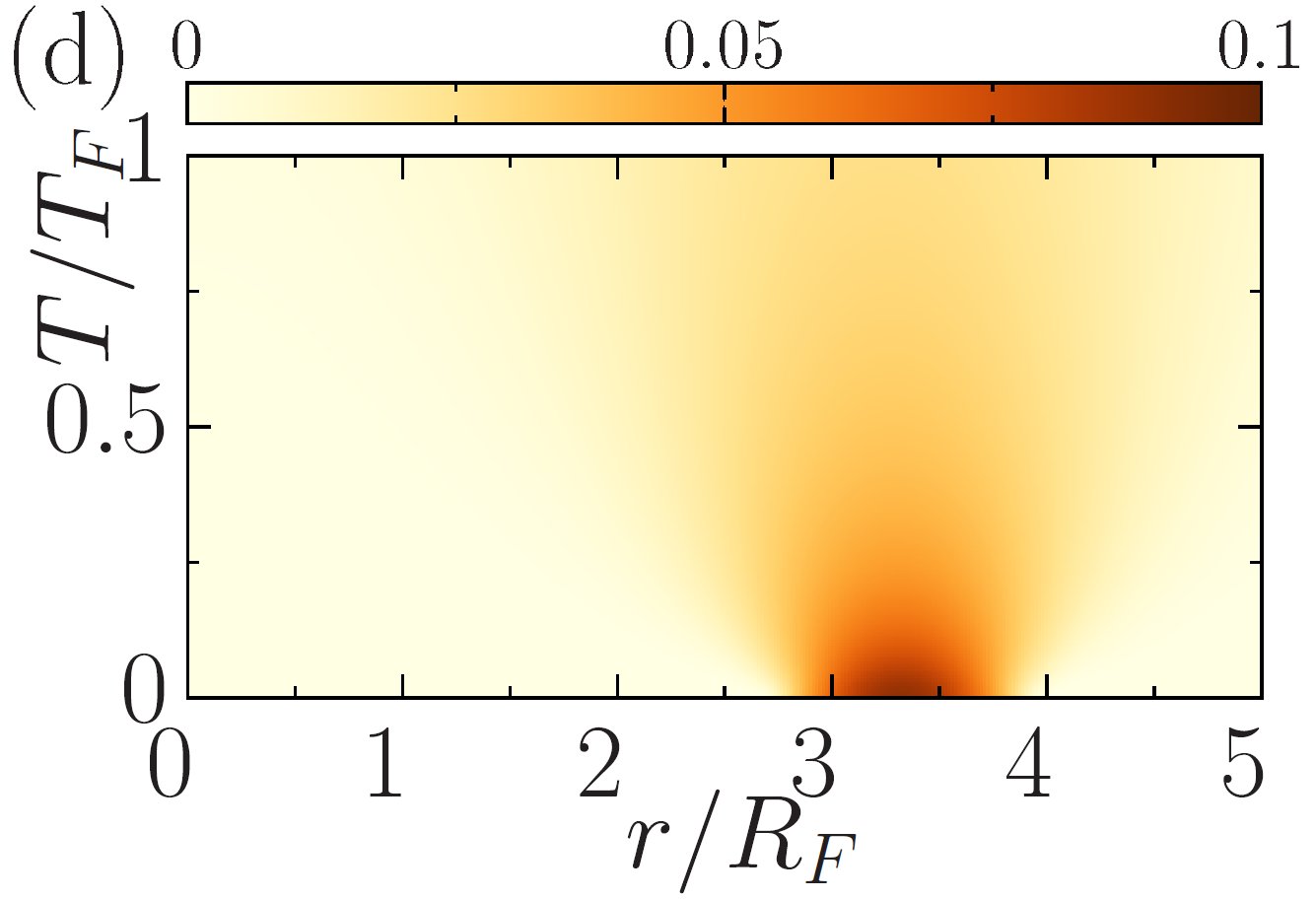}
\caption{(Color online) 
Non-interacting number-density maps at finite $T$. Increasing either $\Omega$ 
or $\alpha$ from 
(a) $\Omega=0.5 \omega$ and $\alpha=1E_F/k_F$, the disk-shaped profile
transform into a ring-shaped annulus for 
(b) $\Omega=0.8 \omega$ and $\alpha=1E_F/k_F$ and
(c) $\Omega=0.5\omega$ and $\alpha=3E_F/k_F$. 
While the depletion of the central density that is clearly visible 
at low $T$ is eventually blurred by the thermal broadening, it can be 
sustained at very high $T$ with increasing either $\Omega$ and/or $\alpha$. 
This is shown in (d) for $\Omega=0.8 \omega$ and $\alpha=3E_F/k_F$.  
\label{fig:Den_T}
} 
\end{figure}
\subsection{Thermal effects}
\label{sec:NIT}

Before moving to the effects of interactions, here we conclude the non-interacting 
Fermi gas section by addressing how much of our zero-temperature trap profiles 
survives at finite $T$. For this purpose, we fix $\Omega$ and $\alpha$ in Fig.~\ref{fig:Den_T}, 
and plot the number-density maps in the entire trap as a function of $T$. 
It is clearly shown that the thermal-broadening effects on the number density 
are most significant in the ring-shaped regions, as the system eventually recovers 
the disk-shaped profile at sufficiently high $T$. However, it is encouraging to see that, 
by increasing either $\Omega$ and/or $\alpha$, a visible ring-shaped annulus may 
still form at very high $T$ that is of the order of a Fermi temperature $T_F=E_F/k_B$, 
making its experimental observation quite feasible. We note that the number density 
first appears as nearly flat in a wide region, the width of which is of the order of $R_F$, 
at some intermediate $T$, and then it ultimately attains the usual Gaussian shape 
at high $T$.

Having convincingly shown that the interplay between the effects of Rashba coupling 
on the LDOS and the Coriolis effects caused by rotation gives the number 
density of a non-interacting Fermi gas a characteristic ring-shaped annulus form
that survives at experimentally-accessible temperatures, we next analyze how this 
interplay effects the trap profiles of an interacting Fermi gas, and the associated 
SF properties.

\section{Interacting Fermi gas} 
\label{sec:MFT}

Armed with a thorough understanding of the generic properties of a non-interacting 
Fermi gas with Rashba coupling under adiabatic rotation, we are ready to discuss 
the effects of interaction as characterized by the two-body binding energy $E_b > 0$
in vacuum. The SF ground state of an interacting Fermi gas is protected by an energy 
gap in the low-energy excitation spectrum, and the gap is directly related to the 
order parameter $\Delta_\br$ of the underlying Cooper pairs that are made of 
time-reversed particles. Thus, while this order parameter acts as an energy barrier 
and protects the pairs, by breaking the time-reversal symmetry, rotating a SF Fermi 
gas may energetically favor a state with broken pairs beyond a critical rotation 
frequency. Our primary objective here is to study such a pair-breaking mechanism 
that is induced by the Coriolis effects on superfluidity, where we calculate the critical 
rotation frequencies both for the onset of pair breaking and for the complete destruction 
of superfluidity in the system. In particular, by comparing the results of 
fully-quantum-mechanical BdG approach with those of semi-classical LDA one, 
we construct extensive phase diagrams consisting of non-rotating gapped SF, 
partially-rotating gSF and rigidly-rotating N regions. These diagrams allow us to 
predict all sorts of phase profiles in the trap for a wide-range of parameter regimes, where the 
interplay between Rashba coupling and adiabatic rotation may favor, e.g., an outer 
N edge that is completely phase separated from the central SF core by vacuum. 

\begin{figure} 
\includegraphics[scale=.46]{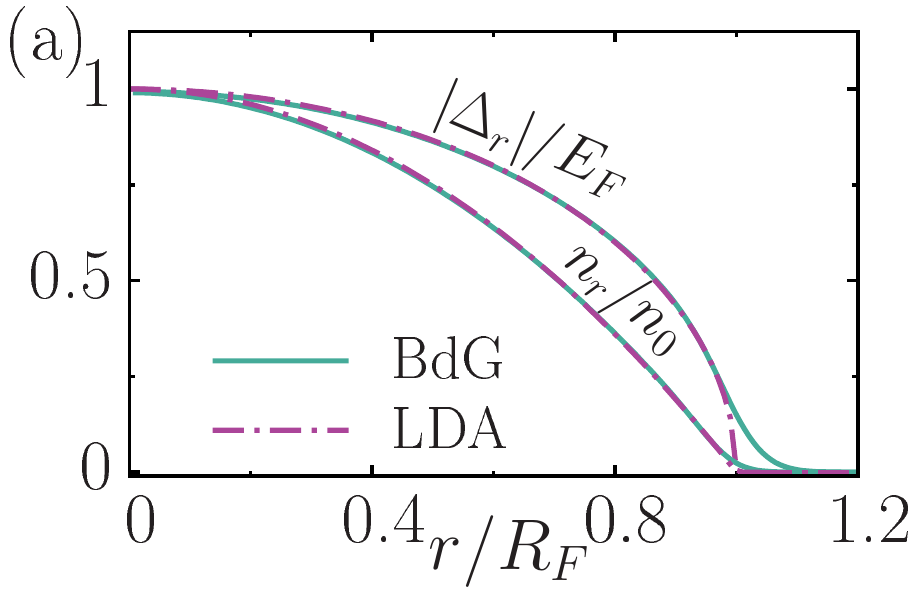}
\includegraphics[scale=.46]{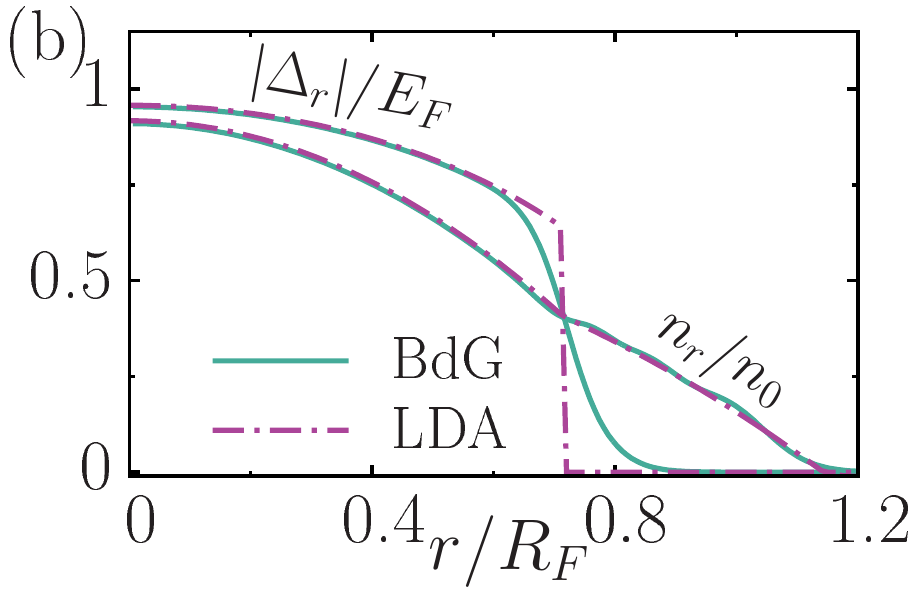}
\includegraphics[scale=0.46]{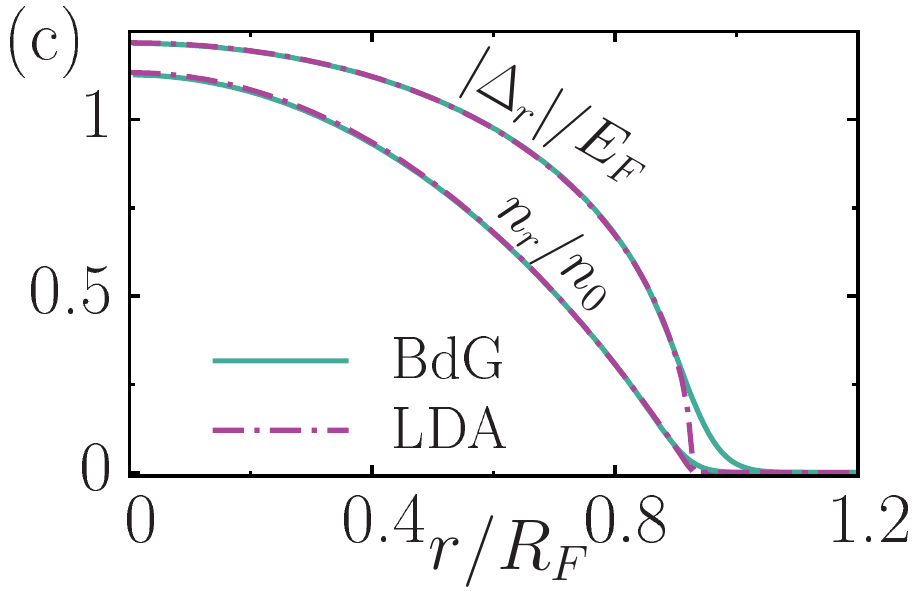}
\includegraphics[scale=0.46]{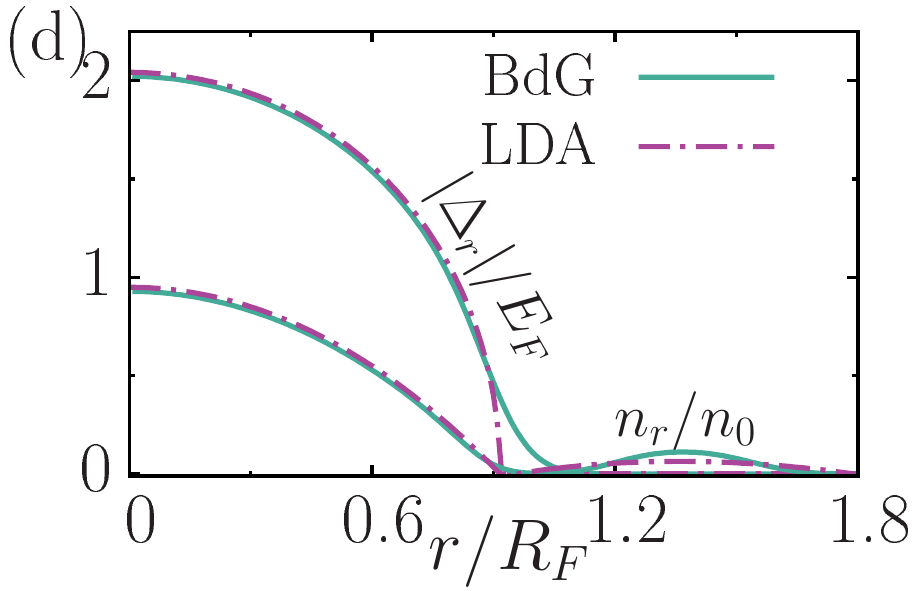}
\caption{(Color online) 
The radial order-parameter and number-density profiles at $T = 0$ showing an excellent
agreement between the LDA and BdG results, where
(a) $E_b=0.5 E_F$, $\alpha=0$ and $\Omega=0$, 
(b) $E_b=0.5 E_F$, $\alpha=0$ and $\Omega=0.7\omega$,
(c) $E_b=0.5 E_F$, $\alpha=2 E_F/k_F$ and $\Omega=0$ with $N=500$ particles, and
(d) $E_b=2 E_F$, $\alpha=2 E_F/k_F$ and $\Omega=0.7\omega$ with $N=100$ particles.  
\label{fig:LDAvsBdG}
} 
\end{figure}

Similar to Sec.~\ref{sec:NI}, we again rely mostly on the LDA approach throughout 
this section, given that the LDA results are in very good agreement with those of 
BdG ones for a wide-range of parameter regimes and with the additional advantage 
that they permit analytical insights into the limiting cases. 
For instance, we benchmark our LDA and BdG results in Fig.~\ref{fig:LDAvsBdG},
where we set $N=500$ particles with $E_b=0.5E_F$ and $E_c\sim 27 E_F$ 
in Fig.~\ref{fig:LDAvsBdG}(a),~\ref{fig:LDAvsBdG}(b) and~\ref{fig:LDAvsBdG}(c),
and $N=100$ particles with $E_b=2E_F$ and $E_c\sim 60 E_F$ in 
Fig.~\ref{fig:LDAvsBdG}(d). It is not surprising that the LDA approach works really 
well in the regions where the changes in $|\Delta_\br|$ and $n_r$ are slow. 
To gain as much insight as possible into the basic properties of an interacting
Fermi gas, we again discuss these quantities first in a few analytically-tractable limits, 
prior to the presentation of our numerical results for the generic case with 
arbitrary $\Omega$ and $\alpha$.

\subsection{Trapped Fermi gas 
\\ ($\omega\protect\neq0$, $\alpha=0$ and $\Omega=0$) \label{sec:I}}
\label{sec:Itrap}

The first analytically-tractable limit is a usual 2D Fermi gas with neither Rashba 
coupling nor rotation, for which case the gas becomes a SF as soon as $E_b \neq 0$. 
While the energy gap of the local excitation spectrum is given by $|\Delta_r|$ 
and it is located at $k=\sqrt{2M\mu_r}$ in the local regions with $\mu_r\geq0$, 
it gradually moves towards the origin with decreasing $\mu_r$, where it ultimately 
changes to $\sqrt{\mu_r^{2}+|\Delta_{r}|^{2}}$ at $\mathbf{k} = \mathbf{0}$ in the 
local regions with $\mu_r<0$. Thus, the $\mu_r=0$ point signals a critical change 
from a BCS- to BEC-like state in the so called BCS-BEC crossover.

By integrating the order-parameter and number equations given in Eqs.~(\ref{eq:Gap})
and~(\ref{eq:Number}), it is possible to obtain $\mu=E_{F}-E_{b}/2$,
$
n_r=M(E_{F}-M\omega^{2}r^{2}/2)/\pi,
$
and
$
|\Delta_r|=\sqrt{2E_{F}E_{b}}\sqrt{1-r^{2}/R_{F}^{2}}.
$
We note that not only the dependences of $\mu$ on $E_b$ have exactly the 
same form in both trapped and uniform 2D systems~\cite{Randeria1989}, 
but also $n_r$ is independent of $E_b$ as can be seen in Sec.~\ref{sec:NItrap}. 
These peculiar results follow from the LDA approach for trap under the BCS mean-field 
approximation for pairing~\cite{He2008}. Since $|\Delta_r| \ne 0$ in the regions
where $n_r \ne 0$, the entire trapped gas is a disk-shaped SF with gapped 
excitations, and its edge is located at  $R_{O}=R_{F}$ for any $E_b$.

\subsection{Trapped Fermi gas with Rashba coupling 
\\ ($\omega\protect\neq0$, $\alpha\protect\neq0$ and $\Omega=0$)}
\label{sec:Isoc}

The second semi-analytically-tractable limit is a 2D Fermi gas with Rashba 
coupling, for which case the main effect of this coupling on the excitation 
spectrum is similar to that of the non-interacting problem. In particular, 
by shifting the minimum of the energy spectrum to finite momentum for 
the lower-energy $-$ branch, it increases the low-energy LDOS leading to
an enhanced pairing. To see this effect we first note that the spectrum of the 
$-$ branch can have one or two minima at finite momentum depending on 
$\alpha$ and $E_b$. For instance, while the energy gap $|\Delta_r|$ of the 
local spectrum is at
$
k=M \alpha+\sqrt{M^2 \alpha^{2}+2M\mu_r}
$ 
in the local regions with $\mu_r>0$, and an additional gap $|\Delta_r|$ 
also opens at
$
k=M\alpha-\sqrt{M^2\alpha^{2}+2M\mu_r}
$
in the local regions with $-M\alpha^{2}/2 < \mu_r<0$, they eventually merge 
at $k=M\alpha$ with decreasing $\mu_r$ where the gap becomes 
$
\sqrt{(\mu_r+M\alpha^{2}/2)^{2}+|\Delta_r|^{2}}
$
in the local regions with $\mu_r<-M\alpha^{2}/2$. 

\begin{figure}
\includegraphics[scale=.46]{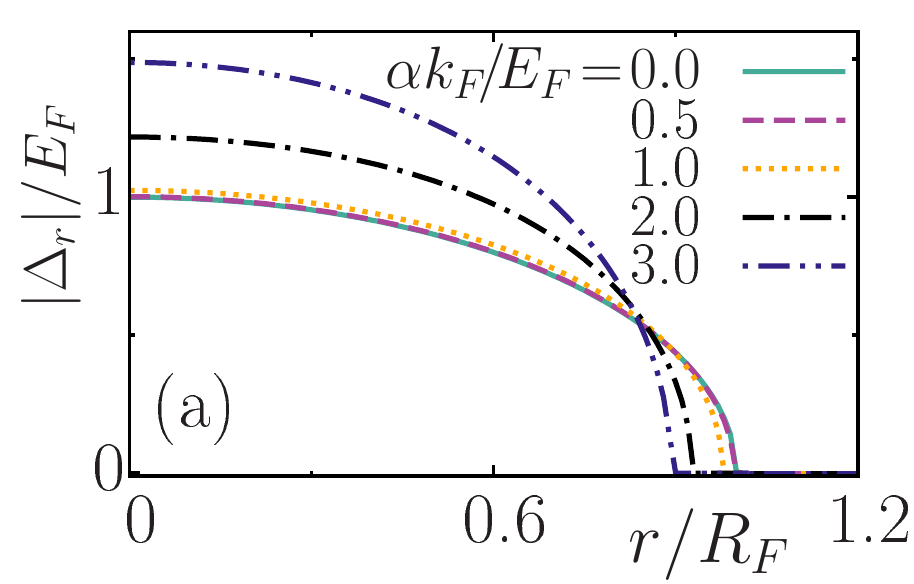}
\includegraphics[scale=0.46]{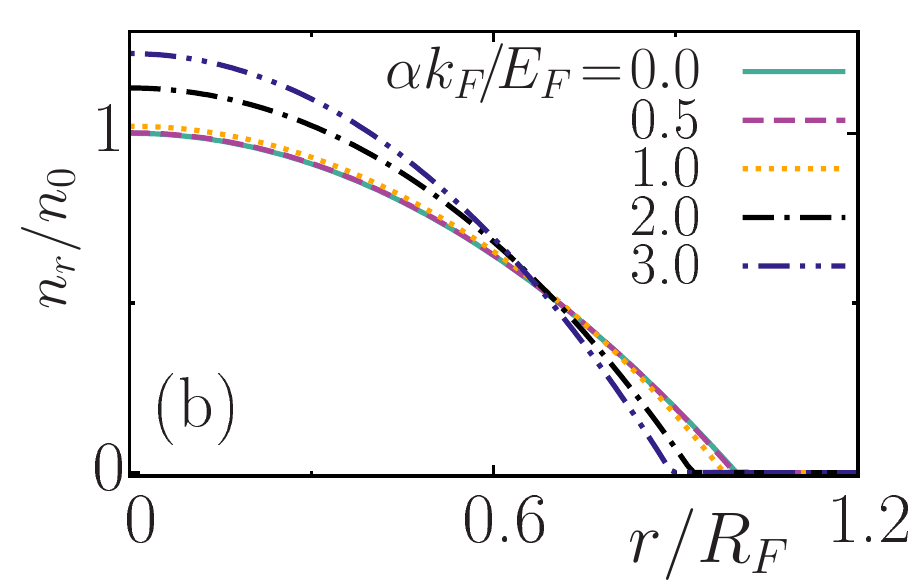}
\caption{(Color online) The radial
(a) order-parameter and 
(b) number-density 
profiles at $T = 0$ when $E_b=0.5 E_F$. 
While the radial profiles remain unchanged for $\alpha \lesssim E_F/k_F$, 
increasing $\alpha$ eventually increases both the order parameter and 
the number density at the trap center.
\label{fig:SOC_trap}
} 
\end{figure}

Even though it is not possible to obtain a closed-form analytic expression for the 
order-parameter and number equations for arbitrary $\alpha$, 
we perturbatively find
$
\mu\simeq E_{F}-E_{b}/2-M\alpha^{2}
$
and
$
|\Delta_r|\simeq\Delta_0\sqrt{1-r^{2}/R_{F}^{2}}
$
with
$
\Delta_0=\sqrt{2E_{F}E_{b}}\sqrt{1+2M^{2}\alpha^{4}/3(E_{b}+2E_{F})^{2}}
$
in the $M\alpha^{2}\ll E_{b}+2E_{F}$ limit~\cite{Chen2012}.
This suggests that neither $|\Delta_r|$ nor $n_r$ are affected much by weak 
Rashba coupling as clearly illustrated in our numerical solutions presented
in Figs.~\ref{fig:SOC_trap}(a) and~\ref{fig:SOC_trap}(b) for $E_b=0.5 E_F$ 
and $\alpha \lesssim 1E_F/k_F$. Increasing $\alpha$ energetically favors 
more and more the finite-angular-momentum states in the $-$ branch 
leading to an increased low-energy LDOS. 
When $M\alpha^{2} \sim E_{b}+2E_{F}$, we see that the monotonic 
contraction of the gas towards the trap center with $R_{O}<R_{F}$, which is 
similar to what happens in the non-interacting case, monotonically increases 
$|\Delta_r|$. Thus, the Rashba coupling alone enhances pairing in general, 
and the entire gas remains to be a disk-shaped SF with gapped excitations.

\subsection{Trapped Fermi gas with adiabatic rotation
\\ ($\omega\protect\neq0$, $\alpha=0$ and $\Omega\protect\neq0$)}
\label{sec:Irot}

Another analytically-tractable limit is a 2D Fermi gas with adiabatic rotation, 
for which case the main effect of this coupling is to break some of the 
Cooper pairs that are made of time-reversed particles within the BCS
mean-field approximation for pairing. Note that since the vortices are 
assumed not to be excited by rotation and that the gapped SF cannot 
carry any angular momentum, these broken pairs carry the extra 
angular momentum. 

\begin{figure} 
\includegraphics[scale=.35]{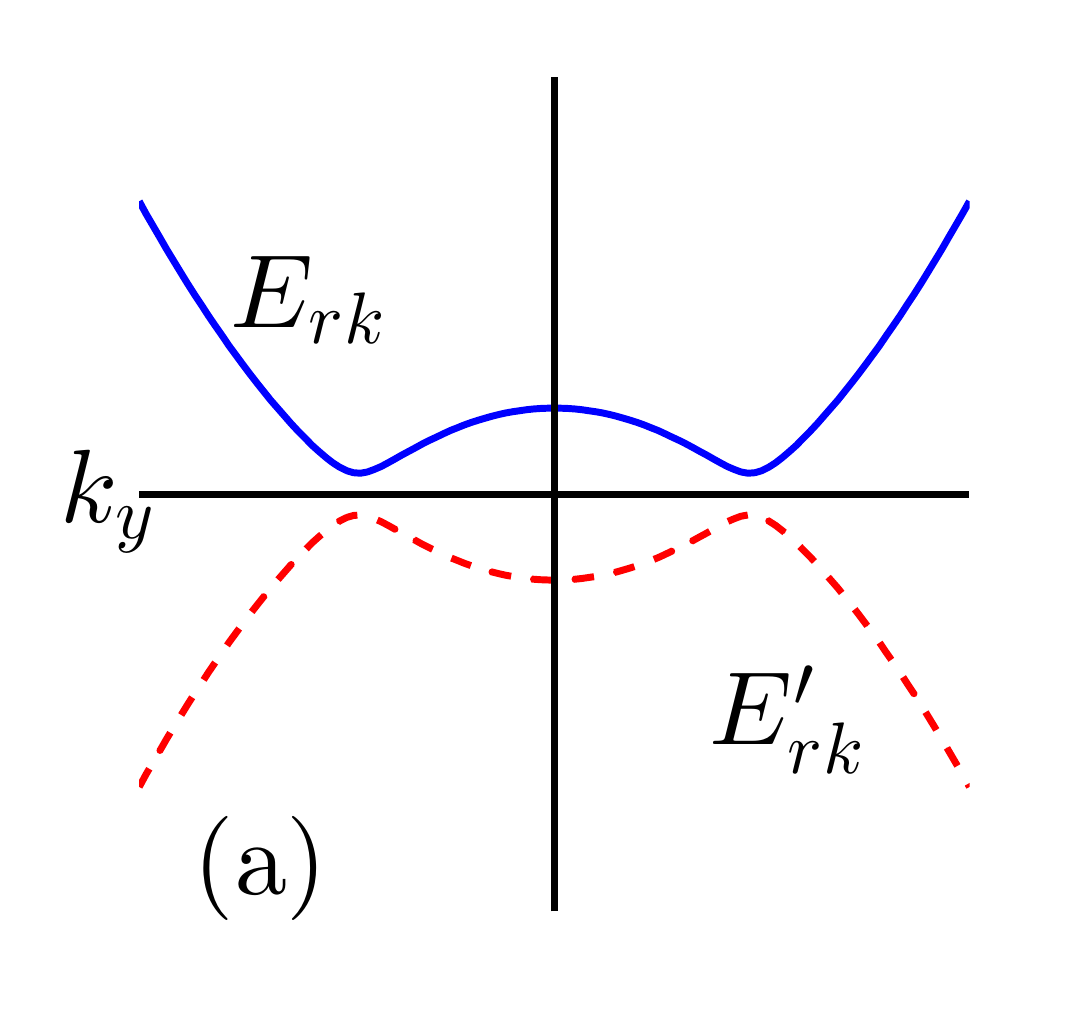}
\includegraphics[scale=0.35]{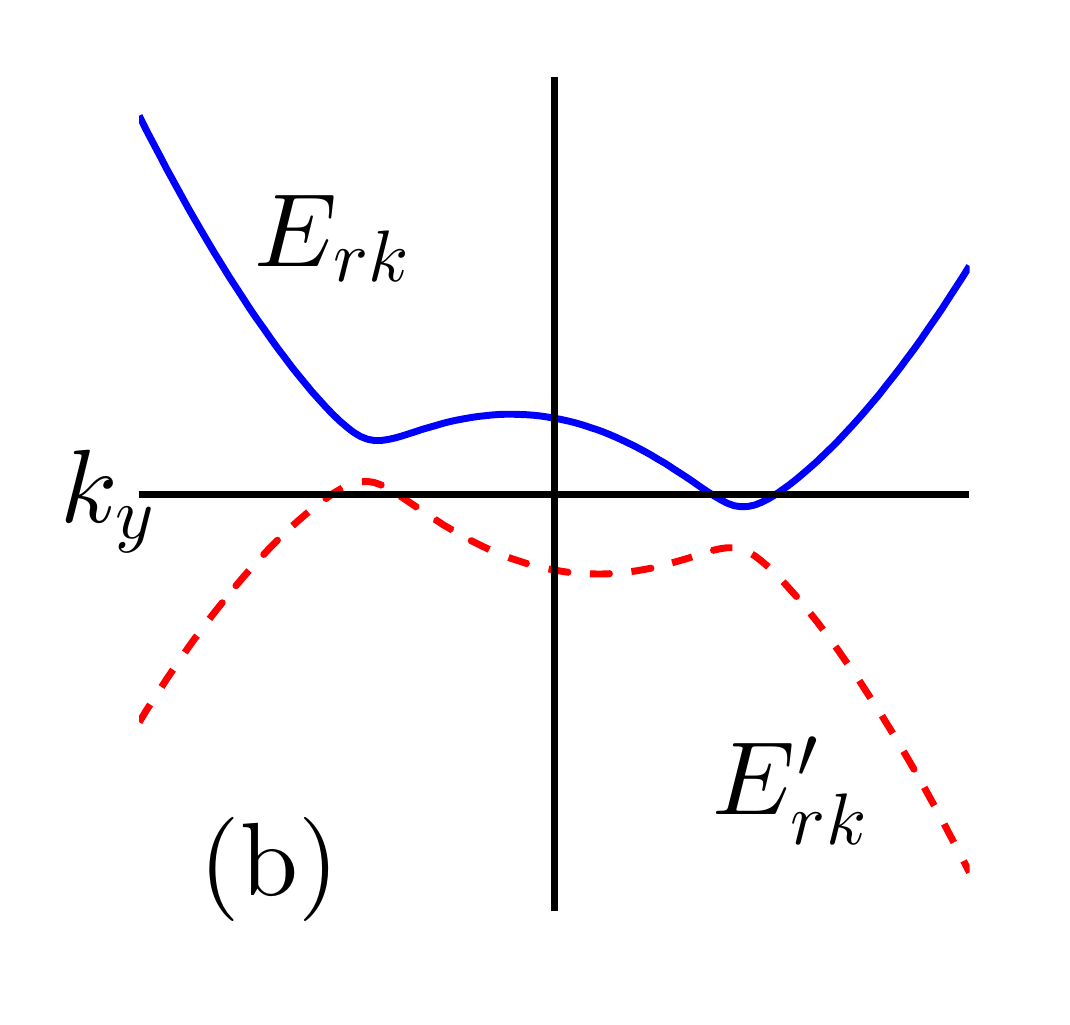}
\includegraphics[scale=.35]{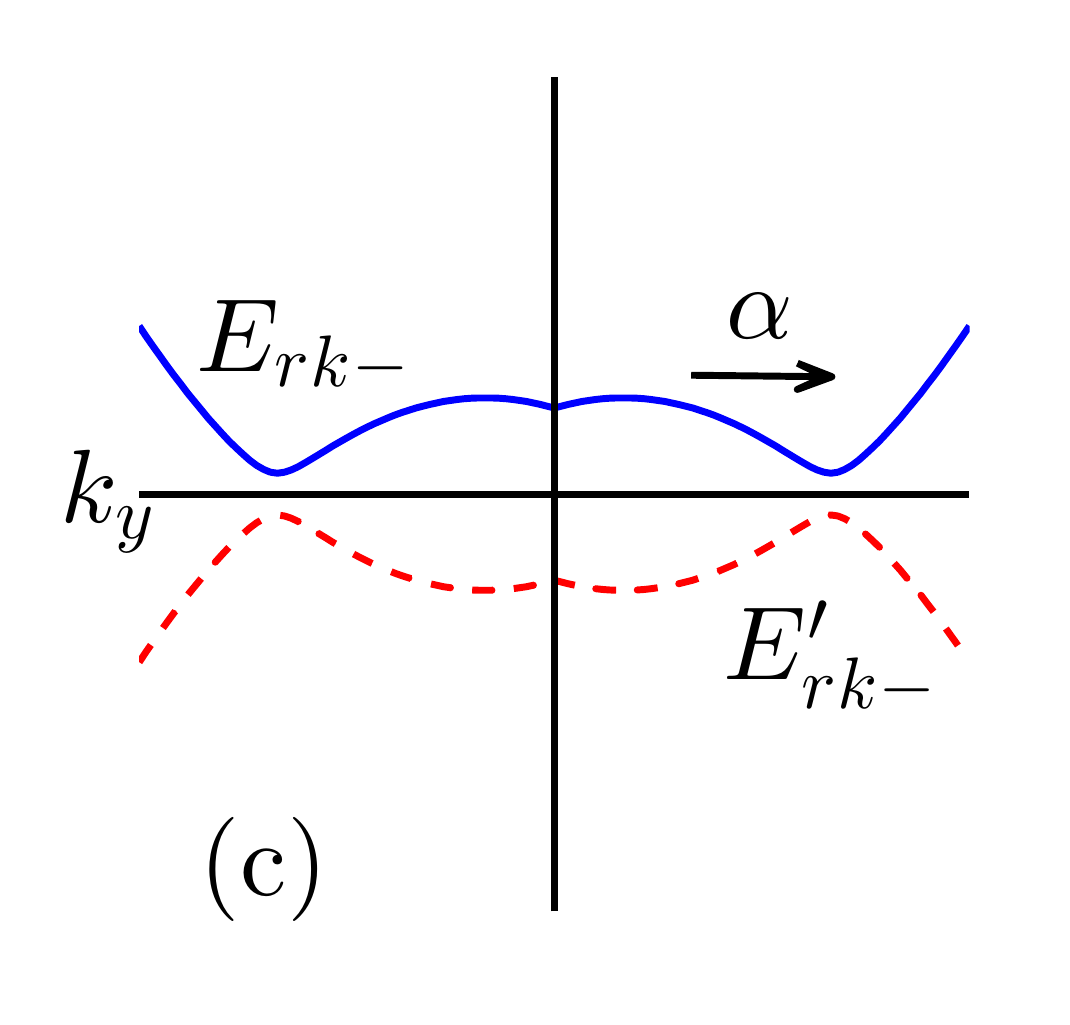}
\includegraphics[scale=0.35]{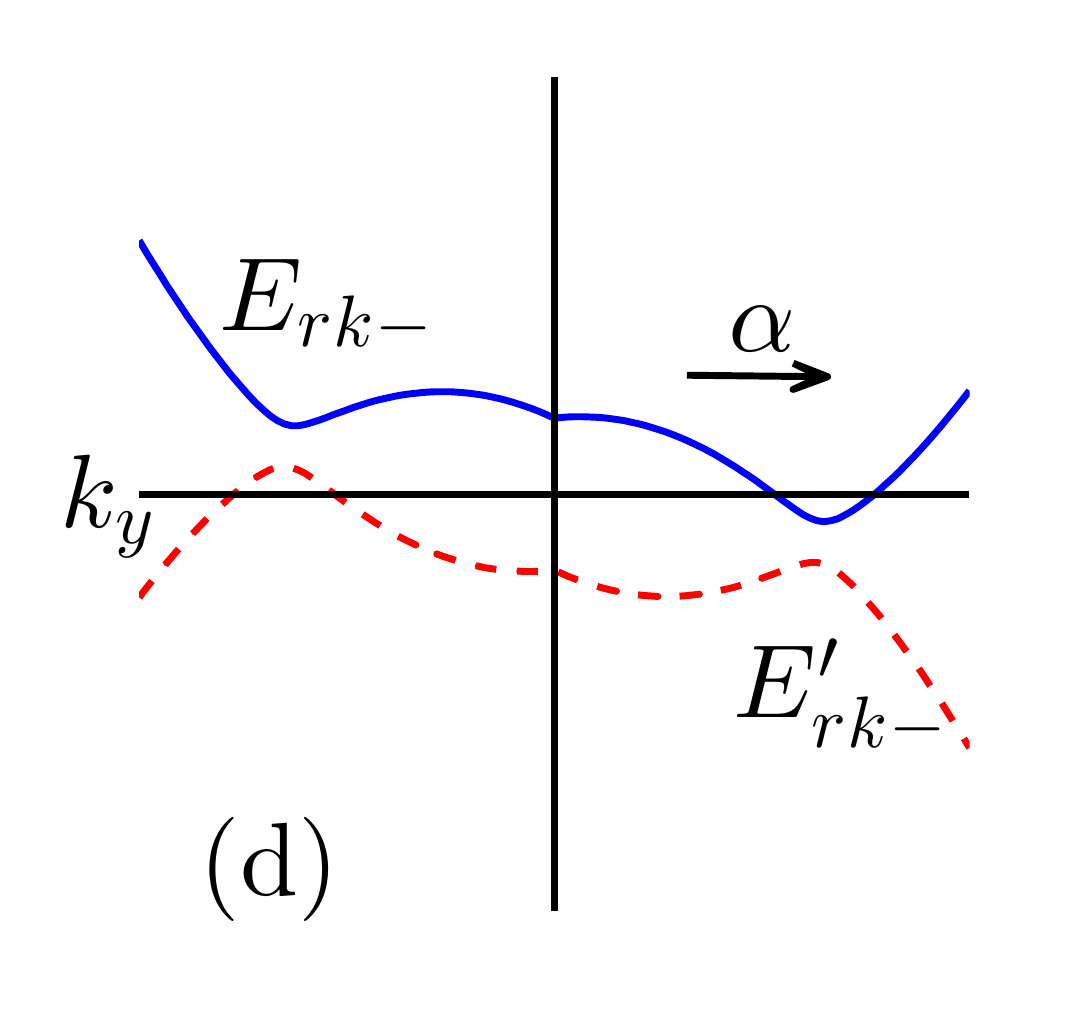} 
\caption{(Color online) 
Schematic diagrams showing the excitation spectrum $E_{r\bk}$ with $\Omega\neq 0$ for 
(a) a gapped SF at $r=0$ for $\alpha=0$, 
(b) a gapless SF at $r>0$ for $\alpha=0$, 
(c) a gapped SF at $r=0$ for $\alpha>0$, and
(d) a gapless SF at $r>0$ for $\alpha>0$.  
The broken Cooper pairs occupy $\mathbf{k}$-space regions with negative/positive 
quasi-particle/quasi-hole energies.
\label{fig:Eph}
} 
\end{figure}

The pair-breaking mechanism is based on the Coriolis effects and it can be 
analyzed by looking at the excitation spectra shown in Fig.~\ref{fig:Eph}.
In the $\Omega \to 0$ limit shown in Fig.~\ref{fig:Eph}(a), both the quasi-particle 
$
E_{r\bk}=\sqrt{\xi_{r\bk}^{2}+|\Delta_r|^{2}}-\Omega L_{r\bk}^{z}
$
and quasi-hole  
$
E'_{r\bk}=- \sqrt{\xi_{r\bk}^{2}+|\Delta_r|^{2}}-\Omega L_{r\bk}^{z}
$ 
excitation energies are particle-hole symmetric around the zero-energy 
axis along the $k_y$-direction, corresponding to an ideal situation for 
the formation of Cooper pairs with zero center of mass momentum. 
On the other hand, while $\Omega \neq0$ still preserves the particle-hole 
symmetry, it breaks the symmetry between the time-reversed pairing 
states $(\bk,\uparrow; -\bk,\downarrow)$, leading to asymmetric excitation 
energies that depend on the direction of momentum as shown in 
Fig.~\ref{fig:Eph}(b). Increasing $\Omega$ increases this asymmetry, and it 
eventually leads to negative/positive quasi-particle/quasi-hole energies 
and broken pairs in the ground state, i.e., the $\bk$-space regions with 
$E_{r\bk}<0$ and $-E'_{r,-\bk}>0$ are not occupied by pairs but by single 
particles. These $\mathbf{k}$-space regions $k_{1}<k<k_{2}$ are found 
by setting $E_{r \bk}=0$, leading to
$
k_{1,2}^{2}=2M \mu_r+2M^2\Omega^{2}r^{2} \pm 2M\sqrt{\mathrm{A}_r}
$
with 
$
\mathrm{A}_r=2M\Omega^{2}r^{2} \mu_r + M^2\Omega^{4}r^{4} - |\Delta_r|^{2}.
$
Thus, $\mathrm{A}_r\geq 0$ is a necessary condition for the emergence 
of local phases with gapless excitations, i.e., a gapless SF (gSF) or N phase. 

This LDA analysis suggests that the Cooper pairs are robust for sufficiently 
slow rotations, and low $\Omega$ has no effect whatsoever on $\mu$, $n_r$ 
and $|\Delta_r|$, for which case the entire gas remains to be a disk-shaped SF 
with gapped excitations, and its edge is located at $R_{O}=R_{F}$.  
Once the critical rotation frequency $\Omega_c$  for the onset of 
pair breaking is reached, the gapless excitations naturally appear at the edge of the gas, 
i.e., $\mathrm{A}_{r=R_{F}}=0$, suggesting that the radius of the N gas having 
the same $\mu$ coincides with the Thomas-Fermi radius at $\Omega_c$, 
i.e., $
R_{O}^{0}(\alpha=0,\Omega=\Omega_{c})=R_{F}.
$  
Thus, transitions from SF to gSF and gSF to N phase first emerge at the edge 
of the system, and then the SF region contracts toward the trap center as a 
function of increasing $\Omega$. 
In contrast to the 3D case, we are able to obtain an analytic expression 
$
\Omega_{c}=\omega\sqrt{E_{b}/(2E_{F})}
$
for arbitrary $E_b$, and given the upper bound on $\Omega\leq\omega$, 
it follows that pairs are robust against rotation for $E_{b}>2E_{F}$. The fact that 
$\mu$ also changes sign at $E_{b}=2E_{F}$ is not a lucky coincidence, since 
the pairs are known to be robust for $\mu<0$ in earlier works on a 3D Fermi 
gas~\cite{Urban2008, Iskin2009}.

When $\Omega > \Omega_c$, we reach a generic conclusion that the trap profile 
consists of three regions, where the central SF core and the outer N edge are 
connected by a coexistence region gSF phase in between. 
Unlike the N region where $|\Delta_r| = 0$ and the associated mass-current 
density $J^{\theta}_r=Mn_r\Omega r$ is exactly of the form of a rigidly-rotating gas,
the gSF region is characterized by $|\Delta_r|>0$ and a partially-rotating gas with
$J^{\theta}_r < Mn_r\Omega r$.
While the N region expands both inwards and outwards as a function of increasing 
$\Omega$, the SF and gSF regions survive around the trap center even in 
the $\Omega\to\omega$ limit since the trap center is immune to the direct-effects of rotation.
We note in passing that sufficiently fast rotations may cause a kink in $n_r$ right 
at the SF-N interface (not shown), which is a direct consequence of the competition 
between the curvature of $n_r$ in the SF region which is not effected by rotation 
and that of the N region which increases with increasing $\Omega$.

\subsection{Trapped Fermi gas with Rashba coupling and rotation 
\\ ($\omega\protect\neq0$, $\alpha\protect\neq0$ and $\Omega\protect\neq0$)}
\label{sec:Igen}

Having shown analytically that the Rashba coupling and adiabatic rotation 
have competing effects on superfluidity, we are ready to discuss the generic 
case with arbitrary $\Omega\neq0$ and $\alpha\neq0$, for which case the main 
effect of their interplay is to form an outer ring-shaped N edge that is completely phase
separated from the central SF core by vacuum. This is clearly a remnant of 
the characteristic ring-shaped density profile that is found in Sec.~\ref{sec:NIgen}
considering a non-interacting Fermi gas.

For a given $\alpha$, since the entire SF gas is robust against rotation up to 
again a critical $\Omega<\Omega_{c}$, all of the physical quantities remain 
the same as the $\Omega = 0$ case discussed above in Sec.~\ref{sec:Isoc}.
Furthermore, the rigidity of the SF phase can also be used to determine $\Omega_{c}$ 
using the following recipe. We first remark that $\Omega_c$ is the lowest 
$\Omega$ satisfying the inequality condition
$
R_{O}^{0}(\alpha,\Omega=\Omega_{c}) \ge R_{O}(\alpha,\Omega=0),
$
i.e., the radius of the rotating N phase becomes equal or greater to the radius of the non-rotating 
SF phase. Then, we observe that the emergent outer N edge is connected (disconnected) 
to (from) the SF phase by an intermediate gSF region (vacuum) when this equality 
(inequality) condition is satisfied. Assuming $\Omega>\Omega_{c}$, the former profile is 
realized for $\alpha^{2}+2\mu/M>0$ with a non-rotating SF core that is characterized 
by $|\Delta_r|\neq0$ and $J^{\theta}_r=0$ near the trap center, 
an outer N edge that is characterized by $|\Delta_r|=0$ rotating rigidly with 
$J^{\theta}_r=Mn_r\Omega r$, and a gSF region in between that is characterized by 
$|\Delta_r|\neq0$ rotating partially with $J^{\theta}_r < Mn_r\Omega r$. 
See Fig.~\ref{fig:gapless_phase_diag}(b) for such a trap profile.  
On the other hand, assuming again $\Omega>\Omega_{c}$, the latter profile may 
be realized for $\alpha^{2}+2\mu/M<0$ (this condition is necessary but not sufficient) 
with a non-rotating central SF core and a rigidly-rotating ring-shaped N annulus. 
See Fig.~\ref{fig:LDAvsBdG}(d) for such a trap profile, where the region with 
$|\Delta_r|=0$ fully overlaps with the one with $J^\theta_{r}=Mn_r\Omega r$. 

Increasing $\Omega$ beyond $\Omega_c$ leads ultimately to the complete expulsion
of superfluidity from the entire trap at a higher critical rotation frequency $\Omega_s$. 
In contrast to the $\alpha \to 0$ limit discussed in Sec.~\ref{sec:Isoc} where the trap 
center remains a SF even at $\Omega = \omega$ thanks to its immunity to the 
direct-effects of rotation, $\alpha \ne 0$ allows such an expulsion since the ring-shaped 
N annulus that is formed by broken pairs is energetically more favorable than the gapped 
SF at the trap center. 

Next we calculate the critical rotation frequencies both for the onset of pair breaking 
and for the complete destruction of superfluidity in the system, and construct extensive 
phase diagrams and trap profiles for a wide-range of parameter regimes.

\begin{figure} 
\includegraphics[scale=0.59]{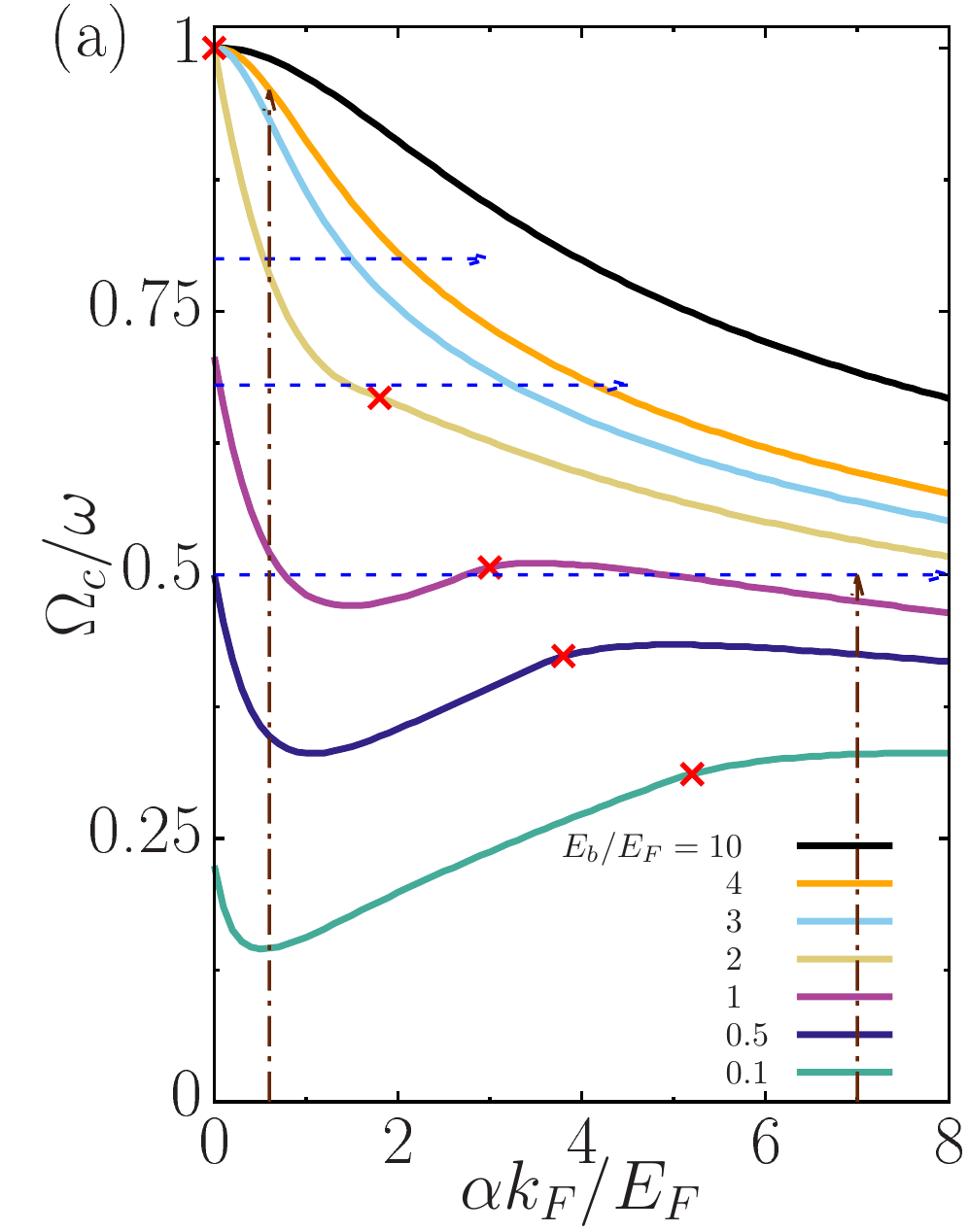}
\includegraphics[scale=0.59]{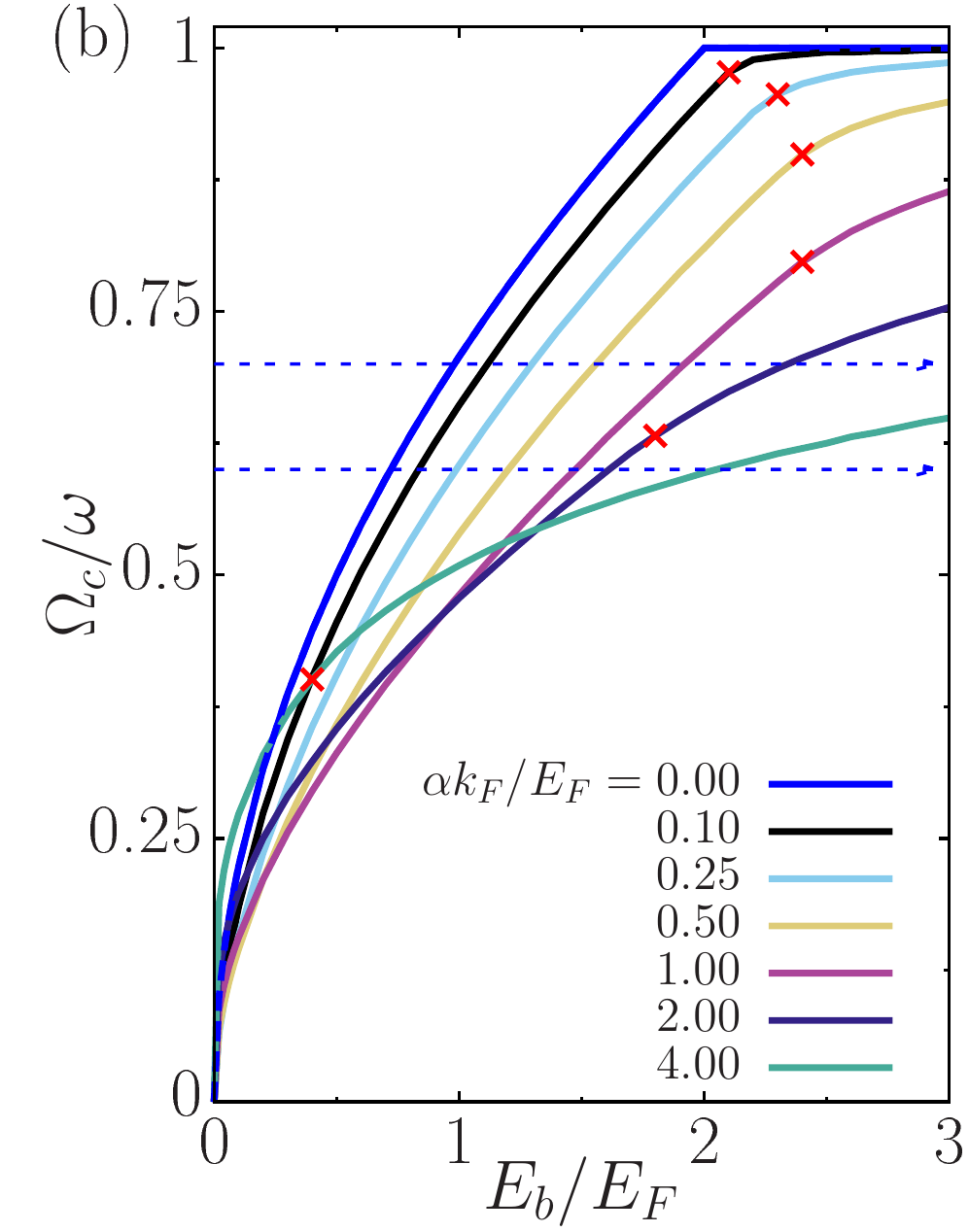} 
\caption{(Color online) 
The critical rotation frequency for the onset of pair breaking as a function of 
(a) Rashba coupling, and 
(b) binding energy
at $T = 0$.
The ``x'' marks indicate the point beyond which the ring-shaped N annulus emerges 
disconnectedly from the central SF core by vacuum (see the main text for details).
The radial phase profiles along the dashed blue (horizontal) and the dash-dotted brown (vertical) lines in (a) are shown, respectively, 
in Figs.~\ref{fig:n_alpha_r} and~\ref{fig:n_omega_r}. Similarly, the radial phase profiles along the dashed blue (horizontal) lines 
in (b) are shown in Fig.~\ref{fig:n_eb_r}.
\label{fig:omc}
} 
\end{figure}
\section{Numerical Analysis and Discussion} 
\label{sec:NAD}

In comparison to the non-interacting phase diagram presented in Fig.~\ref{fig:Om_c0}, 
here we show that the complex interplay between Rashba coupling, adiabatic rotation 
and interaction gives rise to much richer phase diagrams and trap profiles. 
For the onset of pair breaking and the associated emergence of an outer N edge, 
the inequality condition discussed above in Sec.~\ref{sec:Igen}, i.e.,
$
R_{O}^{0}(\alpha,\Omega=\Omega_{c}) \ge R_{O}(\alpha,\Omega=0),
$ 
turns out to be a very convenient one for determining $\Omega_{c}$. 
This is because even though $R_{O}(\alpha,\Omega=0)$ is still obtained through the 
numerical solutions of the self-consistency equations in this inequality, such an implicit
calculation is much more effective than an explicit one requiring self-consistent 
solutions of the trap profiles in the entire parameter space of interest. 
For the complete destruction of the central SF core, by noting that this is linked to 
the depletion of the central density as the trap center is immune to the direct-effects of 
rotation, solving simultaneously the conditions $n_{r = 0} \to 0$ and $\Delta_{\br = 0} \to 0$
turns out to be a very convenient approach for determining $\Omega_{s}$. 
We achieved this by first setting $\Delta_{\br=0} \to 0$ in the order-parameter 
equation and obtain $\mu$, and then extract $\Omega_s$ from the number equation 
by substituting $\mu$. Let us first construct $\Omega_c$ and $\Omega_s$ phase 
diagrams based on these two implicit conditions, and then verify their validity by 
looking explicitly at the trap profiles.

\subsection{Phase diagrams}
\label{sec:PD}

In Fig.~\ref{fig:omc}, we show $\Omega_{c}$ as a functions of $\alpha$ and $E_b$.  
In contrast to the $\alpha\to0$ limit discussed in Sec.~\ref{sec:Irot} where pairs 
are shown to be robust against rotation for $E_{b}>2E_{F}$, Fig.~\ref{fig:omc}(a) 
shows that $\alpha \ne 0$ eventually leads to pair breaking at some $\Omega_{c}$ 
no matter what $E_b$ is.
For instance, when $E_b\gtrsim2E_F$ as illustrated by the top three curves in this 
figure, $\Omega_c=\omega$ first remains unchanged up to some low but finite
$\alpha$ threshold, and then it decreases monotonically as pair breaking starts at 
lower $\Omega$ with increasing $\alpha$. In this strongly-interacting regime, 
the outer N edge always emerges disconnectedly from the central SF core by vacuum 
for $\Omega>\Omega_c$. 

On the other hand, when $E_{b}\lesssim2 E_F$ as illustrated by the bottom three 
curves in the same figure, $\Omega_{c}$ first decreases up to some critical $\alpha$ 
threshold, and then it increases with a minimum in between. This is a result 
of the competing Rashba effects discussed in Sec.~\ref{sec:Isoc}. 
The dominant effect at low $\alpha$ is that, by shifting the excitation minima 
to higher momentum states some of which are more susceptible to rotation, 
the interplay of Rashba coupling and adiabatic rotation makes pair breaking easier. 
In sharp contrast, increasing $\alpha$ causes two additional effects, 
i.e., it not only increases $|\Delta_r|$ and $n_r$ near the trap center but also 
decreases the radius of the gas, making pair breaking more difficult. 
Once these latter effects dominate beyond some intermediate $\alpha$ 
then $\Omega_{c}$ increases with $\alpha$ exhibiting a minimum in between. 
The location of the minimum shifts to higher $\alpha$ with increased $E_b$ 
as the latter effects become significant only at relatively higher $\alpha$. 

Even though it is not possible to obtain a closed-form analytic expression for $\Omega_c$ 
for arbitrary $\alpha$, we approximate the initial drop of $\Omega_{c}$ for low 
$\alpha$ using the following recipe. By neglecting the secondary effect of Rashba coupling 
on the radius of the gas, i.e., taking $R_{O}\approx R_{F}$, and inserting $\mu$ that 
is derived in Sec.~\ref{sec:Isoc} for a perturbative $\alpha$ into the dispersion relation, 
the value of $\Omega$ for which the dispersion relation becomes zero at the edge 
of the gas then gives
$
\Omega_{c}\approx \omega \sqrt{\alpha^{2}k_{F}^{2}/(2E_{F}^{2})
+E_{b}/(2E_{F})}-\omega \alpha k_{F}/(2E_{F}).
$ 
We checked the validity of this expression with the numerical data given in 
Fig.~\ref{fig:omc}(a), and find an excellent agreement between the two in the low 
$\alpha$ regime. For even higher values of $\alpha$, since the N edge is favored 
under adiabatic rotation, it is pushed to longer distances away from the trap center. 
This changes the mechanism of suppressing the SF phase by directly 
favoring previously unoccupied unpaired states and alters the behavior of 
$\Omega_{c}$ curve. Beyond the points marked by ``x'' in Fig.~\ref{fig:omc}, 
the curve first goes through a maximum and then decreases with increasing 
$\alpha$. In this regime, the ring-shaped N annulus emerges disconnectedly 
from the central SF core by vacuum. Increasing $E_b$ lowers the location of this
point in the $\alpha$ axis because the condition given above and the fact that $\mu$ 
decreases with increasing $E_b$, in such a way that it ultimately approaches to 
$\alpha=0$ and $\Omega=\omega$ for $E_b\gtrsim 2.5E_F$.

In Fig.~\ref{fig:omc}(b), we plot $\Omega_{c}$ as a function of $E_b$, showing that
$\Omega_c$ increases monotonically with $E_b$ until it saturates at $\Omega_c=\omega$. 
In addition, we see that increasing $\alpha$ shifts $\Omega_c$ curves upwards 
(downwards) in the low- (high-) $E_b$ regime as the Rashba coupling favors 
(supports) pairing (pair breaking). 
The points indicated by ``x'' again mark the critical $E_b$ threshold beyond which 
the ring-shaped N annulus emerges disconnectedly from the central SF core by vacuum. 
There is only one ``x'' mark up to $E_b \lesssim 2.5 E_F$, indicating that a gSF 
region never appears for $E_b\gtrsim 2.5E_F$. 
This is in agreement with our analysis given in Sec.~\ref{sec:Igen}, as the condition 
$\alpha^{2}+2\mu/M<0$ is satisfied for any $\alpha$ at higher $E_b$. 
However, there are two ``x''  marks for $E_b \sim 2 - 2.5 E_F$ at different $\alpha$ 
values, indicating that while the N region is initially disconnected from the SF one at 
lower $\alpha$, it first expands with increasing $\alpha$ and connects to the SF 
with an intermediate gSF region in between, and then it re-disconnects from the SF 
at a higher $\alpha$. This explains the structure of ``x'' marks in Fig.~\ref{fig:omc}(b) 
for $E_b \sim 2 - 2.5E_F$ curves, but it is not shown in Fig.~\ref{fig:omc}(a).

\begin{figure}
\includegraphics[scale=0.45]{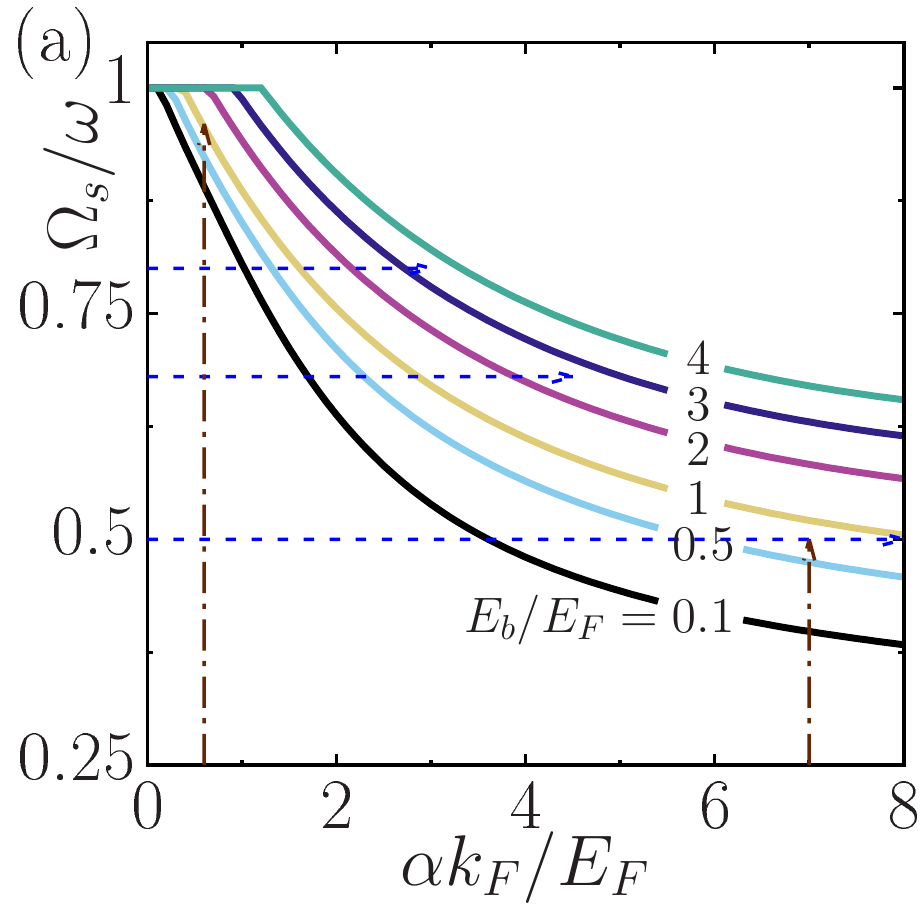} 
\includegraphics[scale=0.45]{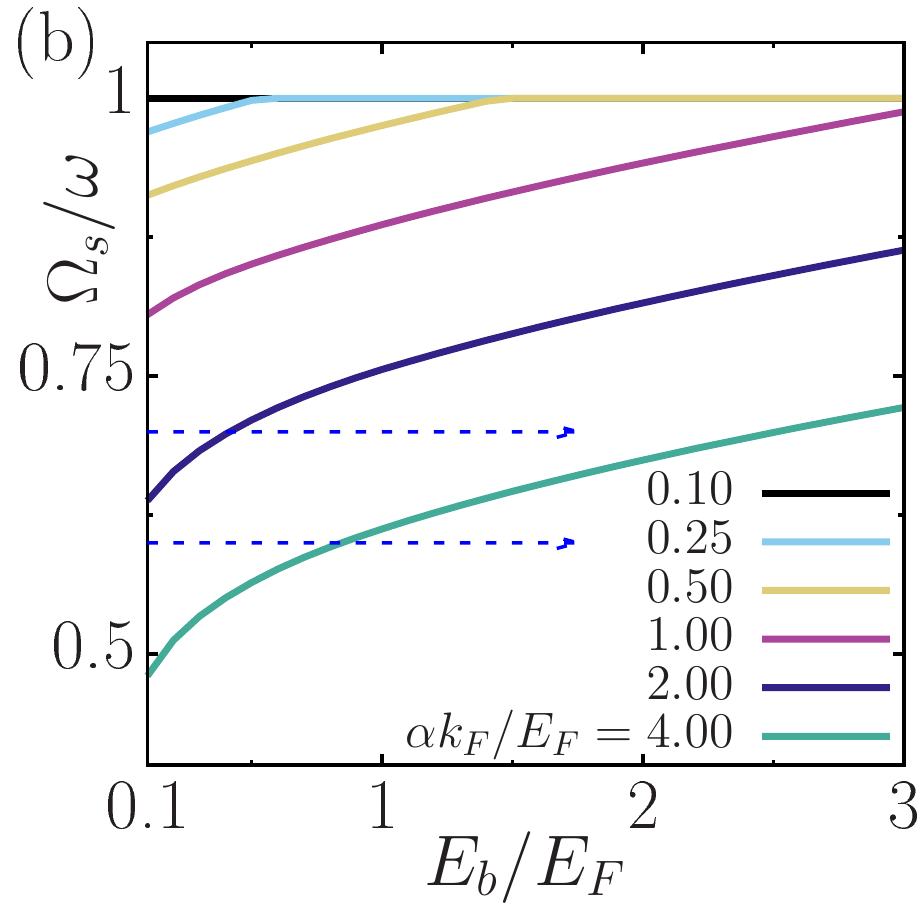}
\caption{(Color online) 
The critical rotation frequency for the complete destruction of superfluidity as a function of 
(a) Rashba coupling, and 
(b) binding energy
at $T = 0$.
The complete expulsion of SF core from the trap is accompanied by the depletion 
of the central density, requiring the interplay of Rashba coupling and adiabatic rotation. 
The horizontal and vertical lines are the same as the ones drawn in Fig.~\ref{fig:omc}.
\label{fig:omc2}
} 
\end{figure}

Lastly, in Fig.~\ref{fig:omc2}, we show $\Omega_{s}$ as functions of $\alpha$ and $E_{b}$.  
Increasing $E_b$ shifts $\Omega_s$ curves upwards in Fig.~\ref{fig:omc2}(a) 
as the complete destruction of the SF core is expected at higher $\Omega$. 
In addition, $\Omega_c=\omega$ remains unchanged at first up to some 
low but finite $\alpha$ threshold, and then it decreases monotonically as pair 
breaking starts at lower $\Omega$ with increasing $\alpha$.
This can also be seen in Fig.~\ref{fig:omc2}(b), where the critical curves move 
downward with increasing $\alpha$, and monotonically increase with increasing 
$E_b$. In addition, all of the curves saturate at $\Omega_s=\omega$ beyond 
the critical $E_b$ threshold once $E_b$ is high enough to protect the pairs against 
the effects of maximally-allowed rotation frequency $\omega$.

The phase diagrams shown in Figs.~\ref{fig:omc} and~\ref{fig:omc2} are some of our
most important contributions in this paper, as they can be used to predict all sorts of 
phase profiles in the trap for a wide-range of parameter regimes. Next we demonstrate this along 
several lines drawn in these figures by directly solving the self-consistency equations
for the trap profiles in the entire trap.

\begin{figure}
\includegraphics[scale=0.6]{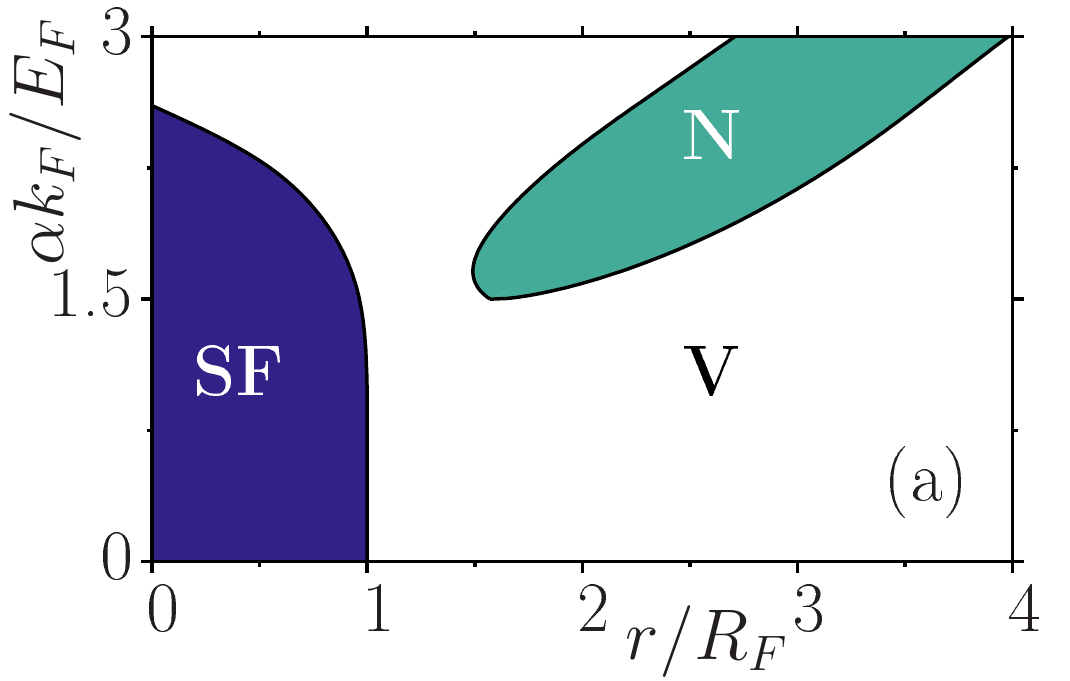}
\includegraphics[scale=0.6]{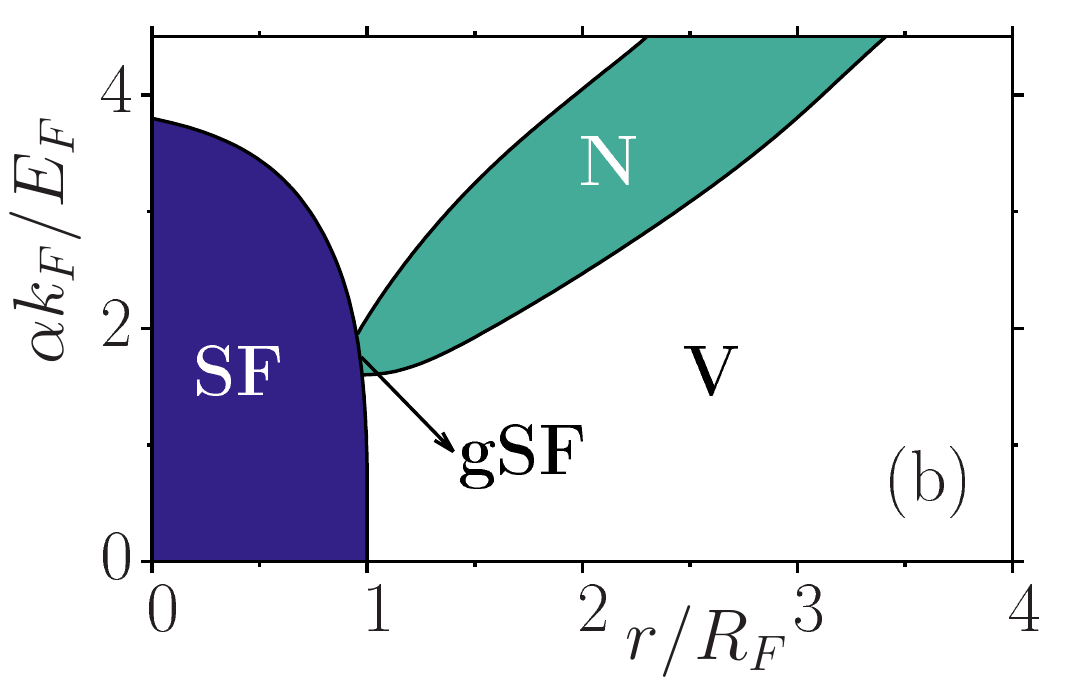}
\includegraphics[scale=0.6]{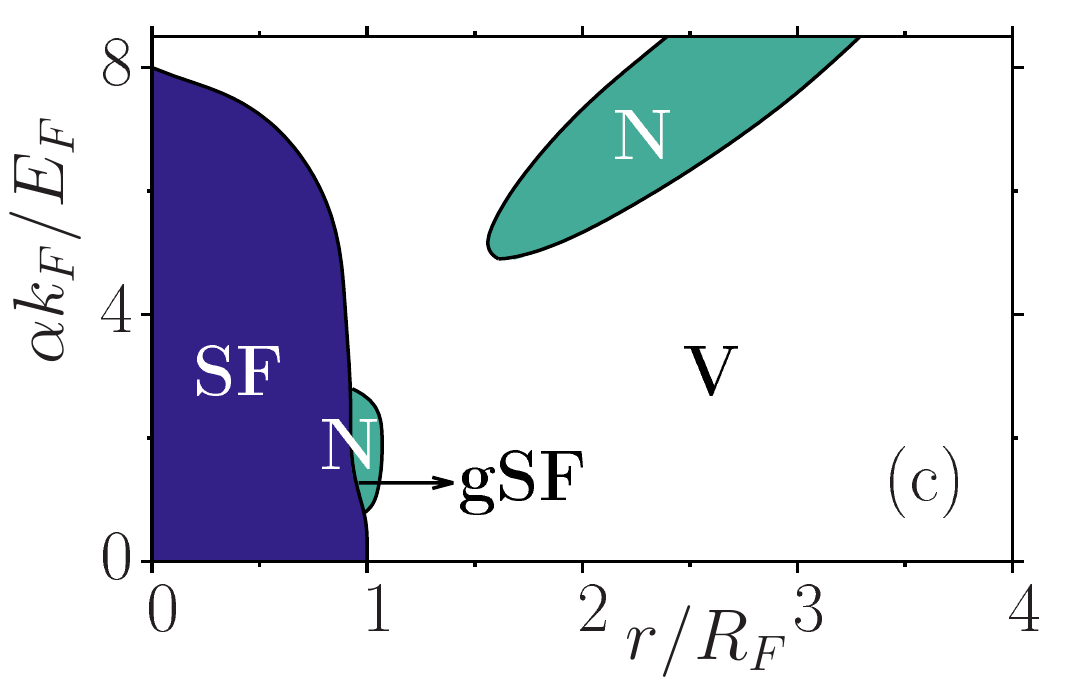}
\caption{ (Color online) 
Radial phase profiles at $T = 0$ with changing Rashba coupling for
(a) $E_{b}=3E_{F}$ and $\Omega=0.8\omega$, 
(b) $E_{b}=2E_{F}$ and $\Omega=0.68\omega$, and
(c) $E_{b}=1E_{F}$ and $\Omega=0.5\omega$.  
Here, the dark blue regions are gapped superfluid (SF), 
the green regions are normal (N) and the white regions are vacuum (V). 
The gapless superfluid (gSF) at the SF-N boundaries is not visible on this scale.
\label{fig:n_alpha_r}
} 
\end{figure}
\subsection{Trap profiles}
\label{sec:TP}

To demonstrate the practicality of $\Omega_{c}$ and $\Omega_{s}$ phase diagrams 
shown, respectively, in Figs.~\ref{fig:omc}(a) and~\ref{fig:omc2}(a), we first show 
the resultant phase profiles in Fig.~\ref{fig:n_alpha_r} along the dashed blue (horizontal) 
lines drawn in Figs.~\ref{fig:omc}(a) and~\ref{fig:omc2}(a). This figure explicitly
shows the emergence and disappearance of N and/or gSF regions in the trap with 
varying $\alpha$. 

For instance, we set $E_b = E_F$ and $\Omega = 0.5\omega$ in Fig.~\ref{fig:n_alpha_r}(c), corresponding to the bottom horizontal line in Fig.~\ref{fig:omc}(a), and illustrating 
an exemplary phase profile where the N region appears at the edge of the gas 
beyond a critical $\alpha$ threshold, as the first intersection point of the bottom 
horizontal line with $\Omega_c$ curve is to the left of the corresponding ``x'' mark.
There is a very thin gSF layer connecting SF and N regions but it is hardly visible 
in this scale. This $\alpha$ threshold is consistent with the first intersection point 
of the bottom horizontal line with $\Omega_c$ curve in Fig.~\ref{fig:omc}(a).
Increasing $\alpha$ in Fig.~\ref{fig:n_alpha_r}(c) first expands and then contracts
the N region. The disappearance of the N edge is due to the interplay
of competing Rashba effects discussed in Sec.~\ref{sec:PD}, and its $\alpha$ threshold
is again consistent with the second intersection point of the bottom horizontal line 
with $\Omega_c$ curve. Increasing $\alpha$ further, we see that an N region 
that is disconnected from central SF core reappears, forming a ring-shaped annulus
in the trap, with an increasing width as the SF region is gradually suppressed 
by $\alpha$. The reappearance $\alpha$ threshold is also consistent with the third 
intersection point of the bottom horizontal line with $\Omega_c$ curve. 
Lastly, Fig.~\ref{fig:n_alpha_r}(c) shows that the complete destruction of the 
SF region occurs beyond $\alpha\approx 8 E_F/k_F$, and its $\alpha$ threshold
is again consistent with the intersection points of the bottom horizontal line with 
$\Omega_s$ curve in Fig.~\ref{fig:omc2}(a). 

\begin{figure}
\includegraphics[scale=0.71]{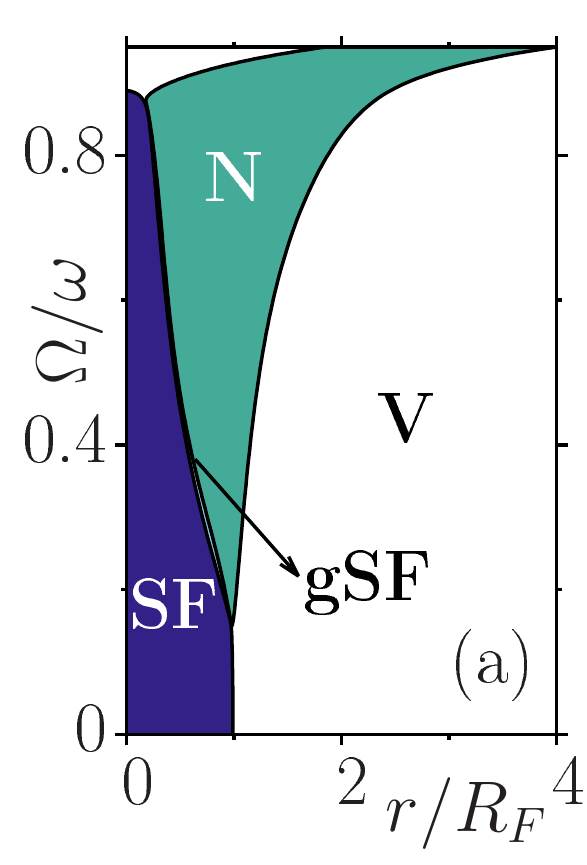}
\includegraphics[scale=0.71]{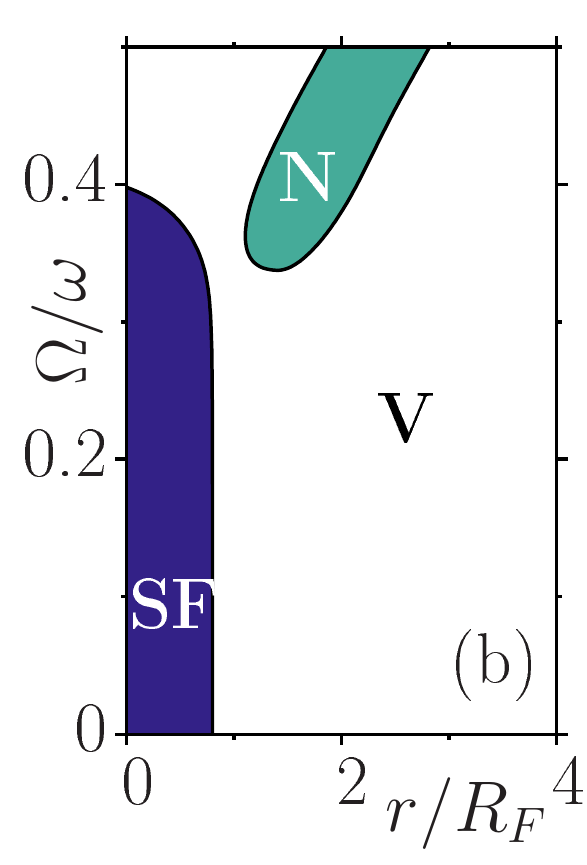}
\caption{ (Color online)
Radial phase profiles at $T = 0$ with changing rotation frequency for
(a) $E_{b}=0.1E_{F}$ and $\alpha=0.6 E_F/k_F$, and
(b) $E_{b}=0.1E_{F}$ and $\alpha=7 E_F/k_F$.  
Here, the dark blue regions are gapped superfluid (SF), 
the green regions are normal (N) and the white regions are vacuum (V). 
The gapless superfluid (gSF) at the SF-N boundary is hardly visible in (a) on this scale.
\label{fig:n_omega_r}
} 
\end{figure}

Similarly, we set $E_b = 2E_F$ and $\Omega = 0.68\omega$ in Fig.~\ref{fig:n_alpha_r}(b), corresponding to the middle horizontal line in Fig.~\ref{fig:omc}(a), and illustrating 
an exemplary phase profile where the N region first appears at the SF edge 
of the gas beyond a critical $\alpha$ threshold, as the intersection point of the middle 
horizontal line with $\Omega_c$ curve is to the left of the corresponding ``x'' mark, 
and then it separates and moves away from the SF region with an increasing 
width as the SF region is gradually suppressed by $\alpha$. 
However, we set $E_b = 3E_F$ and $\Omega = 0.8\omega$ in the remaining 
Fig.~\ref{fig:n_alpha_r}(a), corresponding to the top horizontal line in Fig.~\ref{fig:omc}(a), 
and illustrating an exemplary phase profile where the N region first appears 
away from the SF region beyond a critical $\alpha$ threshold, as the intersection point
of the top horizontal line with $\Omega_c$ curve is to the right of the corresponding 
``x'' mark, and then it moves further away from the SF region with again an increasing 
width as the SF region is gradually suppressed by $\alpha$. In both of these figures, 
the $\alpha$ thresholds for the appearance of the N edge are consistent with the 
only intersection points of the middle/top horizontal lines with $\Omega_c$ curves 
in Fig.~\ref{fig:omc}(a). In addition, the $\alpha$ thresholds for the complete 
destruction of the SF regions are again consistent with the intersection points 
of the middle/top horizontal lines with $\Omega_s$ curves in Fig.~\ref{fig:omc2}(a). 

\begin{figure}
\includegraphics[scale=0.6]{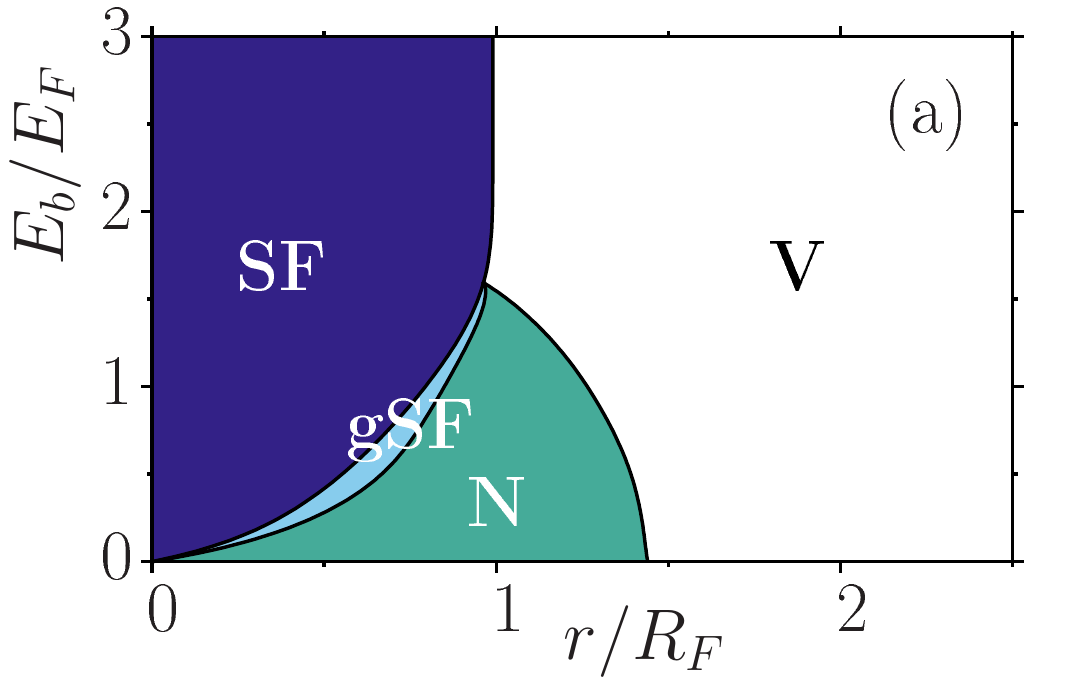}
\includegraphics[scale=0.6]{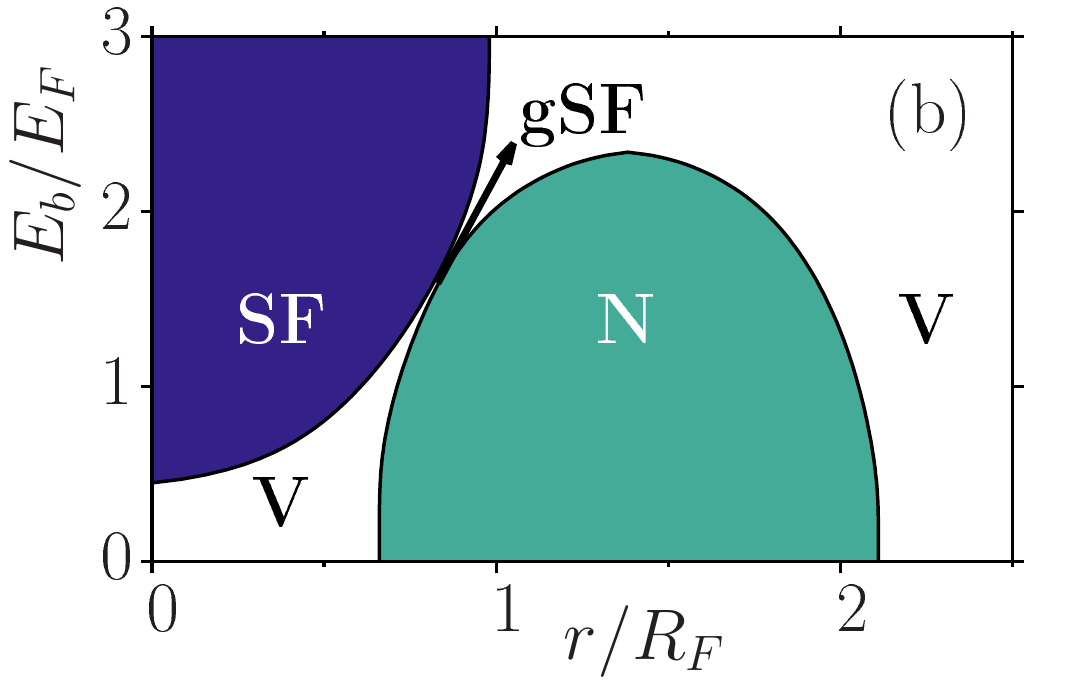}
\includegraphics[scale=0.6]{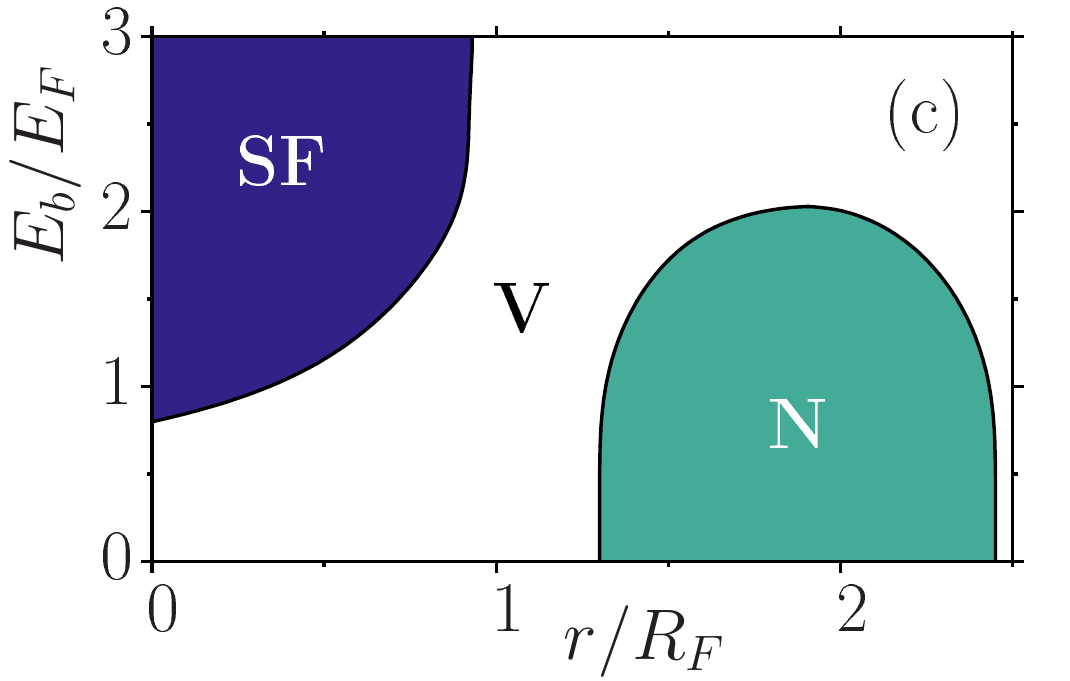}
\caption{ (Color online) 
Radial phase profiles at $T = 0$ with changing binding energy for 
(a) $\alpha=0.5 E_F/k_F$ and $\Omega=0.7\omega$, 
(b) $\alpha=2 E_F/k_F$ and $\Omega=0.7\omega$, and 
(c) $\alpha=4 E_F/k_F$ and $\Omega=0.6\omega$.  
Here, the dark blue regions are gapped superfluid (SF), the light blue regions are gapless 
superfluid (gSF) (not visible in (b)), the green regions are normal (N) and the white regions are vacuum (V). 
\label{fig:n_eb_r}
} 
\end{figure}

We next verify the consistency of our $\Omega_{c}$ and $\Omega_{s}$ phase 
diagrams along the dash-dotted brown (vertical) lines drawn in Figs.~\ref{fig:omc}(a) and~\ref{fig:omc2}(a)
with the resultant phase profiles shown in Fig.~\ref{fig:n_omega_r}. 
This figure explicitly shows the emergence and disappearance of N and/or gSF 
regions in the trap with varying $\Omega$. 
For instance, Fig.~\ref{fig:n_omega_r}(a) exemplifies the low $\alpha < 2 E_F/k_F$ 
and/or low $E_b<2 E_F$ regimes, where the N and gSF regions appear simultaneously 
at the edge of the gas at a critical $\Omega$ threshold, as the intersection point
of the left vertical line with $\Omega_c$ curve is to the left of the corresponding 
``x'' mark, beyond which increasing $\Omega$ ultimately disconnects the N edge 
from the SF core which is accompanied by the disappearance of the gSF region. 
However, in the high $\alpha$ and/or high $E_b$ regimes, the N region first 
appears away from the SF region beyond a critical $\Omega$ threshold, 
as the intersection points of the right vertical line with $\Omega_c$ curve is to 
the right of the corresponding ``x'' mark, and then the N edge moves further away 
from the SF region with an increasing width as the SF region is gradually 
suppressed by $\Omega$. In addition, the $\Omega$ thresholds for the complete 
destruction of the SF regions are again consistent with the intersection points 
of the left/right vertical lines with $\Omega_s$ curves in Fig.~\ref{fig:omc2}(a). 

Similarly, to demonstrate the practicality of $\Omega_{c}$ and $\Omega_{s}$ phase 
diagrams shown, respectively, in Figs.~\ref{fig:omc}(b) and~\ref{fig:omc2}(b), we next 
show the resultant phase profiles in Fig.~\ref{fig:n_eb_r} along the dashed blue (horizontal) 
lines drawn in Figs.~\ref{fig:omc}(b) and~\ref{fig:omc2}(b). This figure explicitly
shows the emergence and disappearance of SF and/or gSF regions in the trap with 
varying $E_b$. For instance, Fig.~\ref{fig:n_eb_r}(a) exemplifies the low $\alpha$ 
and/or low $\Omega$ regimes, where the SF and gSF regions appear simultaneously 
at the trap center at a critical $E_b$ threshold, as the intersection point
of the top horizontal line with $\Omega_c$ curve is to the left of the corresponding 
``x'' mark, beyond which increasing $E_b$ ultimately turns the entire gas into a SF, 
i.e., once $E_b$ is high enough to protect all of the pairs against the effects of $\Omega$, 
which is accompanied by the disappearance of the gSF region. 
In addition, Fig.~\ref{fig:n_eb_r}(a) shows that the complete destruction of the 
N region occurs beyond $E_b\sim 1.6 E_F$, and its $E_b$ threshold
is again consistent with the intersection point of the top horizontal line with 
$\Omega_c$ curve in Fig.~\ref{fig:omc}(b). However, in the high $\alpha$ and/or 
$\Omega$ regimes, the SF region first appears away from the N region beyond a 
critical $E_b$ threshold, as the intersection points of the top horizontal line 
with $\Omega_c$ curve is to the right of the corresponding ``x'' mark. 
Note that since the $\alpha$ and $\Omega$ parameters of Figs.~\ref{fig:n_eb_r}(b) 
and~\ref{fig:n_eb_r}(c) are above the $\Omega_c^{0}(\alpha)$ curve shown 
in Fig.~\ref{fig:Om_c0}, while the entire gas forms a ring-shaped annulus in the 
$E_b \to 0$ limit, we see that a SF core that is disconnected from the outer 
N edge appears with an increasing width as the N region is gradually 
suppressed by $E_b$. In addition, the $E_b$ thresholds for the complete 
destruction of the N regions are again consistent with the intersection points 
of the top/bottom horizontal lines with $\Omega_c$ curves in Fig.~\ref{fig:omc}(b). 

\begin{figure}
\includegraphics[scale=0.6]{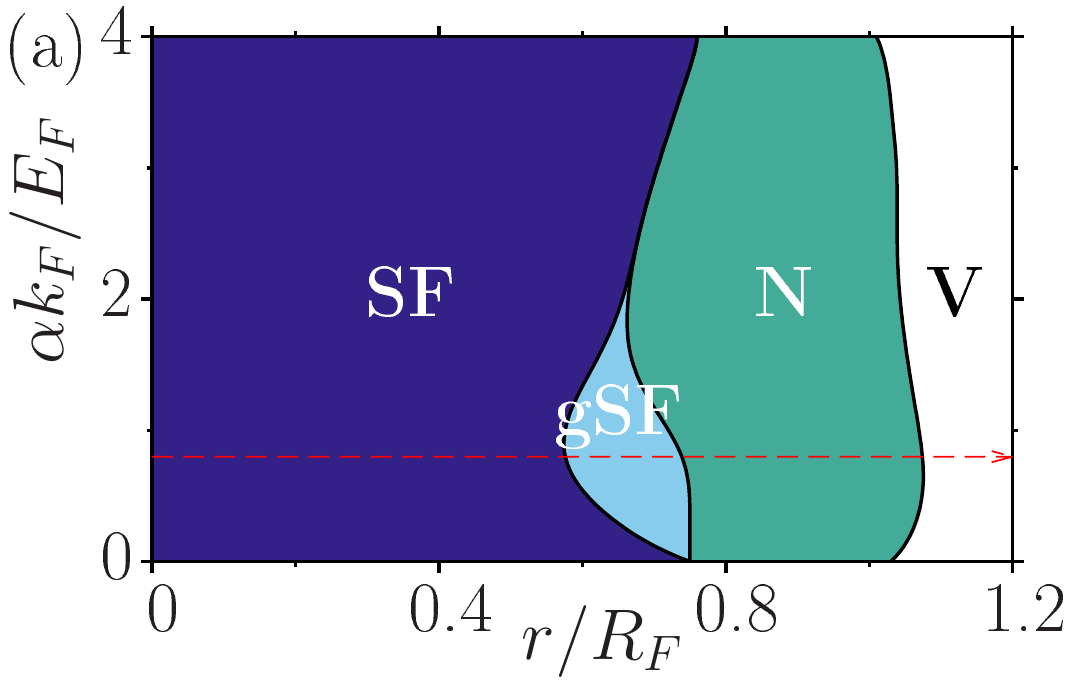} 
\includegraphics[scale=0.6]{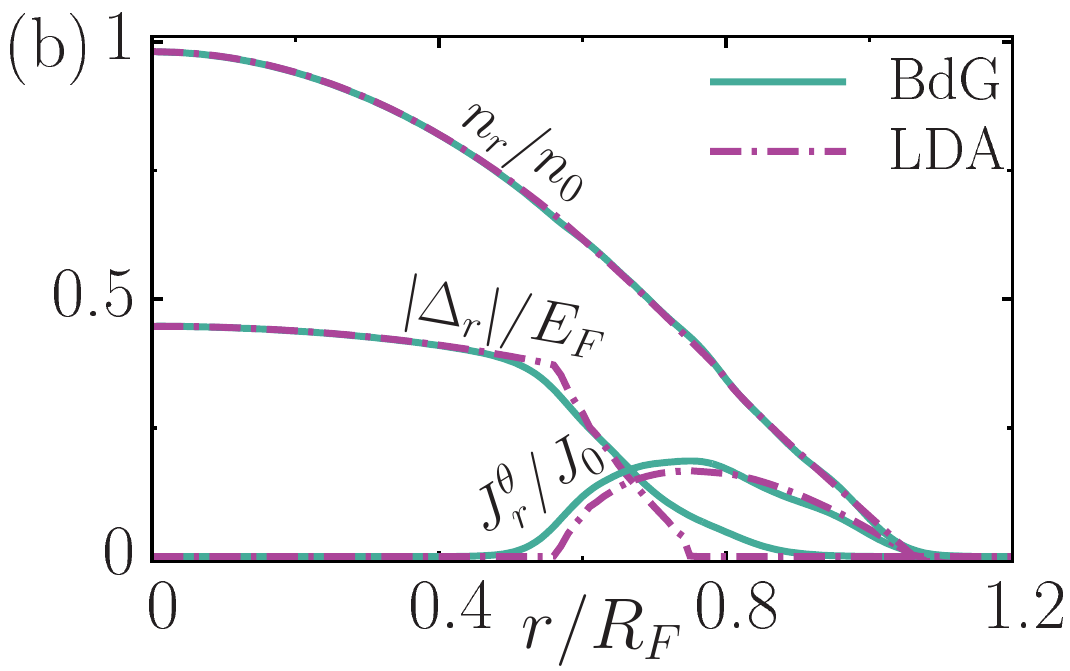}
\caption{(Color online) 
(a) Radial phase profiles at $T = 0$ with changing Rashba coupling for $E_{b}=0.1E_{F}$ 
and $\Omega=0.3\omega$. The interplay of Rashba coupling and adiabatic rotation 
increases the gSF region.
(b) An exemplary radial trap profile is plotted along the red dashed line in (a)
where $\alpha=0.8 E_F/k_F$, comparing the results of LDA (dot-dashed) and 
BdG (solid lines, $N=500$) approaches for the order parameter, number density 
and mass-current density. 
\label{fig:gapless_phase_diag}
} 
\end{figure}

We remark here that the interplay between the Rashba coupling and adiabatic rotation 
may also favor a much wider gSF region in the trap, especially in the low $E_b$ regime
with an intermediate $\alpha$. For instance, we set $E_b = 0.1E_F$ and 
$\Omega=0.3\omega$ in Fig.~\ref{fig:gapless_phase_diag}(a), illustrating an exemplary
phase profile where the emerging gSF region is sandwiched between the central 
SF core and the outer N edge. In addition, the radial trap profiles are also shown in 
Fig.~\ref{fig:gapless_phase_diag}(b) along the horizontal line drawn in 
Fig.~\ref{fig:gapless_phase_diag}(a), where we compare the results of LDA approach
with those of BdG one for $|\Delta_r|$, $n_r$ and $J_r^\theta$.
Unlike the N region where $|\Delta_r| = 0$ and the associated $J^{\theta}_r=Mn_r\Omega r$ 
is exactly of the form of a rigidly-rotating gas, while the SF region is characterized by 
$|\Delta_r|>0$ and $J^{\theta}_r=0$, the gSF region is characterized by $|\Delta_r|>0$ 
and a partially-rotating gas with $J^{\theta}_r < Mn_r\Omega r$. In comparison to the LDA results, 
we find that the BdG ones exhibit a somewhat wider gSF region in the trap, but such
finite-size effects are expected to become more and more negligible with increasing $N$. 
See also Sec.~\ref{sec:MFT} for further comparison.

Having accomplished our primary objective, i.e., the exploration the trap profiles of a 2D 
Fermi gas in the presence of a Rashba coupling and under an adiabatic rotation, 
next we end the paper with a brief summary of our main findings and an outlook 
for further research.

\section{Conclusions and Outlook}
\label{sec:conclusions}

To conclude, here we considered a harmonically-trapped 2D Fermi gas in the 
presence of a Rashba coupling and under an adiabatic rotation. By adopting 
the BCS mean-field approximation for local pairing, and BdG and LDA approaches 
for the isotropic trap, we not only extended our earlier LDA analysis to a wider 
parameter regime but also compare its results with those of BdG approach 
showing a perfect agreement for the most parts. For instance, we first analyzed 
a non-interacting system and showed that the competition between the 
effects of Rashba coupling on the LDOS and the Coriolis effects caused by 
rotation gives rise to a characteristic ring-shaped density profile that survives 
at experimentally-accessible temperatures. Furthermore, we also showed that 
the Rashba splitting of the Landau levels takes the density profiles on a ziggurat 
shape in the rapid-rotation limit. We then analyzed an interacting system, 
and studied the pair-breaking mechanism that is induced by the Coriolis effects 
on superfluidity, where we calculated the critical rotation frequencies both for 
the onset of pair breaking and for the complete destruction of SF regions 
in the system. We also constructed extensive phase diagrams consisting of
non-rotating gapped SF, partially-rotating gSF and rigidly-rotating N regions, and 
used these diagrams to predict all sorts of phase profiles in the trap for a wide-range 
of parameter regimes, where the aforementioned competition may, e.g., favor an 
outer N edge that is completely phase separated from the central SF core by vacuum. 

This problem offers many extensions for future research. For instance, the interplay 
between Rashba coupling and adiabatic rotation in a population-imbalanced 
Fermi gas is a promising one, as these systems manifest topologically 
non-trivial SF phases in the non-rotating limit~\cite{Zhou2011a}. Since a finite 
population imbalance is analogous to a perpendicular Zeeman field, we expect 
not only rich spin-polarization textures reminiscent of skyrmions but also diverse 
density profiles including the formation of successive ring-shaped regions.
Another promising direction is to analyze the effects of real- and/or momentum-space
anisotropies on the trap profiles, i.e., trapping potential and/or SOC. In the case 
of an anisotropic SOC, we again expect exotic density profiles including not only 
the ring-shaped ones with more than one local maxima in general but also 
isolated pocket-shaped ones (like a cut through the ring-shaped density) 
in the 1D SOC limit, i.e., an equal-weight combination of Rashba and Dresselhaus 
couplings.

\section{Acknowledgments}
\label{sec:ack}

This work is supported by the TUBITAK Grant No. 1001-114F232 and BAGEP award 
of the Turkish Science Academy, and E. D. is partially supported by a TUBITAK-2215 
Ph.D. Fellowship.

\appendix
\onecolumngrid

\section{Expansion of BdG equations in a simple-harmonic-oscillator basis}
\label{sec:angmombasis}

Using the conservation of total angular momentum about the axis of rotation, we first 
decompose the BdG eigenvectors into $|l|+1/2$ sectors with $|l|\geq 0$, and then expand 
the wave functions in terms of the angular-momentum basis of a 2D harmonic oscillator as 
\begin{eqnarray}
u_{\br \uparrow  \eta}&=& u^{l}_{\br \uparrow \mathfrak{m} } 
                       = \sum_{n} u^{l}_{n  \uparrow  \mathfrak{m}}   R_{r n}^{|l|}   \ee^{\ii l\theta}, \quad
\,\, u_{\br \downarrow\eta}= u^{l+1}_{\br \downarrow \mathfrak{m}}
                       = \sum_{n} u^{l+1}_{n+1\downarrow\mathfrak{m}}R_{r n+1}^{|l|+1} \ee^{\ii( l+1)\theta}, \\
v_{\br \downarrow\eta}&=& v^{l}_{\br \downarrow \mathfrak{m}}
                       = \sum_{n} v^{l}_{n  \downarrow\mathfrak{m}}   R_{r n}^{|l|}     \ee^{\ii l\theta}, \quad  
v_{\br \uparrow  \eta}= v^{l+1}_{\br \uparrow \mathfrak{m}}
                       = \sum_{n} v^{l+1}_{n+1\uparrow  \mathfrak{m}}R_{r n+1}^{|l|+1} \ee^{\ii (l+1)\theta},
\end{eqnarray}
where the harmonic-oscillator wave functions are given by
\begin{equation}
\braket{ r\theta | nl} \equiv  R_{r n}^{|l|} \ee^{\ii l\theta} 
=(-1)^{(n-|l|)/2} \sqrt{\frac{[(n-|l|)/2]!}{\pi a_0^2 [(n+|l|)/2]!}} e^{il\theta} \left( \frac{r}{a_0} \right)^{|l|} e^{-r^2/(2a_0^2)}L^{|l|}_{(n-|l|)/2}(r^2/a_0^2).
\label{eq:psinl}
\end{equation}
Here, $a_0=1/\sqrt{M\omega}$ with $\hbar=1$ is the characteristic length scale for the harmonic-oscillator,
and the associated Laguerre polynomials $L^{|m|}_n(x)$ can be generated from the recursion relation
\begin{equation}
L^{|m|}_{n + 1}(x) =  \frac{1}{n + 1} \left[ (2n + 1 + |m|-x)L^{|m|}_n(x) - (n + |m|)L^{|m|}_{n - 1}(x) \right],
\label{eq:Lm} 
\end{equation}
where $n>1$, $L^{|m|}_0(x)=1$ and $L^{|m|}_1(x)=1+|m|-x$.

Using the orthogonality of the basis states, we obtain the following matrix-eigenvalue equation for 
each $|l|+1/2$ sector,
\begin{equation}
\label{eqn:bdg.nl}
\sum_{n'} \left( \begin{array}{cccc}
K_{nn'}^{l} & S_{n,n'+1}^{-,l+1} & 0 & \Delta_{nn'}^{l'} \\
 S_{n+1,n'}^{+,l} & K_{n+1,n'+1}^{l+1} & -\Delta_{n+1,n'+1}^{l+1} & 0 \\ 
0 & -(\Delta_{n+1,n'+1}^{l+1})^{*} & -K_{n+1,n'+1}^{-l-1} & -S_{n+1,n'}^{+,-l} \\ 
(\Delta_{nn'}^{l})^{*}& 0 &  -S_{n,n'+1}^{-,-l-1}  & -K_{nn'}^{-l'} 
\end{array} \right)
\left( \begin{array}{l}
u^{l}_{n'\uparrow \mathfrak{m}}\\
u^{l+1}_{n'+1\downarrow \mathfrak{m}}  \\
v^{l+1}_{n'+1\uparrow \mathfrak{m}} \\
v^{l}_{n'\downarrow \mathfrak{m}}
\end{array} \right) 
= E^{l}_\mathfrak{m}
\left( \begin{array}{l}
u^{l}_{n\uparrow \mathfrak{m}}\\
u^{l+1}_{n+1\downarrow \mathfrak{m}}  \\
v^{l+1}_{n+1\uparrow \mathfrak{m}} \\
v^{l}_{n\downarrow \mathfrak{m}}
\end{array} \right)
\end{equation}
with the matrix elements
\begin{eqnarray}
K_{nn'}^{l}       &=& \braket{nl| K_\br-\Omega L^z_\br| n'l} = [\omega(n+1)-\mu-\Omega l] \delta_{nn'}, \\
\Delta_{nn'}^{l}  &=& \braket{nl| \Delta_r| n'l} = 2\pi \int_{0}^{\infty}r dr \Delta_r R_{rn}^{|l|} R_{rn'}^{|l|}, \\
S_{nn'}^{-,l}     &=& \frac{\alpha i}{2 a_0} \braket{ nl-1|S_{\br}| n'l}  \nonumber 
                   =  \frac{\alpha i}{2 a_0} \int d^2\br R^{|l|-1}_{rn}\ee^{-\ii (l-1)\theta}S_\br R_{rn'}^{|l|}\ee^{\ii l\theta} \\
                  &=& \frac{\alpha i}{2 a_0}[\sqrt{(n'+l)/2}\delta_{n,n'-1}-\sqrt{(n'-l)/2+1}\delta_{n,n'+1}]=-S_{n'n}^{+,l-1}.
\end{eqnarray}
Recall that we restrict our numerical calculations to rotationally-symmetric solutions for $\Delta_r$. 
Similarly, expanding the order parameter, number density and mass-current density equations, we obtain
\begin{align}
  \Delta_r&=g \sum_{l \mathfrak{m}} \left( \sum_{n} u^{l}_{n\uparrow\mathfrak{m}} R_{rn}^{|l|} \sum_{n'} v^{*l}_{n'\downarrow \mathfrak{m}}R_{rn'}^{|l|} h^{l}_\mathfrak{m}+
 \sum_{n} u^{l+1}_{n+1\downarrow \mathfrak{m}}R_{rn+1}^{|l|+1} \sum_{n'}v^{*l+1}_{n'+1\uparrow \mathfrak{m}} R_{rn'+1}^{|l|+1} f^{l}_\mathfrak{m} \right), \\
n_r&=\sum_{l \mathfrak{m}} \left[ \left(\left|\sum_n u^{l}_{n\uparrow \mathfrak{m}}R_{rn}^{|l|}\right|^2+
\left|\sum_n u^{l+1}_{n+1\downarrow \mathfrak{m}}R_{rn+1}^{|l|+1}\right|^2 \right)f^{l}_\mathfrak{m}+\left( \left|\sum_n v^{l+1}_{n+1\uparrow \mathfrak{m}}R_{rn+1}^{|l|+1} \right|^2+
\left|\sum_n v^{l}_{n\downarrow \mathfrak{m}}R_{rn}^{|l|}\right|^2 \right) h^{l}_\mathfrak{m}  \right], \\
	J^\theta_r&=\sum_{l \mathfrak{m}}\left\{ \left(\frac{l}{r}\left|\sum_n u^{l}_{n\uparrow \mathfrak{m}}R_{rn}^{|l|}\right|^2+
	\frac{l+1}{r}\left|\sum_n u^{l+1}_{n+1\downarrow \mathfrak{m}}R_{rn+1}^{|l|+1}\right|^2 \right)f^{l}_\mathfrak{m}
	-\left( \frac{l+1}{r}\left|\sum_n v^{l+1}_{n+1\uparrow \mathfrak{m}}R_{rn+1}^{|l|+1}\right|^2+\frac{l}{r}\left|\sum_n v^{l}_{n\downarrow \mathfrak{m}}R_{rn}^{|l|}\right|^2 \right) h^{l}_\mathfrak{m} \right. \nonumber \\
	&+\left. 2M\alpha \left[\left|\sum_{n} (u^{l}_{n\uparrow \mathfrak{m}})^*R_{rn}^{|l|}\sum_{n'}u^{l+1}_{n'+1\downarrow \mathfrak{m}} R_{rn'+1}^{|l|+1}\right| f^{l}_\mathfrak{m} 
	+ \left|\sum_{n} (v^{l}_{n\downarrow
\mathfrak{m}})^*R_{rn}^{|l|}\sum_{n'} v^{l+1}_{n'+1\uparrow
\mathfrak{m}}R_{rn'+1}^{|l|+1}\right| h^{l}_\mathfrak{m} \right]\right\},
	\label{eq:currdensnl}
\end{align}
where $f^{l}_\mathfrak{m}=1-h^{l}_\mathfrak{m}=f(E_\mathfrak{m}^l)$. These are alternative 
to the BdG expressions given in the main text.
 
\twocolumngrid

\end{document}